\documentclass{scrartcl} 
\addtokomafont{sectioning}{\rmfamily} 
\usepackage[T1]{fontenc} 
\usepackage[utf8]{inputenc} 
\usepackage{graphicx} 
\usepackage[pdfusetitle]{hyperref} 
\usepackage{cite} 
\usepackage{amsmath} 
\usepackage{amssymb} 
\usepackage[italic]{hepnames} 
\usepackage{braket} 
\usepackage{slashed} 
\usepackage{siunitx} 
\DeclareSIUnit\fb{\femto\barn}
\usepackage{cleveref} 
\usepackage{xspace} 
\usepackage{perpage} 
\MakePerPage{footnote} 
\usepackage{rotating} 

\newcommand{\PphiDM}{\HepParticle{\phi}{}{}\xspace}
\newcommand{\MphiDM}{\ensuremath{m_{\PphiDM}}\xspace}

\newcommand{\mi}{m_{\chi_i}}
\newcommand{\mj}{m_{\chi_j}}

\title{Charming Dark Matter}
\author{Thomas Jubb, Matthew Kirk, Alexander Lenz}

\begin{document}

\begin{titlepage}
\begin{flushright}
IPPP/17/64
\\
JHEP12(2017)010
\end{flushright}
\vskip 1.5cm
\begin{center}

{\Large\boldmath\textbf{
Charming Dark Matter
}}
\vskip 1.3cm 
\textsc{
Thomas Jubb\footnote{thomas.jubb@durham.ac.uk}\(^{,a}\),
Matthew Kirk\footnote{m.j.kirk@durham.ac.uk}\(^{,a}\),
Alexander Lenz\footnote{alexander.lenz@durham.ac.uk}\(^{,a}\)
}
\vskip 0.5cm
\(^a\) IPPP, Department of Physics, Durham University, Durham DH1 3LE, UK
\vskip 5cm

{\large\textbf{Abstract}}
\\[10pt]
\parbox[t]{\textwidth}{
We have considered a model of Dark Minimal Flavour Violation (DMFV), in which a triplet of dark matter particles couple to right-handed up-type quarks via a heavy colour-charged scalar mediator.
By studying a large spectrum of possible constraints, and assessing the entire parameter space using a Markov Chain Monte Carlo (MCMC), we can place strong restrictions on the allowed parameter space for dark matter models of this type.
}

\vfill

\end{center}
\end{titlepage}

\section{Introduction}
\label{sec:intro}

The existence of dark matter (DM) has, since the its early days \cite{1937ApJ....86..217Z}, been established through a wide range of detection techniques, such as galactic velocity curves \cite{1970ApJ...159..379R,Begeman1991,Davis:1996xt,Allen:2007ue}, gravitational lensing \cite{Massey:2010hh}, and its effects on Big Bang Nucleosynthesis (BBN) and the Cosmic Microwave Background (CMB) \cite{Ade:2015xua}.
However the interactions of DM outside of its gravitational influence remain elusive, despite concerted efforts to measure its scattering in terrestrial targets (direct detection), its annihilation or decay products in the galaxy or beyond (indirect detection), or through its direct production in colliders \cite{Patrignani:2016xqp}.

One property of DM that is known to high precision is its abundance in the Universe today.
The evolution of the structure of the Universe is well modelled \cite{Springel:2005nw} and so the starting point for building a model of a particle DM is to consider how its interactions influence its relic abundance.
This leads to the concept of a thermal WIMP (weakly interacting massive particle), in which the DM achieves its relic abundance by decoupling from thermal equilibrium due to its annihilations or decay into standard model (SM) particles.  

Under the assumption of a WIMP particle interpretation of DM, we have no concrete indications of its mass, spin or interactions, which leaves tremendous freedom when building models.
Although many concrete models, e.g.\ supersymmetric theories, predict the existence of a DM candidate, so far these theories remain unverified and the phenomenology is often complicated by the large parameter spaces.
This represents a top-down approach in which DM arises naturally from a UV complete model.

An alternative approach to DM model building is from the bottom up, where a class of simple low energy models or interactions are considered simultaneously.
With no theoretical guiding principle, except gauge symmetry, on which to build such models, one must consider all possible models within a framework of a few assumptions.
This is most easily done using a set of EFT (effective field theory) operators.
Although an EFT may be perfectly valid for low energy experiments such as direct or indirect detection, they face problems with collider searches where the EFT approximation breaks down when heavy (\si{\TeV}) states become energetically accessible.

To ensure the model is valid up to high energies and above the reach of colliders, a commonly used tool is \emph{simplified models}, where often the mediator between the dark sector and the SM is included as a propagating mode.
Simplified models arose first in the context of collider searches for missing energy \cite{Abdallah:2014hon,Alves:2011wf,Alwall:2008ag,Aaboud:2016uro,DeSimone:2016fbz,Goodman:2011jq,Dreiner:2013vla}, but have recently been applied more widely to indirect and direct detection \cite{Abdallah:2014hon,Abercrombie:2015wmb,Boveia:2016mrp}, they allow for a much more broad study since the models themselves are sufficiently simple to contain only a few parameters which dominate the phenomenology of the DM.
This approach is not without criticism, and can at times be too simple, for example neglecting gauge symmetries and perturbative unitarity \cite{Goncalves:2016iyg,Kahlhoefer:2015bea,Englert:2016joy}.

Given the remarkable agreement between the SM and experimentally measured flavour observables it is natural for new physics (NP) models to enforce the \emph{minimal flavour violation} (MFV) assumption to suppress large NP effects \cite{Buras:2000dm,DAmbrosio:2002vsn}.
This assumption limits any quark flavour breaking terms to be at most proportional to the Yukawa couplings, which are responsible for  the small violation of the flavour symmetry in the SM.
This suppresses Flavour Changing Neutral Currents (FCNCs) and avoids strong constraints from rare decays and neutral meson mixing.
Nonetheless, some such observables are not reproduced by SM calculations and hence allow room for violations of MFV, for example \PDzero mixing which we discuss in \cref{Sub:Mixing}.

Some recent studies of simplified models have begun to go beyond the MFV assumptions.
This has been done in the context of down-type couplings \cite{Agrawal:2014aoa}, leptonic couplings \cite{Chen:2015jkt}, and more recently top-like \cite{Blanke:2017tnb}, or top and charm-like couplings \cite{Baek:2017ykw}.
Such models allow a continuous change from the MFV assumption to strong MFV breaking and can quantify the degree of MFV breaking permitted by the flavour constraints.
Similar scenarios have been studied in \cite{Agrawal:2011ze}, taking an overview of both lepton and quark flavoured DM and as well as a more focused study on top DM \cite{Kilic:2015vka}, both in the MFV limit.

Our aim in this paper is to extend the work of \cite{Blanke:2017tnb}, taking a more general approach to these kinds of beyond MFV models -- by placing fewer restrictions on the parameters of the model we include models with dominant up and charm type couplings, which give non-trivially different exclusion regions for different flavours of DM.
We note that a similar scenario, except with scalar dark matter and a fermionic mediator has been studied in \cite{Bhattacharya:2015xha}.
We aim to present statistically robust bounds from the entire parameter space based on a Markov Chain Monte Carlo (MCMC) approach.

We consider the following constraints in detail:
\begin{itemize}
\item Relic Density (\cref{Sec:RD}): We calculate the relic density of all three DM particles, including their widths and important coannihilation effects.

\item Flavour Bounds (\cref{Sec:Flavour}): We provide bounds on the model from neutral charm meson mixing, ensuring that the new physics does not exceed \(1\,\sigma\) of the experimental measurement of the mass difference between the heavy and light state of the \PDzero. We assess the possibility for constraints on rare decays like \HepProcess{\PDp \to \Ppiplus \Plepton \Plepton } but find that the NP is relatively unconstrained compared to mixing.

\item Direct Detection (\cref{sec:dd}): We calculate the event rate for the most excluding DD experiments (LUX and CDMSlite) over a large range of DM masses, including all relevant contributions up to one loop order (including gluon, photon, \PZ and Higgs exchange) and matching to a full set of non-relativistic form factors.

\item Indirect Detection (\cref{sec:id}): We include a large collection of constraints from the literature on the thermally averaged annihilation cross section \(\langle \sigma v \rangle\) for annihilation into various search targets such as photons, electrons, protons. We also include a study of gamma ray line searches, generated at the one-loop level in our model.

\item Collider Searches (\cref{Sec:Collider}): We perform a robust simulation of the dominant signals for a series of monojet, dijet and stop searches for ATLAS and CMS, including the widths of the particles.
\end{itemize}

We also compute constraints coming from electroweak precision observables, and perturbative unitarity.
We calculate the Peskin-Takeuchi parameters \cite{Peskin1990,Peskin1992a}, as these characterise the NP effects in much of the parameter space of our model, and replicate the literature result for a charged singlet scalar \cite{Grimus2008}.
We find that the \(S, T, U\) parameters provide no additional constraints beyond those previously described, and similarly perturbative unitarity calculations prove to be unconstraining and so we make no further mention of them.

Including the various constraints named above we can carry out an MCMC scan in order to identify the parameter space left open to the model -- our results are collected in \cref{Sec:Results}.
We find that current data can be used to restrict the parameter space where DM of this kind can exist, and go beyond the results of \cite{Blanke:2017tnb} by showing how renormalisation group mixing and running can dramatically improve the direct detection constraints, disfavouring attempts to avoid these limits by predominantly coupling to top quarks.

\subsection{The DMFV Model}
\label{Sec:DMFVTheModel}

The SM (without Yukawa couplings) has a flavour symmetry amongst the quarks -- there are no flavour violating effects such as FCNCs at tree level.
\emph{Minimal Flavour Violation} (MFV) is then the statement that the only flavour symmetry breaking terms in the BSM model are the Yukawa terms \cite{DAmbrosio:2002vsn}.

\begin{figure}[t]
\centering
\includegraphics[width=0.45\textwidth]{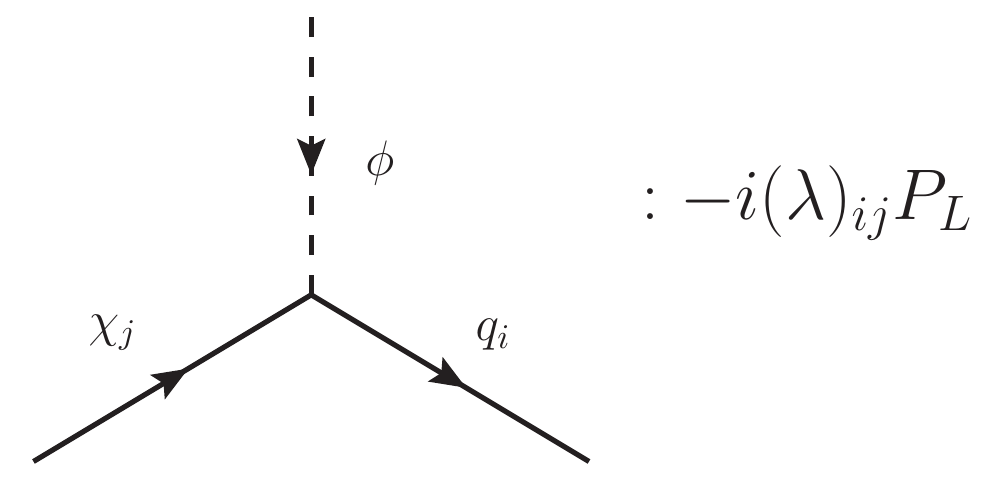}
\includegraphics[width=0.45\textwidth]{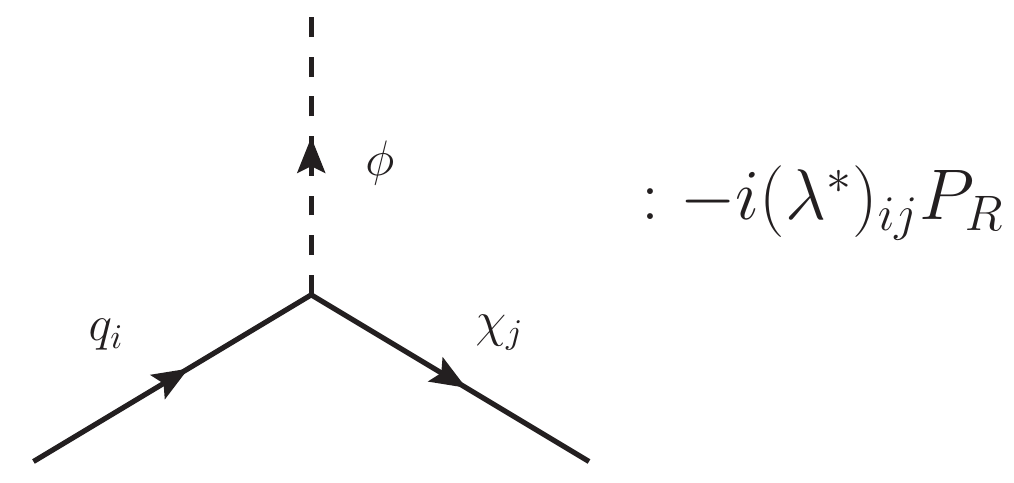}
\caption{Feynman rules for the interaction in \cref{Eq:Lint}}
\label{Fig:FeynmanRulesDMFV}
\end{figure}

In the model of \emph{Dark Minimal Flavour Violation} (DMFV) originally proposed in \cite{Agrawal:2014aoa}, the SM quark flavour symmetry is increased by the inclusion of a U(3) symmetry in the dark sector,
\begin{align}
\mathcal{S}_\text{flavour} = \text{U}(3)_{Q_L} \times \text{U}(3)_{u_R}  \times \text{U}(3)_{d_R}  \times \text{U}(3)_{\chi} \, ,
\end{align}
and the DMFV hypothesis is that this enlarged flavour symmetry is broken only by terms involving the quark Yukawas \emph{and} a new coupling matrix \(\lambda\).
In the original work \cite{Agrawal:2014aoa} \(\lambda\) coupled the DM to right-handed down-type quarks, whereas in this work we couple the DM to up-type right-handed quarks (the choice of right-handed quarks avoids having to introduce any non-trivial \(\text{SU}(2)\) structure).
In this model, we introduce four new particles --  a scalar \PphiDM that is colour and electrically charged, and a flavour triplet \(\chi_i\) that is a singlet under the SM gauge groups (which allows it to have a standard Dirac mass term).
In \cref{Fig:FieldCharges} we detail the behaviour under various gauge and other symmetry groups of the new particles and the coupling matrix -- the transformation of \(\lambda\) under the U(3) flavour symmetries is to be understood in the sense of a spurion field \cite{DAmbrosio:2002vsn}.
The new physics Lagrangian reads
\begin{equation}
\mathcal{L}_\text{NP} = \bar{\chi} (i \slashed{\partial} - m_\chi) \chi + D_\mu \PphiDM (D^\mu \PphiDM)^\dagger - \MphiDM \phi^\dagger \phi  - (\lambda_{ij} \APqu_{R,i} \chi_j \PphiDM + \text{h.c.})\ .
\label{Eq:Lint}
\end{equation}
giving the vertices shown in \cref{Fig:FeynmanRulesDMFV}.
Note that a coupling between the mediator and the Higgs as well as a mediator self-coupling are allowed by the symmetries of the model, but we neglect them in this work.
It was shown in \cite{Agrawal:2014aoa} that coupling matrix can be written in the form
\begin{equation}
\lambda = U_\lambda D_\lambda
\end{equation}
with the matrices \(D_\lambda\) and \(U_\lambda\) parametrised as (defining \(c_{ij} \equiv \cos \theta_{ij}, s_{ij} \equiv \sin \theta_{ij}\))
\begin{align*}
U_\lambda &= 
\begin{pmatrix}
c_{12} c_{13} & s_{12} c_{13} e^{-i \delta_{12}} & s_{13} e^{-i \delta_{13}} \\
-s_{12} c_{23} e^{i \delta_{12}} -c_{12} s_{23} s_{13} e^{i(\delta_{13} - \delta_{23})} & c_{12} c_{23} - s_{12} s_{23} s_{13} e^{i(\delta_{13} - \delta_{12} - \delta_{23})} & s_{23} c_{13} e^{-i \delta_{23}} \\
s_{12} s_{23} e^{i(\delta_{12}  + \delta_{23})} - c_{12} c_{23} s_{13} e^{i \delta_{13}} & -c_{12} s_{23} e^{i \delta_{23}} - s_{12} c_{23} s_{13} e^{i(\delta_{13} - \delta_{12})} & c_{23} c_{13}
\end{pmatrix} \, ,\\
D_\lambda &=
\begin{pmatrix}
D_{11} & 0 & 0 \\
0 & D_{22} & 0 \\
0 & 0 & D_{33} 
\end{pmatrix} \, ,
\end{align*}
where \(\theta_{ij} \in [0, \pi/4]\) to avoid double counting the parameter space, and we require \(D_{ii} < 4 \pi \) for a perturbative theory.

The presence of complex couplings (\(\delta_{ij} \neq 0\)) creates a violation of CP symmetry (note this is also permissible in the MFV assumption, so long as the complex phases are flavour-blind \cite{Isidori:2012ts}).
Due to the stringent constraints from electric dipole moments (EDM) in the presence of CP violation \cite{DAmbrosio:2002vsn} we will set \(\delta_{ij}=0\) throughout.
In total we then have a 10 dimensional parameter space
\begin{equation}
\{m_{\chi,1}, m_{\chi,2}, m_{\chi,3}, m_\phi, \theta_{12}, \theta_{13},  \theta_{23}, D_{11}, D_{22}, D_{33}\} \ .
\end{equation}
Other than those mentioned above, the only other limit we place on our parameters is \(m_\chi, m_\phi \gtrsim \SI{1}{\GeV}\), so that the DM is a conventional WIMP candidate and the mediator is sufficiently heavy to decay to at least the up and charm quarks.

\begin{table}
\centering
\begin{tabular}{| c | c  c  c | c |}
\hline
& \(\text{U}(3)_{u_R}\) & \(\text{U}(3)_\chi\)  & \(\text{U}(3)_c\) & \(\text{U}(1)_Q\) \\
\hline
\hline
\(u_{R}\) & \(\mathbf{3}\) & \(\mathbf{1}\) & \(\mathbf{3}\)& \(2/3\)
\\
\(\chi\) & \(\mathbf{1}\) & \(\mathbf{3}\) & \(\mathbf{1}\) & \(0\) \\
\(\PphiDM\) & \(\mathbf{1}\) & \(\mathbf{1}\) & \(\mathbf{3}\)& \(2/3\) \\
\hline
\(\lambda\) & \(\mathbf{3}\) & \(\mathbf{\bar{3}}\) & \(\mathbf{1}\) & \(0\) \\
\hline
\end{tabular}
\caption{The representation for the relevant symmetries of the particles introduced in the DMFV model, along with the coupling matrix \(\lambda\) and the SM right-handed quarks.}
\label{Fig:FieldCharges}
\end{table}

Although the masses of the DM fields and mediator field are in principle arbitrary free parameters, one must impose \(m_{\chi,\text{min}} < m_\phi + m_q\) (where \(m_q\) is the lightest quark to which \(m_{\chi,\text{min}}\) couples) to ensure \(\chi\) cannot decay.
Similarly we must have \(m_\phi > m_{\chi, \text{min}} + m_q\), which ensures the mediator has at least one decay channel and prevents it obtaining a relic abundance itself.

It can be shown additionally that a residual \(\mathbb{Z}_3\) symmetry exists in the model \cite{Agrawal:2014aoa, Batell:2011tc}, which prevents either \(\chi\) or \PphiDM decaying into purely SM particles.
This useful symmetry argument ensures the relic DM (the lightest of the three) is completely stable even once non-renormalisable effects are considered.
It is possible for the heavier \(\chi\) fields to decay to the lightest \(\chi\) (DM) -- in fact the rate of such decays are always large enough to totally erase the relic density of the heaviest two DM.

Finally, we briefly mention some interesting behaviour of the widths of our new particles.
First, the mediator width \(\Gamma_{\PphiDM}\) can be shown to be very narrow, with \(\Gamma_{\PphiDM} / \MphiDM \leq \frac{9}{128 \pi} \lesssim \SI{1}{\percent} \) even in the limit of non-perturbative couplings.
Secondly for small mass splittings (\(\mi = \mj (1 + \epsilon)\)) the decay rate \(\chi_i \to \chi_j + \Pquark \APquark\) scales as \(\epsilon^5\), which is important when we consider the relic abundance of the different DM species.

\section{Relic Density}
\label{Sec:RD}

\subsection{Relic Density with Coannhilations}
\label{sub:RD_Intro}

As mentioned in the introduction, the relic density (RD) of DM is currently measured to a very high accuracy by the Planck collaboration \cite{Ade:2015xua}, and this must be reproduced by any self-respecting DM model.
We will assume that dark matter is produced thermally via a freeze-out mechanism, but the resulting constraints may be alleviated via non-thermal mechanisms as in asymmetric dark matter \cite{Petraki:2013wwa,Baer:2014eja}.
We leave this possibility to further studies.

In our model with three possible DM candidates, with potentially almost degenerate masses, we follow the results of \cite{Griest:1990kh} -- Section III in particular deals with the effects of coannihilations (processes with \(\chi_i \chi_j \to \mathrm{SM}, i \neq j\)).
In that work, the authors describe how coannihilations can be very important, and can be included in the ``standard'' computation \cite{Busoni:2014gta,Bertone:2004pz,Gondolo:1990dk} of relic density through the use of an effective annihilation cross-section \(\langle \sigma v \rangle_\text{eff}\), defined in eq.\ (12) of \cite{Griest:1990kh}.
We will not reproduce all the detail from that paper here, but summarise the key results.

To compute the relic density, one first finds the freeze-out temperature \(x_f \equiv m / T_f\) by solving the equation 
\begin{align}
e^{x_f} = \sqrt{\frac{45}{8}} \frac{g_\text{eff} m_{\chi} M_\text{pl} \langle \sigma v \rangle_\text{eff}}{2 \pi^3 g_*^{1/2} x_f^{1/2}} \, ,
\label{Eq:FreezeoutTemp}
\end{align}
with  \(g_\text{eff}\) an effective number of degrees of freedom of the near-degenerate DM candidates, \(M_\text{pl}\) the Planck mass, \(g_*\) the total number of relativistic degrees of freedom at freeze-out.
The relic density itself can then be written 
\begin{align}
\Omega h^2 = 2 \times 1.04 \times 10^9 \frac{x_f}{\sqrt{g^*} M_\text{pl} \left(a_{ii} I_a + 3 b_{ii} I_b / x_f  \right)} \ ,
\label{Eq:TotalOmega}
\end{align}
where \(a_{ii}\) and \(b_{ii}\) are the s-wave and p-wave terms of \(\langle \sigma v \rangle_{ii}\) (the cross section for the relic, plus any particles with degenerate mass), and \(I_{a,b}\) are temperature integrals.

\subsection{The Generation of Mass Splitting}
\label{sub:RenormalizationMassSplitting}

Almost degenerate DM masses mean the mass splittings (\(\Delta m = m_{\chi_i} - m_{\chi_j}\)) between the different \(\chi_i\) are important to determining the true value of the DM relic density.

\begin{figure}[ht]
\centering
\includegraphics[width=0.45\textwidth]{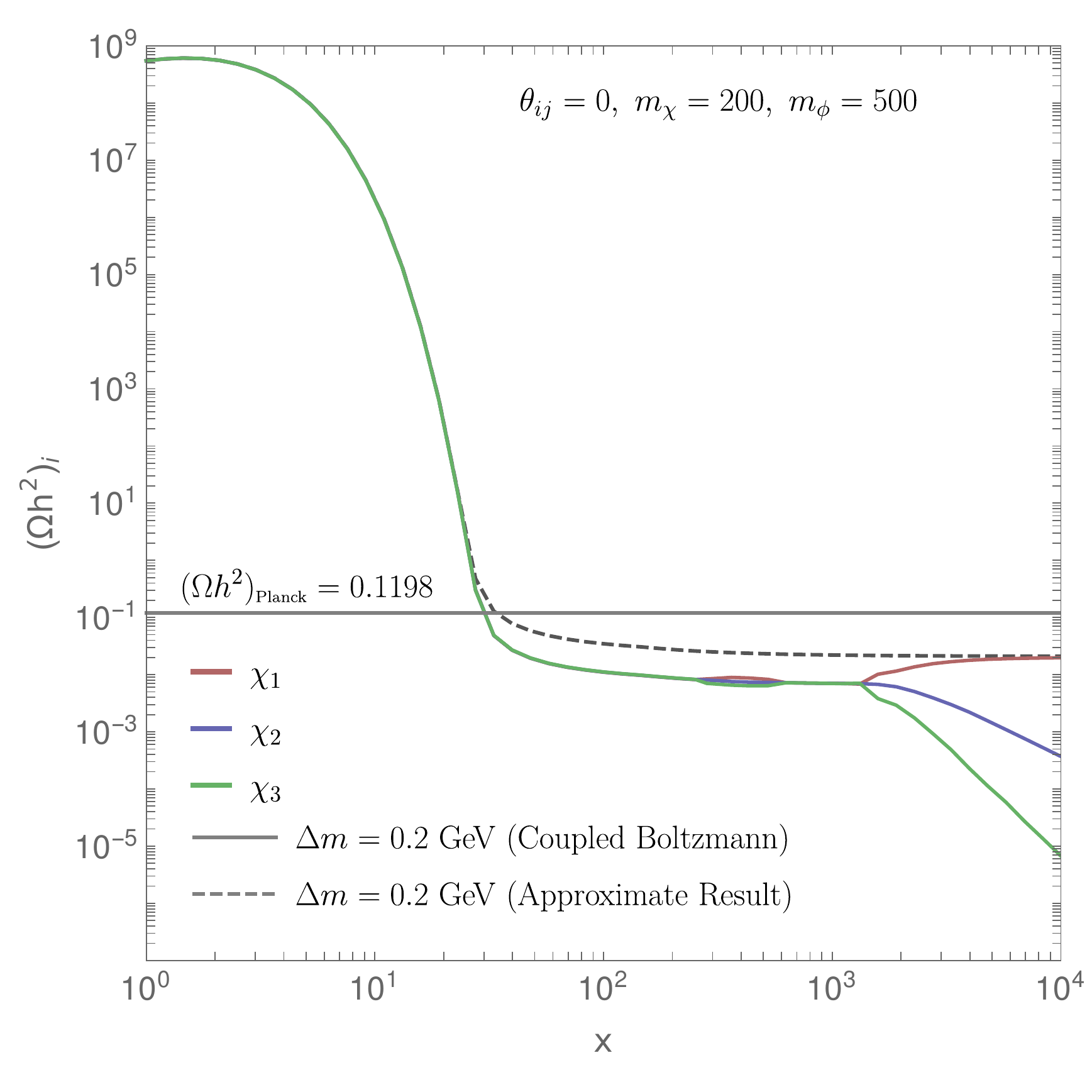}
\includegraphics[width=0.45\textwidth]{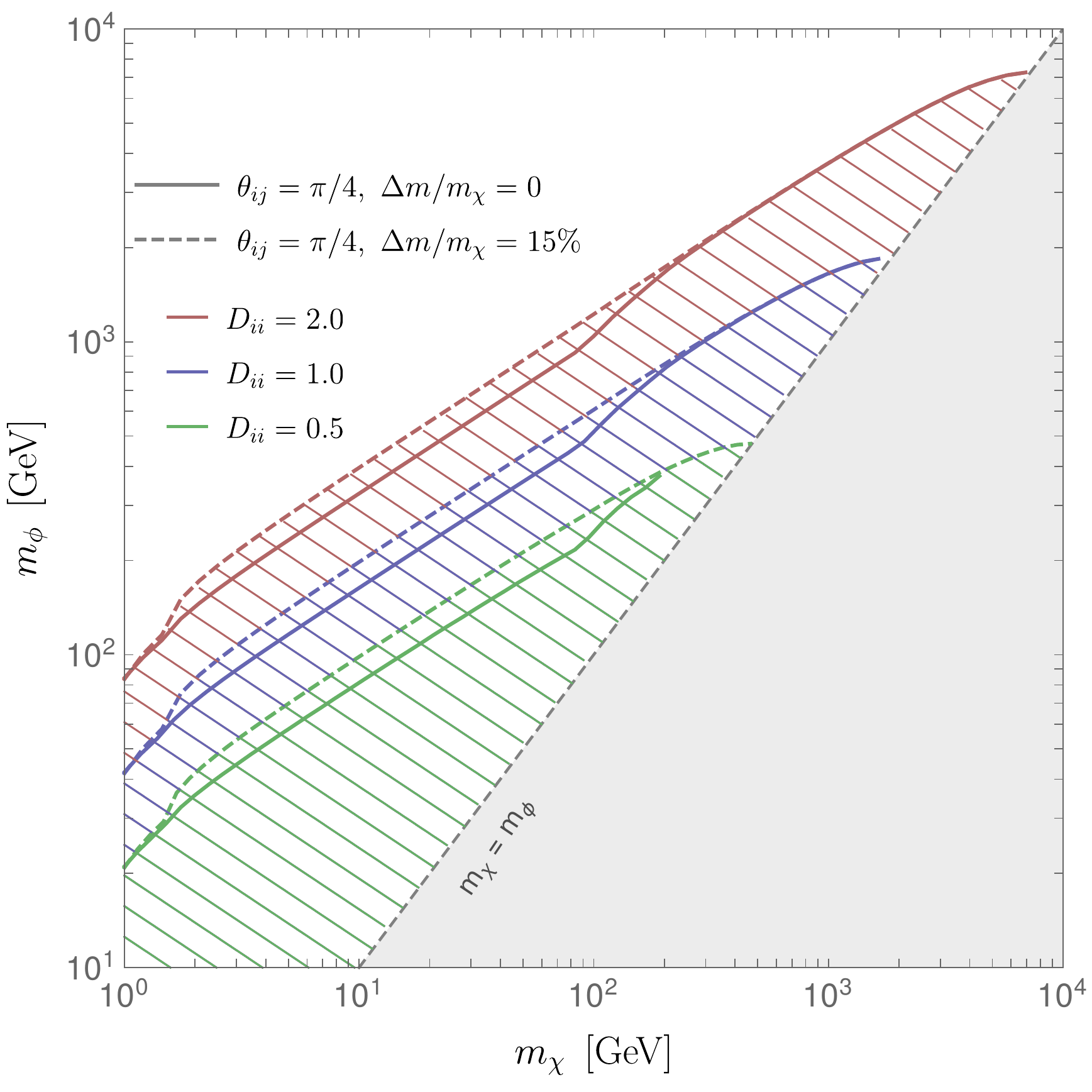}
\caption{Illustration of relic density over time (\(x = m_\chi / T\)) as freeze out occurs (left), and the RD bounds with mass splitting calculated with the effective method mentioned in the main text (hatched regions for which the DMFV models allows the correct relic abundance) (right).}
\label{Fig:RD3FlavourBounds}
\end{figure}

We can follow two regimes which distinguish the various possibilities by the dominant effect on the signals they generate:
\begin{enumerate}
\item The mass splitting is non-zero, the lightest of the \(\chi_i\) survives as the relic. This holds as long as the splitting is large enough to accommodate any kind of decay.
\item The masses are truly degenerate, equivalent to a degeneracy which is sufficiently small to prevent decay, i.e.\ \(\Delta m \leq \SI{4}{\MeV}\). In this case, the three DM particles obtain equal relic abundances, with the total affected primarily by their coannihilations.
\end{enumerate}
The difference between the effective cross-section method mentioned above and a full solution of the coupled Boltzmann equations, and the effect of degenerate masses is shown on the left of \cref{Fig:RD3FlavourBounds}.
We see that the effective cross section approach correctly reproduces the relic density of the lightest candidate at late times, and that relic density constraints are not hugely sensitive to the mass splitting if it is non-zero.

As the final relic density depends sensitively on whether a mass splitting in the candidates exists or not, we briefly talk about how such a splitting can arise.
Splittings can arise from two sources -- a tree-level contribution where \(\mi\) and \(\mj\) are split by mass terms of the form \(\mathcal{O}(1) \times (\lambda^\dagger \lambda)_{ii}\), or a loop-level contribution from renormalisation where the coefficient is instead of the order \(N_c / (16 \pi^2) \log ( \mu^2 /  \Lambda^2 )\) multiplied by the tree level couplings \((\lambda^\dag \lambda)_{ii}\) with \(\Lambda\) some high scale at which the masses are universal, and \(\mu\) a low scale at which we wish to use the mass (e.g.\ for direct detection this could well be the nuclear scale of around \SI{1}{\GeV}).
Explicitly, the resulting shift in the DM mass will be given by
\begin{align}
m_{\chi_i}(\mu) = m_{\chi}(\Lambda) \left( 1 + \frac{N_c}{16 \pi^2} (\lambda^\dagger \lambda)_{ii} \log{\left( \frac{\Lambda}{\mu} \right)}  + \mathcal{O}((\lambda^\dag \lambda)_{ii}^2)\right) \, .
\end{align}
Note that because of our parameterisation of the coupling matrix, \(\lambda^\dagger \lambda\) is diagonal, with elements \(D_{ii}^2\)

Relatively large splittings can be generated this way -- with a high scale of \SI{100}{\TeV}, then the coefficient of \((\lambda \lambda^\dagger)_{ii}\) can be as large as \(\sim 0.35\).
We explore the effect of mass splitting in our work by manually setting the mass splitting (\(\Delta m / m_\chi\)) to a large (\SI{15}{\percent}) and small (\SI{2}{\percent}) value.

\section{Flavour Constraints}
\label{Sec:Flavour}

\subsection{Mixing Observables}
\label{Sub:Mixing}

Since our model introduces couplings to the up-type quarks, we would expect new physics effects in the charm meson sector -- in particular in neutral \PDzero mesons.
Mixing is observed in \PD, \PB, and \PK meson systems, and relates the theoretical quantities \(\Gamma_{12}\) and \(M_{12}\) to the observed decay width differences \(\Delta \Gamma\) and mass differences \(\Delta M\) between the heavy and light mass states of the meson.
For \PD mesons, the current experimental averages from HFLAV are \cite{Amhis:2016xyh},
\begin{equation}
\begin{aligned}
x \equiv \frac{\Delta M}{\Gamma}        &= \SI{0.32 \pm 0.14}{\percent} , \\
y \equiv \frac{\Delta \Gamma}{2 \Gamma} &= (0.69^{+0.06}_{-0.07})\,\% \ .
\label{Eq:experimentalXY}
\end{aligned}
\end{equation}

On the theory side however, things are not so well developed.
There are two possible ways to calculate the mixing parameters -- inclusive, where we assume quark-hadron duality and sum quark level diagrams, or exclusive, where individual decay channels that contribute to \PDzero mixing are calculated. 
In the exclusive approach (e.g. \cite{Falk:2001hx,Falk:2004wg}), values of \(x\) and \(y\) on the order of \SI{1}{\percent} are believed to be possible.
However, currently exclusive \PDzero meson decays cannot be calculated from first principles and the estimates in \cite{Falk:2001hx,Falk:2004wg} were based on phase space arguments and \(\mathrm{SU}(3)_F\) symmetry.

On the inclusive side, we work within the Heavy Quark Expansion (HQE) formalism, see \cite{Lenz:2014jha} for a review, assuming that the charm quark mass is large compared to the hadronic scale. 
For charm mixing the three leading dimension six contributions of the HQE suffer, however, from a huge GIM \cite{Glashow:1970gm} and CKM suppression, leading a prediction that is orders of magnitudes below the experimental values, see e.g.\cite{Golowich:2005pt}, while the individual dimension six contributions are slightly larger than the experimental value.
To decide whether the charm quark is heavy enough to apply the HQE one has to study observables that are not affected by any severe cancellations, a prime example for such an observable are lifetimes.
First studies \cite{Lenz:2010pr,Lenz:2013aua} have suggested that the HQE could hold with corrections of no more than \SI{40}{\percent}.
Assuming now the applicability of the HQE for the charm system we have to find a mechanism that is violating the severe GIM cancellation. In the literature three possibilities for such a breaking are studied. In \cite{Jubb:2016mvq}  it was shown that a small breakdown (\(\mathcal{O}(\SI{20}{\percent})\)) of quark-hadron duality could enhance the predicted value of \(y\) up to its experimental value.
An older idea \cite{Bigi:2000wn} is that the GIM cancellation is much less pronounced for higher orders in the HQE.
A first estimate of  SU(3) breaking dimension nine contributions in the HQE gives \(x \approx 6 \times 10^{-5}, y \approx 8 \times 10^{-6}\) \cite{Bobrowski:2012jf} -- still missing the experimental results by two or three orders of magnitude.
Finally there is the possibility that the GIM suppression is lifted by new physics effects, which we will investigate.
Because of these difficulties we have some freedom in the treatment of the SM contributions to \(\Delta M\) and \(\Delta \Gamma\) when constraining the allowed BSM contribution by comparison to experiment.
One possibility \cite{Golowich:2007ka} is to require that
\begin{align}
x^\mathrm{NP} = \frac{2 | M^\text{NP}_{12} |}{\Gamma_{\PD}} \leq x^\text{exp, upper limit} \, ,
\end{align}
taking the \(1\,\sigma\) upper limit reported by HFLAV (\cref{Eq:experimentalXY}).
This is the limit that would be derived if the NP and SM contributions have roughly the same phase, so that 
\begin{align}
|M_{12}^\text{NP} + M_{12}^\text{SM}| = |M_{12}^\text{NP}| + |M_{12}^\text{SM}| \, ,
\end{align}
since we know \(\Delta M \leq 2 | M_{12} |\).
The NP contribution to \(M_{12}\) is given by
\begin{equation}
M_{12}^\text{NP} = - \frac{f_D^2 B_D M_D}{384 m_\phi^2 \pi^2} \sum_{i,j = 1}^3 F\left(\frac{m_{\chi_i}^2}{\MphiDM^2}, \frac{m_{\chi_j}^2}{\MphiDM^2}\right) \lambda_{1i} \lambda_{1j} \lambda_{2i}^* \lambda_{2j}^*
\end{equation}
where we take the decay constant \(f_D\) from FLAG \cite{Aoki:2016frl,Na:2012iu,Bazavov:2011aa}, the \PD mixing bag parameter \(B_D\) from \cite{Carrasco:2014uya}, and the loop function \(F\) is given by
\[
F(x_i, x_j) = \frac{1}{(1-x_i)(1-x_j)} + \frac{x_i^2 \log x_i}{(x_i-x_j) (1-x_i)^2} - \frac{x_j^2 \log x_j}{(x_i-x_j) (1-x_j)^2} \, .
\]

The important result is that \(M_{12} \propto( (\lambda \lambda^\dag)_{12} )^2\) for degenerate DM masses.
The matrix \((\lambda \lambda^\dag)\) is diagonal if \(D_{ii}\) are all equal, or if \(\theta_{ij} = 0\) (no mixing between quark flavours) and then the flavour constraints disappear.

\begin{figure}
\centering
\includegraphics[width=0.45\textwidth]{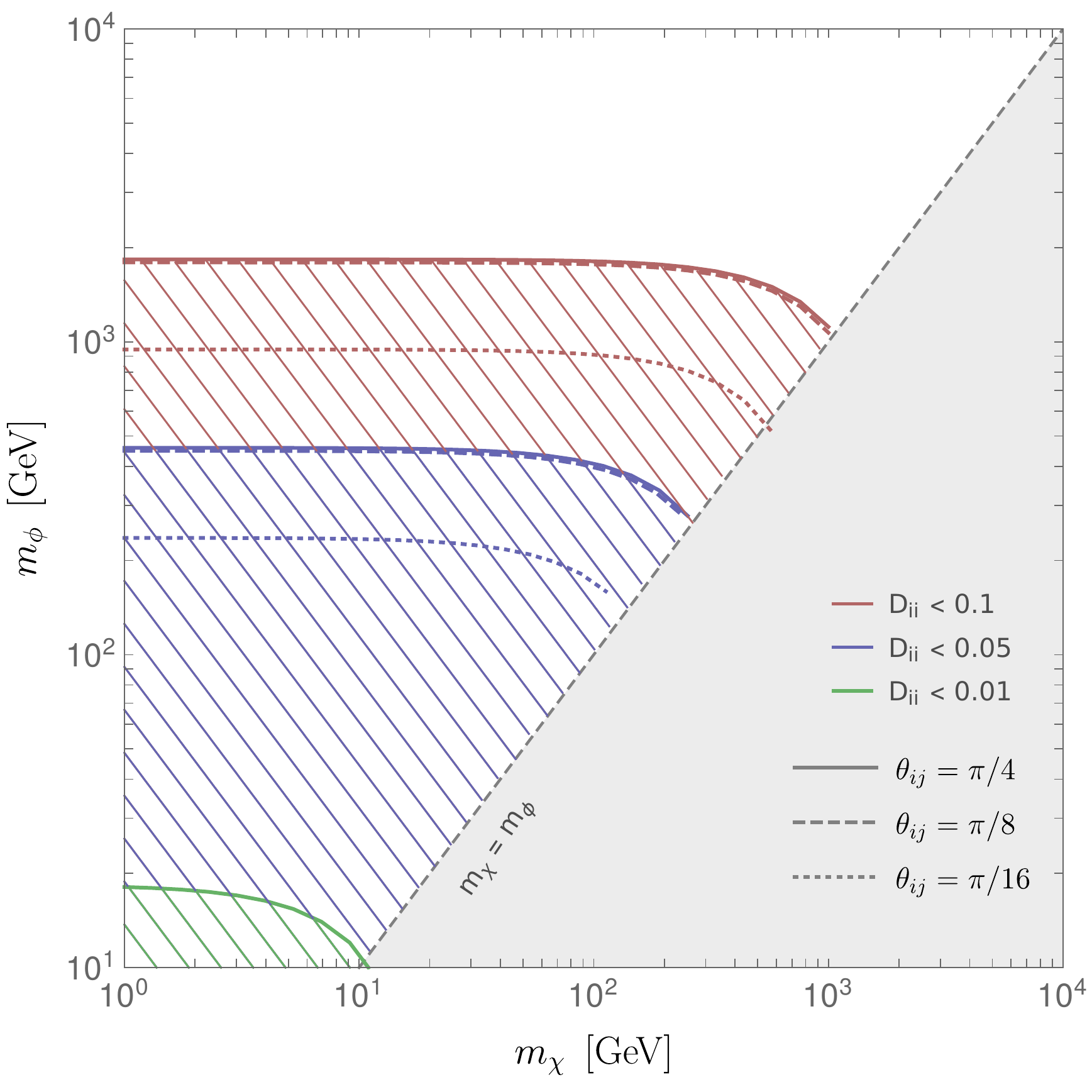}
\includegraphics[width=0.45\textwidth]{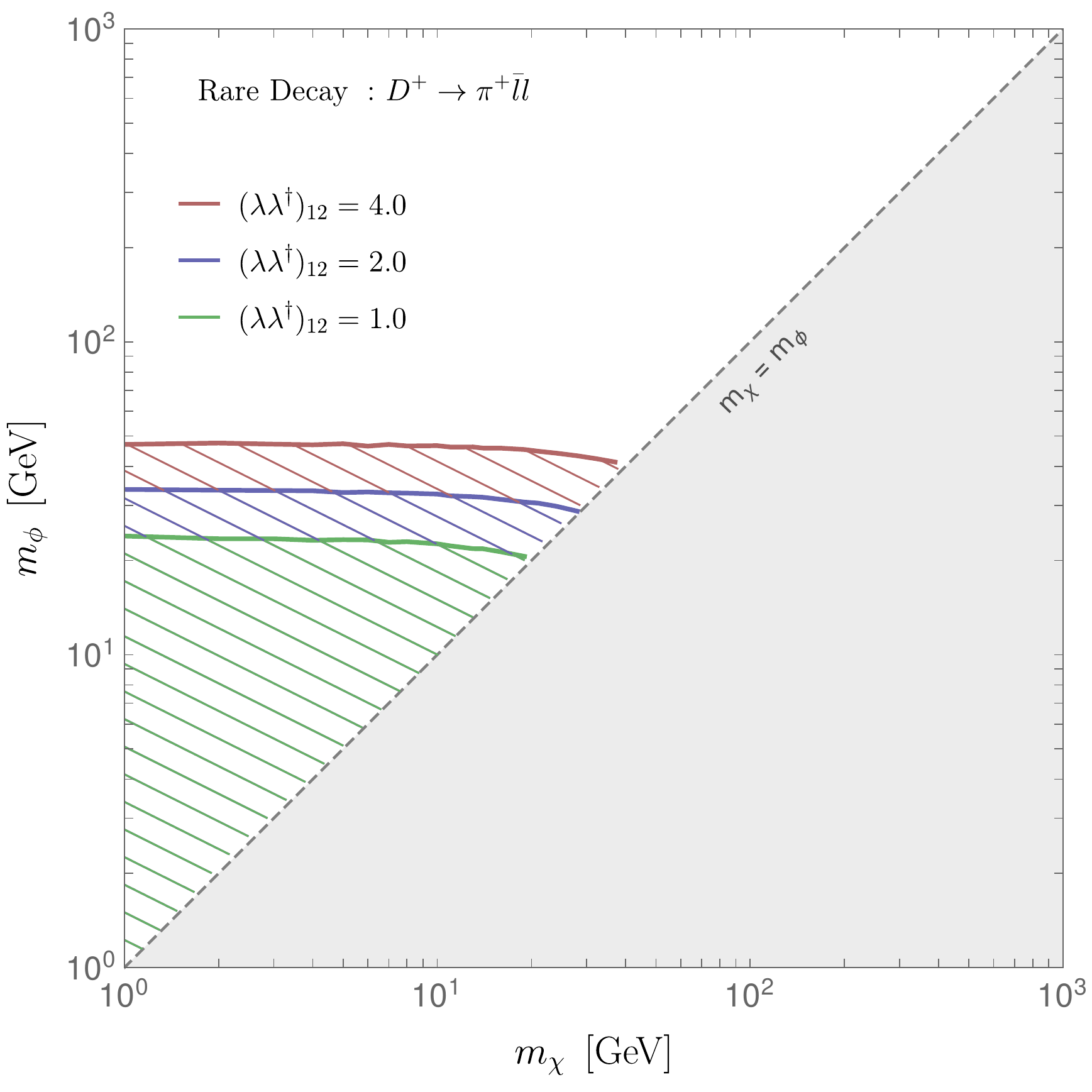}
\caption{Excluded regions (hatched) for which the value of \(\Delta M\) from DMFV diagrams exceeds the \(+1\,\sigma\) contour of the experimental result (left).
The bounds are the most constraining possible given the limits on \(D_{ii}\), and can be made arbitrarily small by adjusting the values (for example with equal values \(D_{ii} = D\)).
The exclusions from \(|C_9^\prime| < 1.3\) varying \((\lambda \lambda^\dag)_{12}\) (right).}
\label{Fig:FlavourBounds}
\end{figure}

Using the upper \(1\,\sigma\) value of the experimentally measured \(x_D\) leads to bounds as shown on the left of \cref{Fig:FlavourBounds}, these bounds can be very strong and significantly exclude almost all masses \(m \lesssim \SI{1}{\TeV}\) for large couplings \(\lambda \gtrsim 0.1\) unless one fine-tunes the model to remove \((\lambda \lambda^\dag)_{12}\).

\subsection{Rare Decays}

We consider the semileptonic decay \HepProcess{\PDp \to \Ppiplus \APmuon \Pmuon}, whose short distance contribution comes from the quark level decay \HepProcess{\Pqc \to \Pqu \APmuon \Pmuon}.
This decay is loop and GIM suppressed in the SM, and so should have good sensitivity to new physics.
In our model contributions are no longer GIM suppressed, coming from electroweak penguin diagrams with our new particles in the loop.

Ref.~\cite{Fajfer:2015mia} examines rare charm decays to provide limits on the Wilson coefficients of an effective theory -- they look at \HepProcess{\PD \to \APmuon \Pmuon} as well as \HepProcess{\PDp \to \Ppiplus \APmuon \Pmuon} and find the latter to place the strongest bounds for the coefficients relevant in our model.
Matching onto their EFT, and neglecting the \PZ penguin since the momentum transfer is small, we find only the \(C_7^\prime, C_9^\prime\) coefficients are non-zero, corresponding to the operators
\begin{equation}
Q_7^\prime = \frac{e m_c}{16 \pi^2} (\APqu \sigma^{\mu \nu} P_L \Pqc) F_{\mu \nu} \quad , \quad Q_9^\prime = \frac{e^2}{16 \pi^2} (\APqu \gamma^\mu P_R \Pqc) ( \APlepton \gamma_\mu \Plepton )\ ,
\label{Eq:RareDecayOperators}
\end{equation}
(our full expressions for the Wilson coefficients can be found in \cref{app:rareDecays}).

Since the SM branching ratios for the \PDzero decay suffer from a strong GIM cancellation, we would expect strong constraints on the flavour breaking terms of the DMFV model.
As with the mixing observables, the rare decay process is primarily sensitive to \((\lambda \lambda^\dag)_{12}\) in the limit of degenerate DM mass.
On the right of \cref{Fig:FlavourBounds} we show the bounds coming from limits on the Wilson coefficients for \((\lambda \lambda^\dag)_{12} = 1, 2, 4\).
The bounds on the individual Wilson coefficients are \(|C_i| \sim 1\) (see Table II of \cite{Fajfer:2015mia}).
Mediators up to \(m_\phi \sim \SI{50}{\GeV}\) can be ruled out for couplings \(D_{ii} \sim (\lambda \lambda^\dag)_{12} \sim \mathcal{O}(1)\). These constraints are therefore substantially weaker than from meson mixing observables.

The rare flavour-changing decays \HepProcess{\Pqt \to \Pqu/\Pqc \Pgamma} have been measured by ATLAS \cite{Khachatryan:2015att}, but we find that the current limits are again not constraining on our model.

\section{Direct Detection Constraints}
\label{sec:dd}

Direct detection experiments are one of the most powerful ways of searching for DM, and operate by searching for DM scattering from atomic nuclei.
The calculation of the scattering rate is done via an effective theory, where all heavy degrees of freedom (save the DM) have been integrated out, and then amplitudes are matched onto four fermion operators.

We choose to examine data from LUX \cite{Akerib:2013tjd,Akerib:2015rjg} and CDMSlite\cite{Agnese:2015nto}, which together provide the best constraints over the range of DM masses we are looking at.
LUX uses liquid xenon as a target, which detects DM with masses above \SI{5}{\GeV} while scattering from DM masses below this is kinematically impossible; CDMSlite is a germanium detector, and best constrains particles with masses between \SI{1.6}{\GeV} and \SI{5.5}{\GeV}.
Details of our exact method can be found in \cref{app:dd} -- for now we merely state that we use a Poisson probability distribution for both, comparing the number of observed events in each bin to our predicted signal plus background.

At tree level, the only EFT operator which arises from our model is given by a diagram with t-channel \PphiDM exchange.
We only consider the scattering amplitudes in which the incoming and outgoing DM (and quark) are the same flavour, as this avoids the computation of (possibly unknown) hadronic matrix elements of quark currents \(\bar{q}_i \Gamma q_j\) for \(i \neq j\).
The operator in question is
\begin{align}
\mathcal{L}_\text{EFT} = C_{ij} (\bar{\chi}_L^i \gamma^\mu \chi_L^i) (\APquark_R^j \gamma_\mu \Pquark_R^j) \;,\;  C_{ij} (\mu \sim m_\phi) = \frac{\lambda_{ji} \lambda_{ji}^*}{2( (m_\chi -m_q)^2 -m_\phi^2 )}
\label{Eq:TreeLevelMatching}
\end{align}
where the Mandelstam variable \(t\) has been replaced by its low velocity expansion and we have performed a Fierz transform (see e.g.\ \cite{Nishi:2004st}).

\begin{figure}
\centering
\includegraphics[scale=0.65]{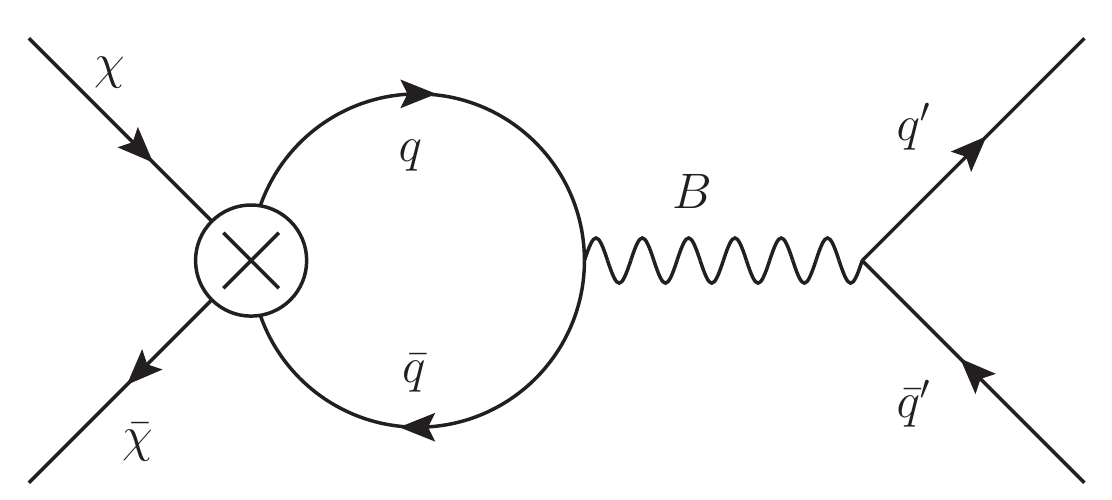}
\\
\includegraphics[scale=0.65]{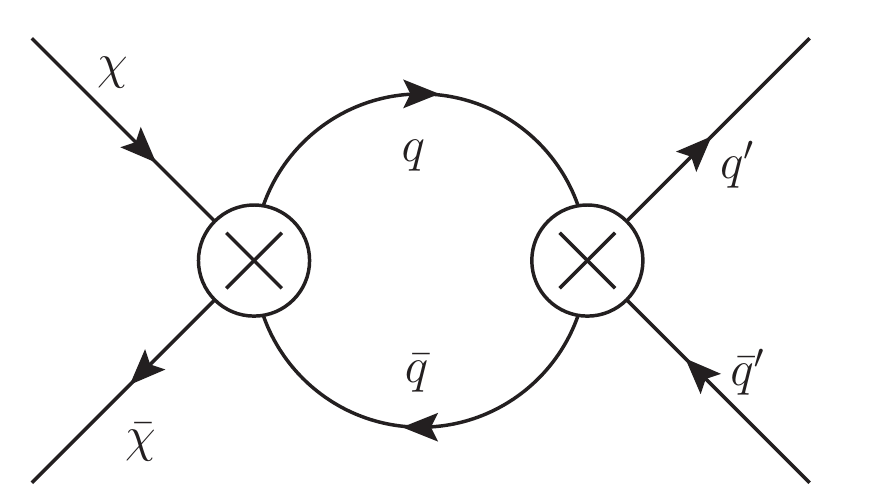}
\includegraphics[scale=0.65]{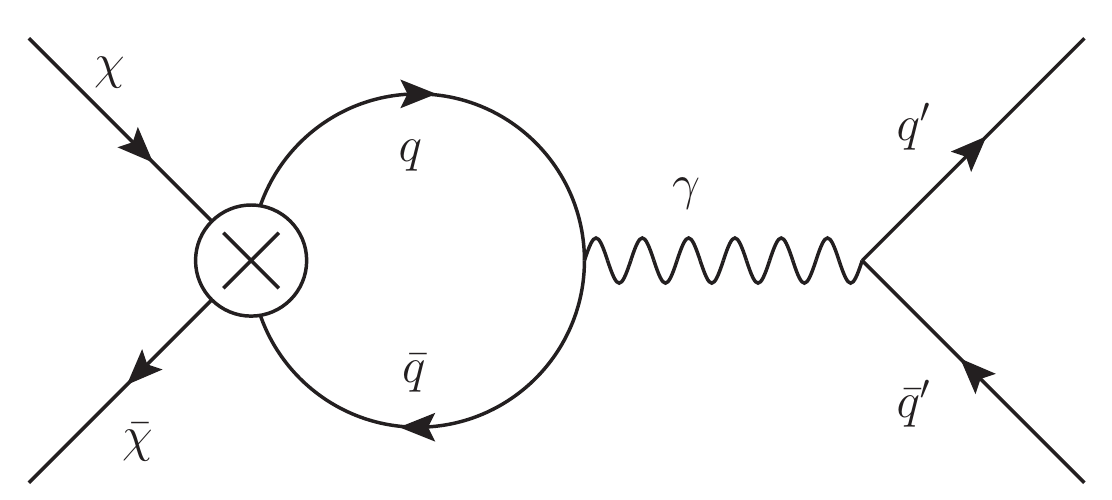}
\caption{The divergent loop diagrams responsible for mixing between the quark vector and axial vector currents ($\bar{\chi} \Gamma \chi \bar{q} \Gamma q$) above the EW scale (top) and below (bottom).
The most important aspect is the mixing of high-scale heavy quark currents $q = c,t$ onto light quark vector currents $q'=u,d$, thus enabling a strong scattering cross section with nuclei.}
\label{Fig:DDMixingDiagrams}
\end{figure}

\begin{figure}
\centering
\includegraphics[width=0.5\textwidth]{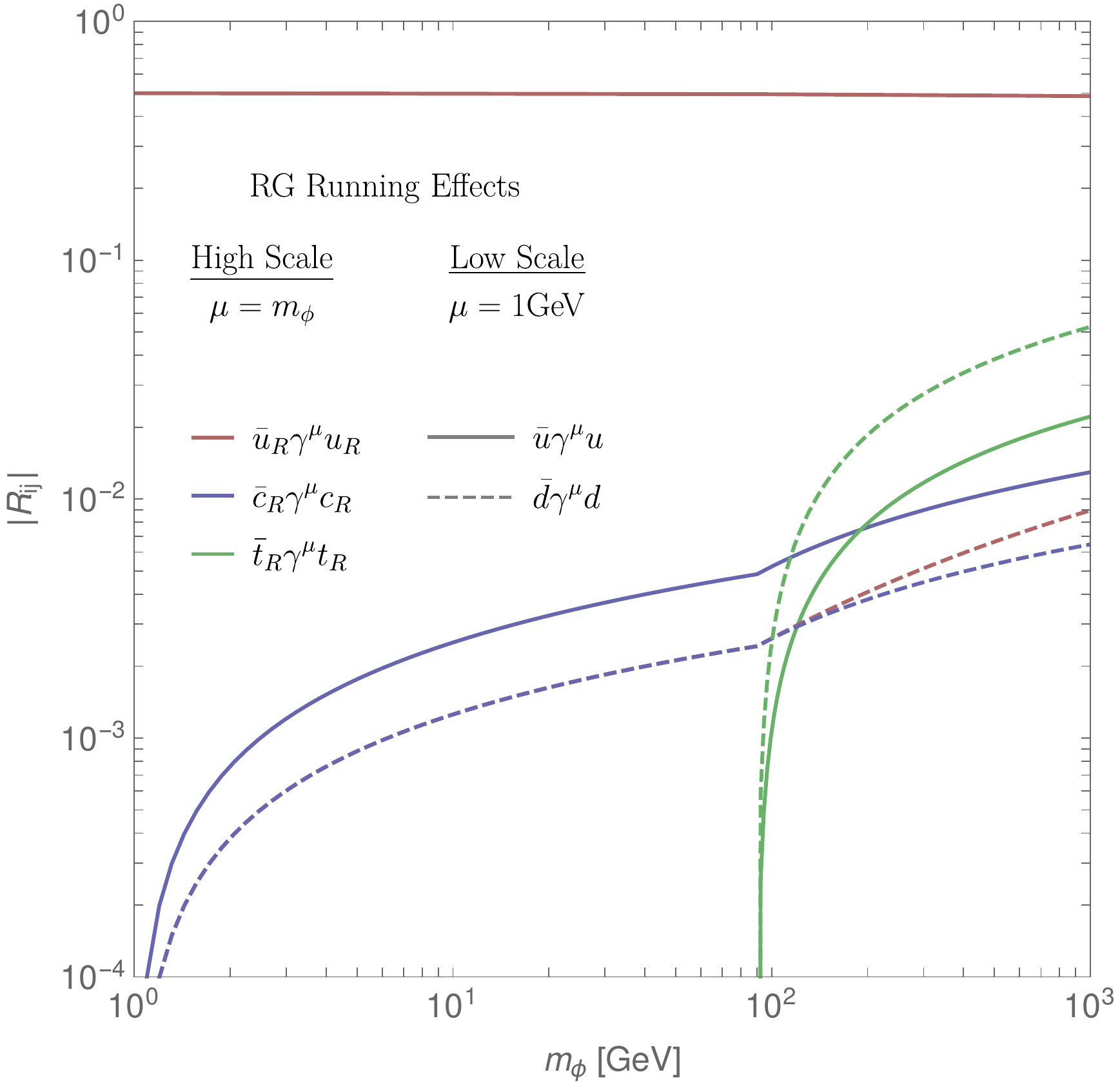}
\caption{The effect of the RG running from a high scale $\Lambda = m_\phi$ down to the nuclear scattering scale \(\mu_N = \SI{1}{\GeV}\).}
\label{Fig:RGrunningEffect}
\end{figure}

\begin{figure}
\centering
\includegraphics[width=0.45\textwidth]{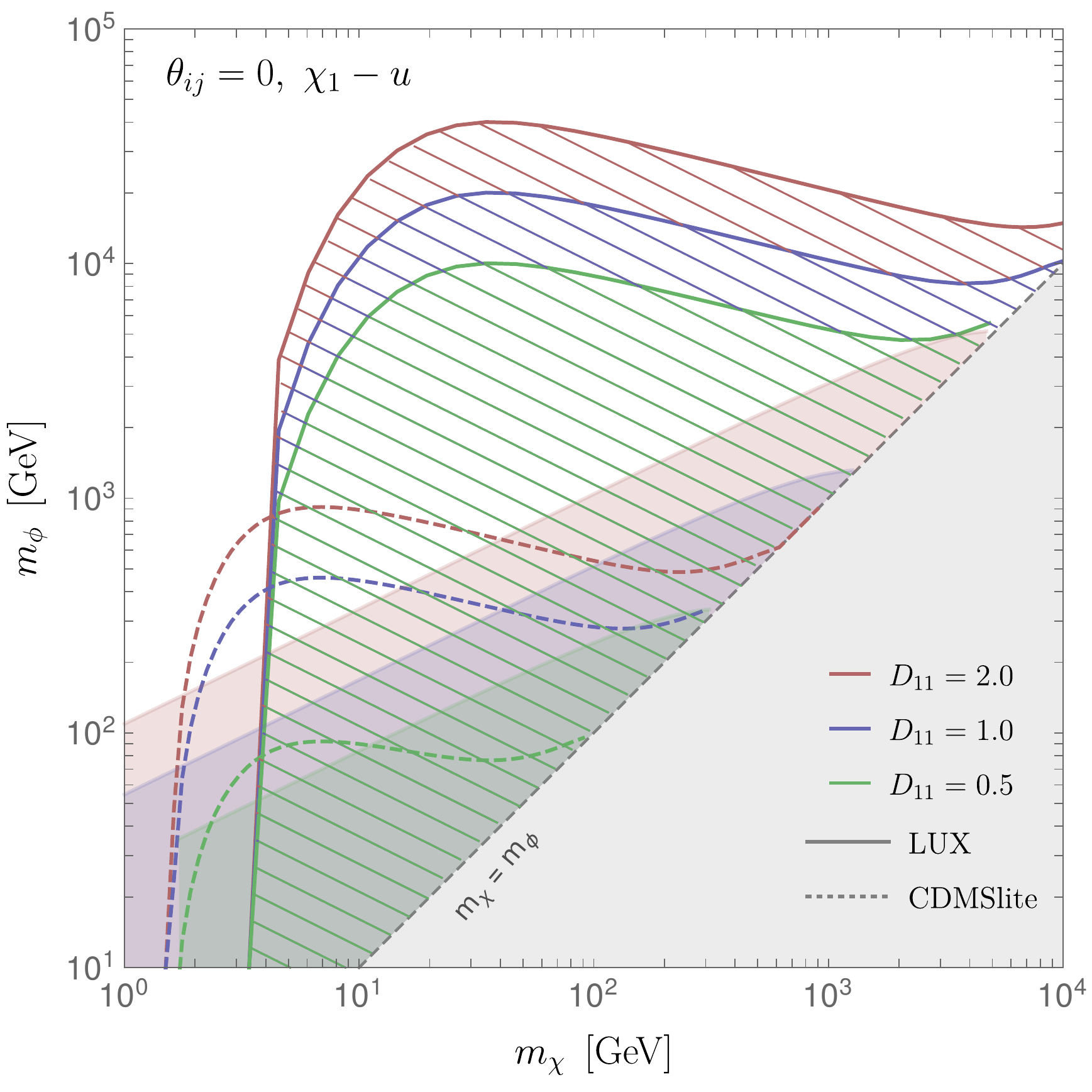}
\includegraphics[width=0.45\textwidth]{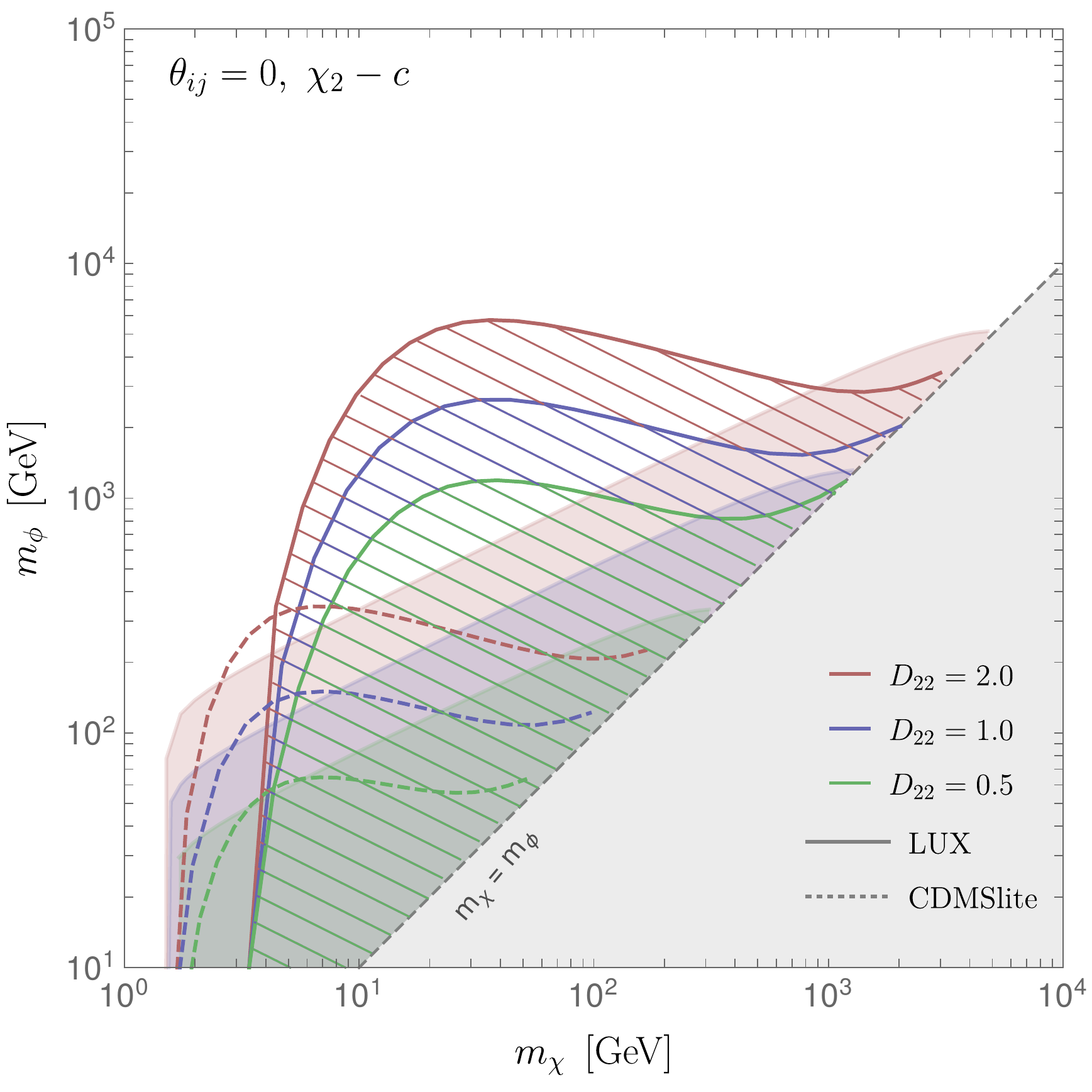}
\includegraphics[width=0.45\textwidth]{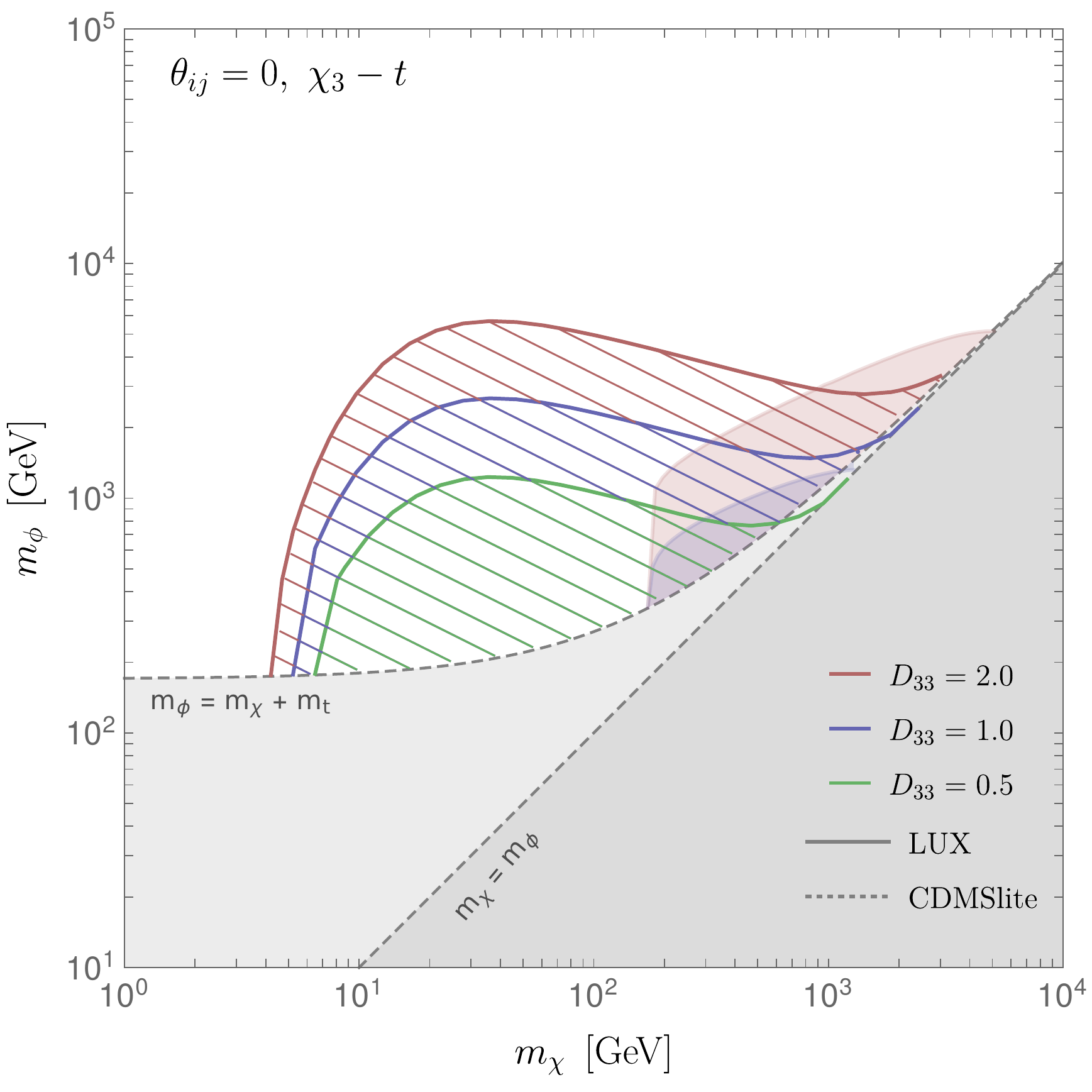}
\caption{The DD bounds for three coupling choices -- \(\chi_1\) exclusively coupling to \Pqu quarks, \(\chi_2\) to \Pqc, and \(\chi_3\) to \Pqt.
Bounds for LUX (CDMSlite) are solid (dashed), and the filled region shows the parameters which give the correct relic abundance.
Constraints are based on the dominant tree level contribution to scattering.}
\label{Fig:DDBounds1}
\end{figure}

Vector and axial-vector currents probe the valence quark content and spin distribution respectively of the scattered nucleon, and so would naively be small for non-valence quarks (i.e.\ \Pqc and \Pqt).
However, there are 1-loop diagrams (see \cref{Fig:DDMixingDiagrams}) that mix operators with heavy quarks into those with up and down quarks, and in the case of heavy mediators RG running down to the direct detection scale (\(\mu \sim \SI{1}{\GeV}\)) also alters the relative coupling to nuclei.
This calculation has been done in \cite{Crivellin:2014qxa,DEramo:2014nmf}, and we find (see \cref{Fig:RGrunningEffect}) that DM that couples to heavy quarks at the mediator scale will mix into up quark coupling at the low scale with up to \SI{10}{\percent} of its high scale coupling strength; tree level scattering is therefore substantial (as can be seen in \cref{Fig:DDBounds1}), even in the case of only coupling to heavy quarks.
The spin-averaged cross section is parametrised by a series of nuclear form factors \(F^{(N, N^\prime)}_{ij}\) \cite{Fitzpatrick:2012ix}, which are functions of the local galactic DM velocity squared \(v^2\) and the momentum transfer \(q^2\),
\begin{equation}
\langle |\mathcal{M}|^2 \rangle \sim \sum_{i, j, N, N^\prime} C_i^{(N)} C_j^{(N^\prime)}  F_{ij}^{(N,N^\prime)}(v^2, q^2)
\end{equation}
where we sum over the form factors and the nucleons \(N, N^\prime = p, n\).
The nucleon coefficients above are related to our Wilson coefficients  by
\begin{align}
C_1^{(p),i} (\mu \sim \SI{1}{\GeV}) &= 4 m_i m_N \sum_{j}  ( 2 R_{ju} + R_{jd} ) C_{ij} (m_\phi) \\
C_1^{(n),i} (\mu \sim \SI{1}{\GeV}) &= 4 m_i m_N \sum_{j}  ( 2 R_{jd} + R_{ju} ) C_{ij} (m_\phi)
\end{align}
where \(R_{ju}\) (\(R_{jd}\)) gives the magnitude of the running of operator \(\APquark^j_R \gamma^\mu \Pquark^j_R\) onto \(\APqu \gamma^\mu \Pqu\) (\(\APqd \gamma^\mu \Pqd\)), and we have quoted the \(i = j = 1\) relation since the corresponding form factor has the dominant scaling behavior.
\(i\) and \(j\) run over the DM and quark flavours respectively.
The dependence of the \(R_{jq}\) parameters on the high scale (which we take to be the mediator mass) is shown in \cref{Fig:RGrunningEffect}.

\begin{figure}[htp]
\centering
\includegraphics[width=0.45\textwidth]{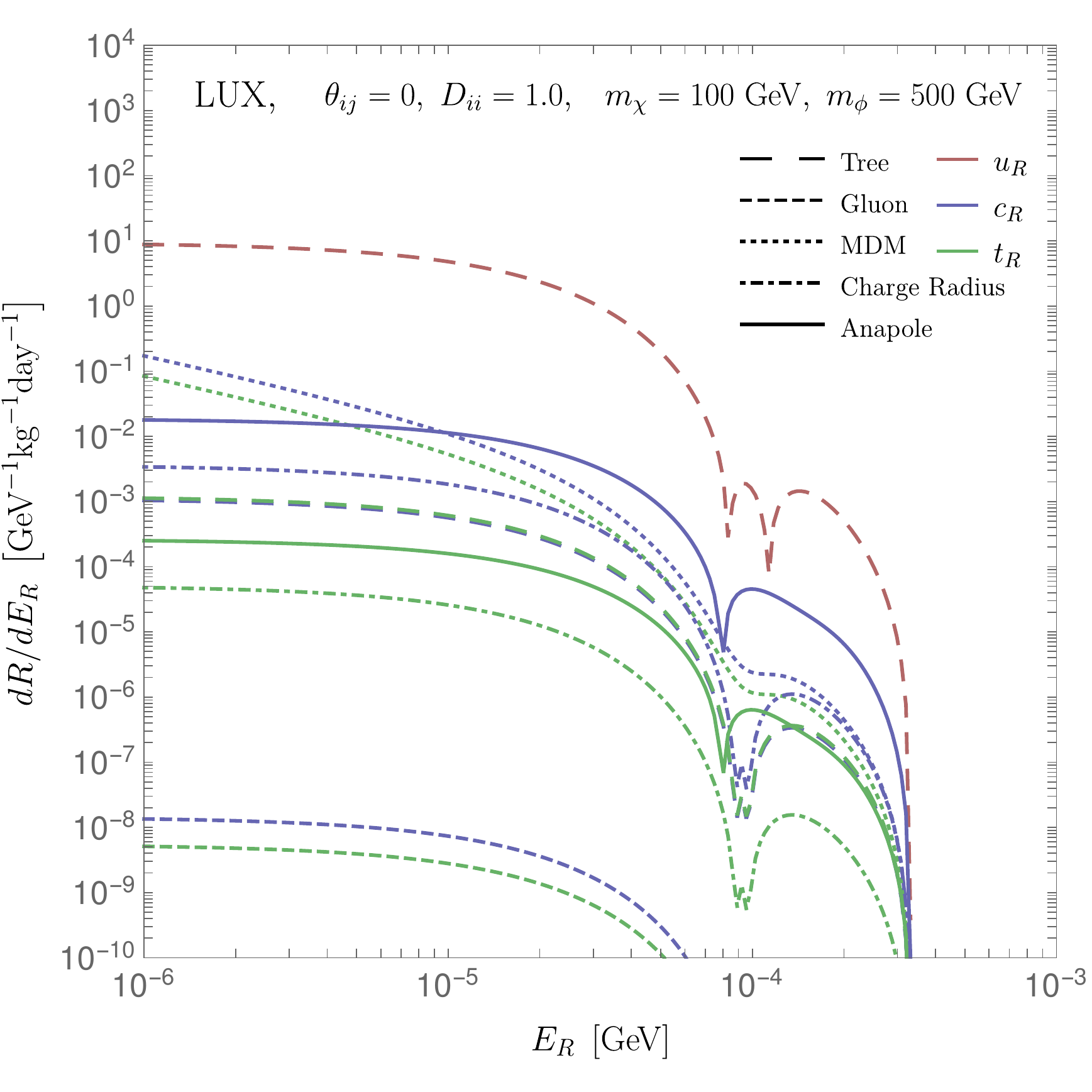}
\includegraphics[width=0.45\textwidth]{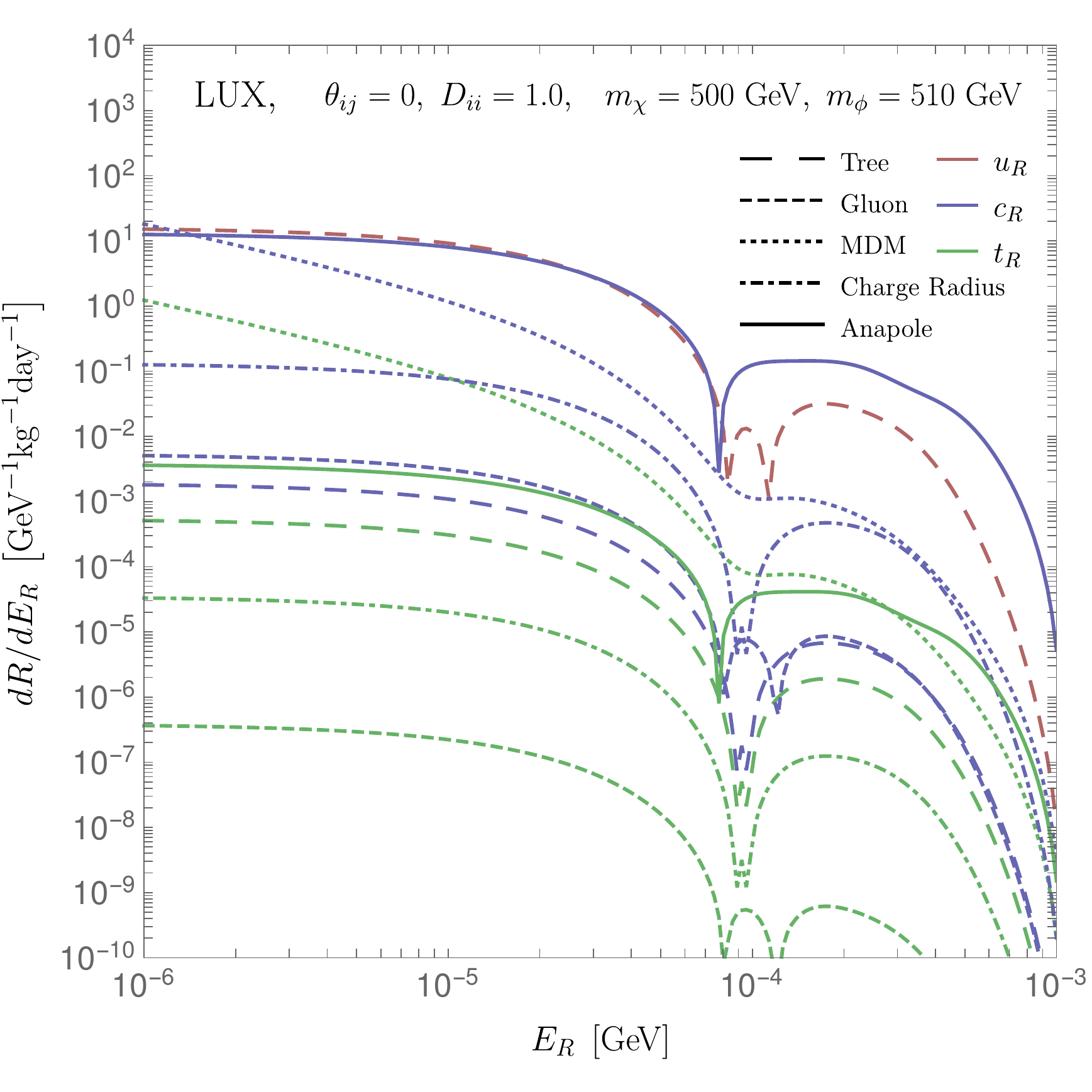}
\caption{The differential scattering rate in recoil energy for DM-nuclear scattering at LUX.
Each of the quark contribution are plotted separately, the rates are also separated according to the way in which they scatter.
The right plot represents a model with almost complete degeneracy between the DM and mediator mass, where the loop level interactions become important.}
\label{Fig:DDsignal}
\end{figure}

At loop-level, there are various new operators that arise -- in general these are highly suppressed, but we include them both because they can become dominant in particular regions of parameter space (see \cref{Fig:DDsignal}) and for completeness.
The operators we consider are photon operators \cite{Ibarra:2015fqa,Kahlhoefer:2016eds} which in the non-relativistic limit correspond to the charge-radius, magnetic dipole moment, and anapole moment, Z penguins \cite{Ibarra:2015fqa}, and those for DM-gluon \cite{Drees:1993bu,Hisano:2015bma,Gondolo:2013wwa}.
We reproduced the quoted literature results as a check.

The very latest null results from XENON1T \cite{Aprile:2017iyp} and PandaX-II \cite{Cui:2017nnn} push the constraining potential of direct detection even further -- nearly an order of magnitude stronger in cross-section, which translates into a factor of \(\sim 2\) in mediator mass.

\section{Indirect Detection Constraints}
\label{sec:id}

\subsection{Basics of Indirect Detection}

Indirect detection experiments looks for signs of annihilating / decaying DM coming from astrophysical sources, typically the centre of galaxies where DM density is largest.
The constraints are based around limits on the annihilation cross-section of DM to SM particles -- in our model the main limits come from annihilation to quark pairs
\begin{equation}
\langle {\sigma v} \rangle_{ \bar{\chi}_i \chi_j \rightarrow \bar{q}_l q_m} \approx \frac{ N_c m_\chi^2}{32 \pi (m_\chi^2 + m_\phi^2)^2} \left( \lambda_{mj} \lambda^*_{li} \right)^2 + \mathcal{O}(v^2) \ .
\label{Eq:TreeLeveAnnXSec}
\end{equation}
There is a bounty of possible search avenues for this annihilation signal; the energetic quarks will hadronize and decay into stable particles (photons, electrons, protons, and their anti-particles, which make up some part of the measured cosmic ray flux), which can be measured directly as they arrive at the earth (in the case of photons especially, which suffer very little energy loss to galactic or inter-galactic material), or indirectly through their influence on cosmic rays (for example photons produced by electrons/protons diffusing through the galaxy).
We also have great freedom in where to look; generally anywhere where there is a cosmic overdensity of dark matter, close to home in the galactic centre or further afield in \emph{dwarf spheroidal} (dSph) galaxies, galaxy clusters or the CMB.

\begin{figure}
\centering
\includegraphics[width=0.4\textwidth]{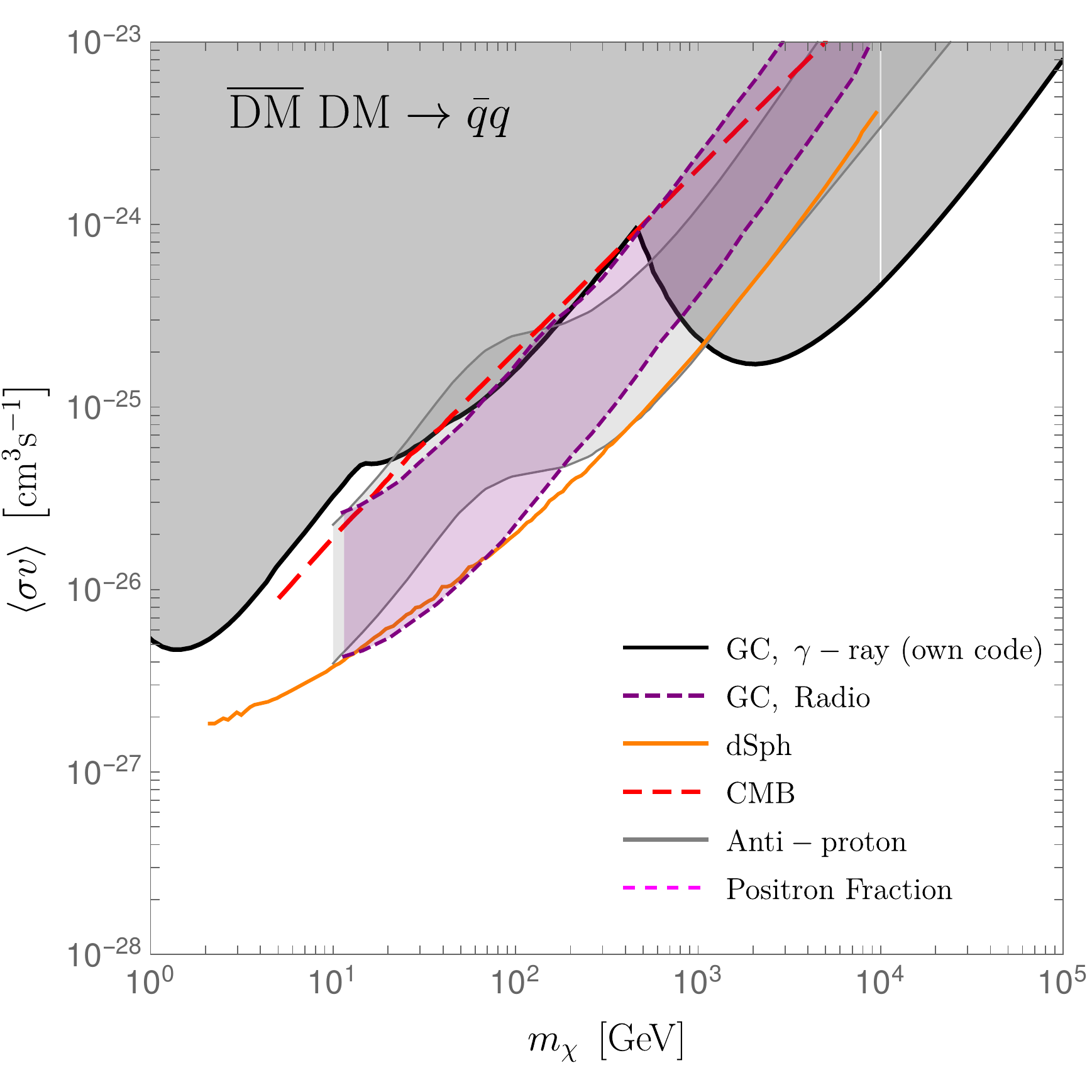}
\includegraphics[width=0.4\textwidth]{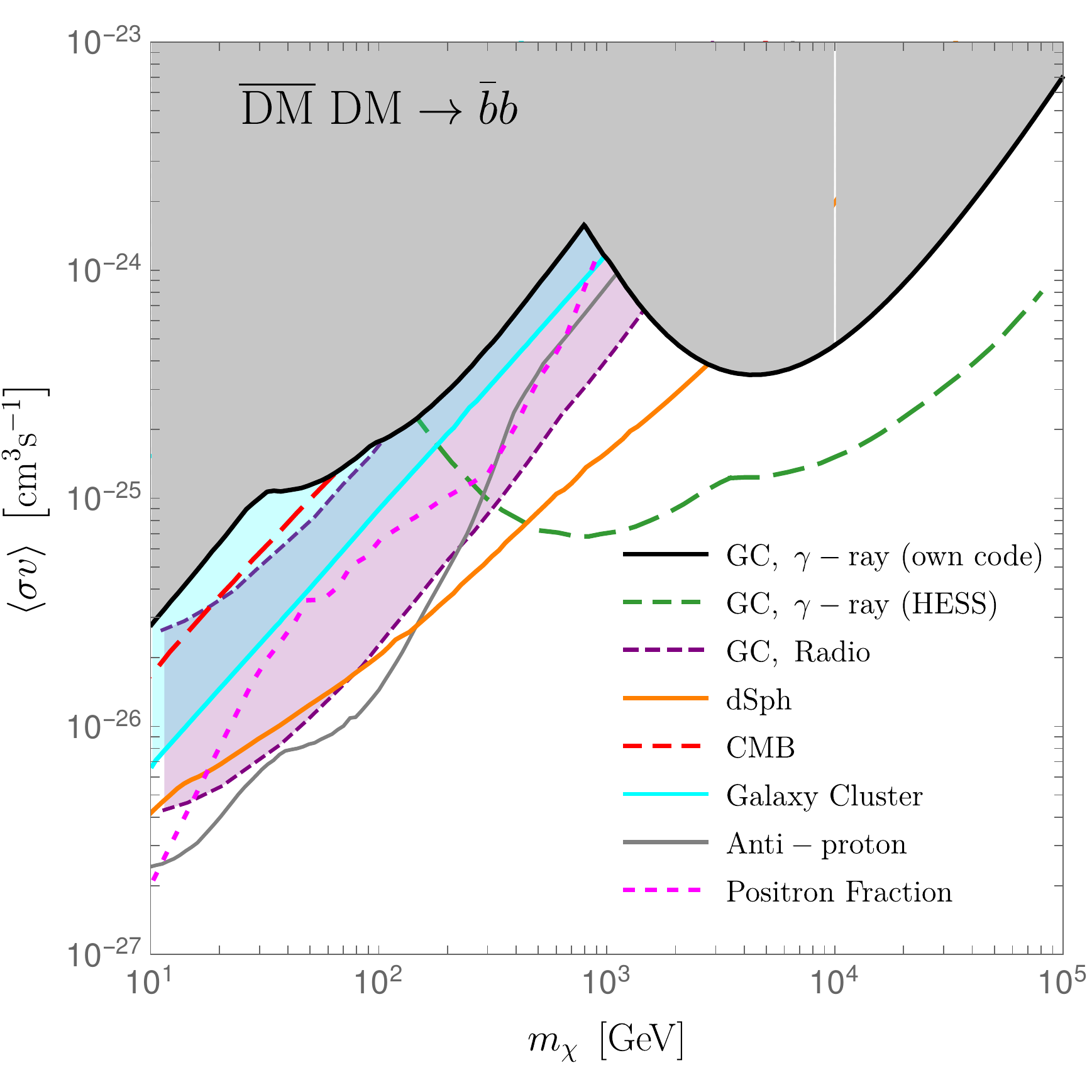}
\caption{The constraints on \(\braket{\sigma v}_{\bar{q} q}\) for \(q=\Pqu,\Pqd,\Pqs\) (left) and \(q=\Pqb\) (right) which is representative of \(q=\Pqc,\Pqt$ for \(m_\chi > m_{c,t}\).
The constraints are taken from many different sources (DSph, galactic centre, clusters) and targets (gamma rays, radio waves, positron, anti-protons).}
\label{Fig:IDExclusions}
\end{figure}

Underlying all these is \cref{Eq:TreeLeveAnnXSec} and so ID constraints are frequently quoted as confidence limits on the thermally averaged annihilation cross section \(\langle \sigma v \rangle_{\bar{f} f}\) into fermions of the same flavour, covering a mass range \(m_\chi \sim \SI{1}{\GeV} - \SI{100}{\TeV}\).
The ID signals from heavy quarks (\(q= \Pqc, \Pqb, \Pqt\)) are very similar (see Fig.~3 and 4 in \cite{Cirelli:2010xx}), and it is uncommon to find constraints on \Pqc, \Pqt final states (more common is the \Pqb).
The primary spectra of electrons, positrons, anti-protons, deuteron and neutrinos are extremely similar between \Pqc, \Pqb, \Pqt quarks, and thus any constraints which look for these particles from DM annihilations will be approximately heavy-flavour independent.
The situation is depicted in \cref{Fig:IDExclusions}.

It should be noted that the relative strength of these constraints is not robust, different authors use different halo profiles, different astrophysical parameters and are subject to varying degrees of uncertainty, some significantly larger than others, it is beyond the scope of this work to accommodate all these effects and compare constraints on a like-for-like basis and so what we present should be taken as representative but not precise.
We will use the \(\APqb \Pqb\) final state as representative for constraints based on dSph \cite{Rico:2015nya} and anti-proton measurements of AMS-02 \cite{Boudaud:2015yda} which dominate other constraints such as those based on other particle targets, such as the positron fraction \cite{DiMauro:2015zba} or neutrinos \cite{Aartsen:2016pfc} and also those based on the galactic centre \cite{Lefranc:2015vza}, or galaxy clusters \cite{Ackermann:2015fdi}.

\subsection{Gamma rays (and other mono-chromatic lines)}
\label{sub:id_gamma}
At the one-loop level, the pair production of quarks from annihilating DM can pair produce photons at a fixed energy \(E_\gamma = m_\chi / 2\) via a box diagram.
We calculate this cross-section using an EFT where the mediator has been integrated out, in which limit only the axial vector operator \( (\bar{\chi} \gamma^\mu \gamma^5 \chi) (\APquark \gamma_\mu \gamma^5 \Pquark) \) contributes to the s-wave annihilation, with cross section
\begin{align}
\langle \sigma v \rangle_{ \gamma \gamma} =  \frac{16 \alpha^2 s}{9468 (m_\chi^2 - m_\phi^2)^2 \pi^4} \left( 1 + 2 m_f^2 C_0 \right)^2
\end{align}
where \(s \approx 2 m_\chi^2\) is the centre of mass energy of the annihilating DM, and \(C_0\) is the scalar integral \(C_0(0,0,s;m_f^2, m_f^2, m_f^2)\) in LoopTools notation \cite{Hahn:1998yk}.

\begin{figure}
\centering
\includegraphics[width=0.45\textwidth]{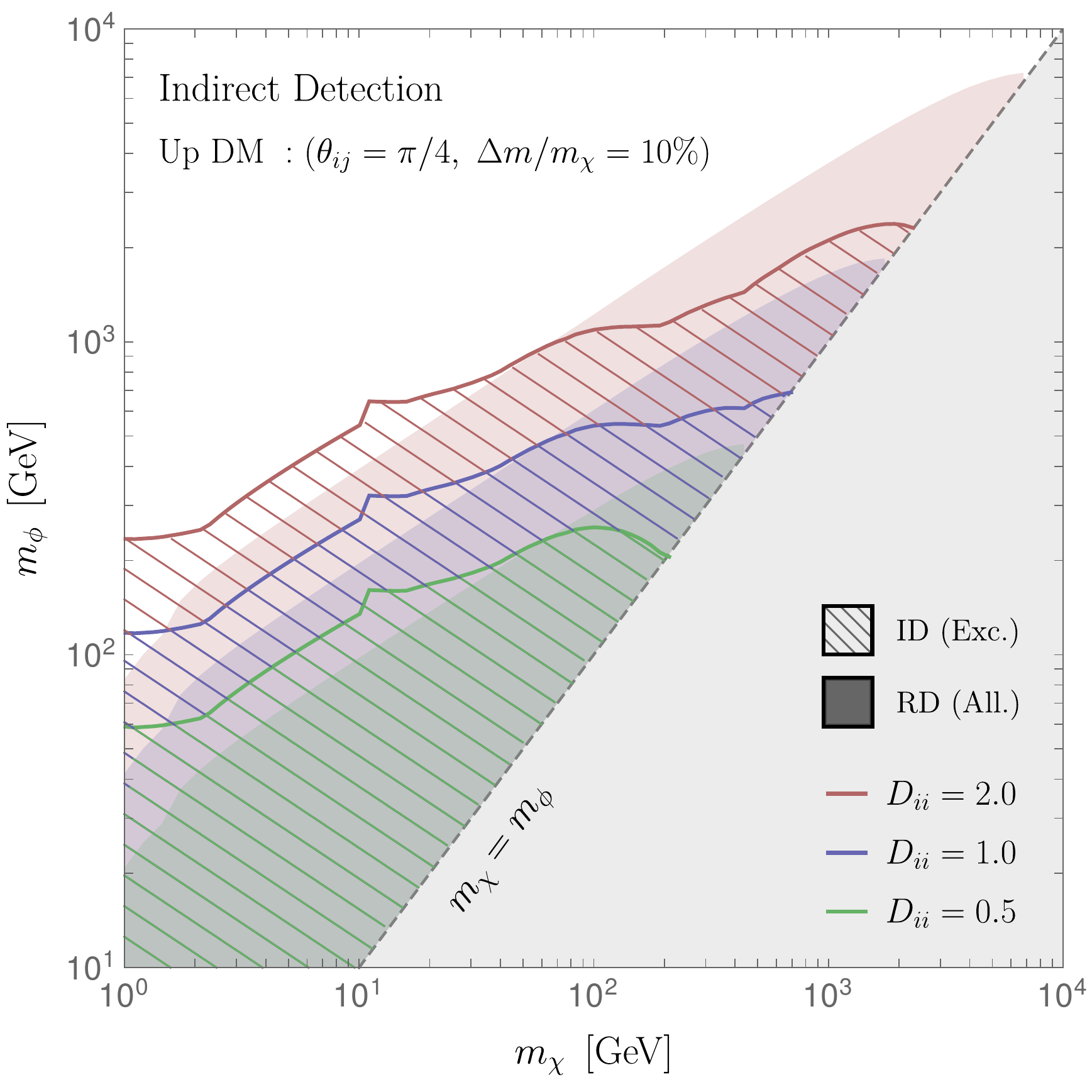}
\includegraphics[width=0.45\textwidth]{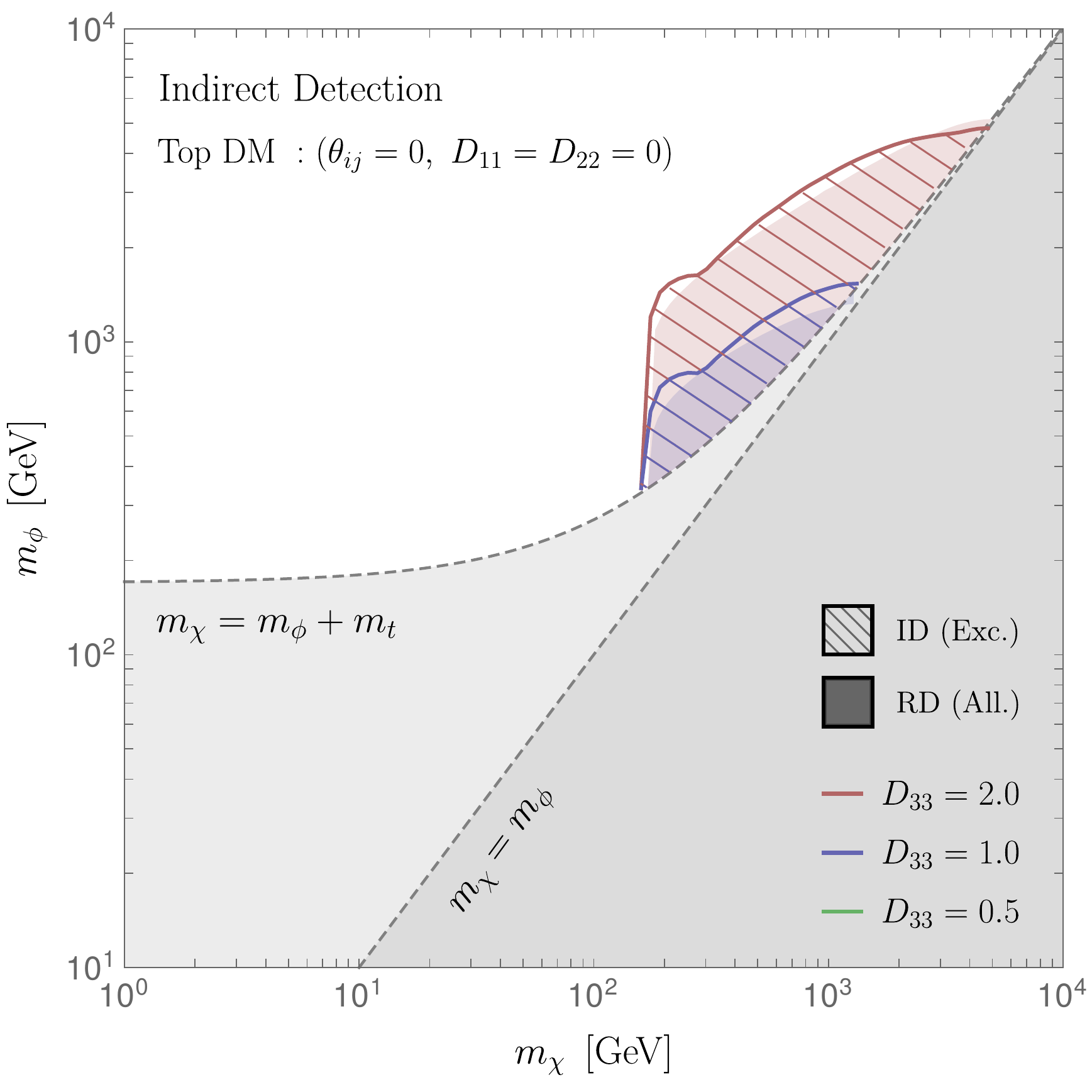}
\caption{The ID constraints on DMFV model, with `maximal' mixing \(\theta_{ij} = \pi / 4\) (left), or for couplings to top quarks only (right), assuming degenerate DM masses.
Bounds are produced on individual final states, and therefore scale with the dominant annihilation channel, somewhat surprisingly the top quark channel gives stronger constraints due to the extremely sensitive $\gamma$-ray search by H.E.S.S \cite{Lefranc:2015vza}.}
\label{Fig:IDExclusionsMassPlane}
\end{figure}

As well as \(\gamma \gamma\) final states, there will be \(\gamma X\) final states where \(X = Z,h\) for example and these also provide constraints. The presence of a massive particle recoiling against the photon shifts the energy to \(E_\gamma = m_\chi (1 - m_X^2 / 4 m_\chi^2)\), but still creates a mono-energetic line signature.
We show some results from the indirect searches in \cref{Fig:IDExclusionsMassPlane} -- we see that indirect searches can be quite powerful, especially in the case of large coupling to top quarks.

\section{Collider Constraints}
\label{Sec:Collider}

Our DMFV model contains a new particle with colour charge, and so we expect there to be significant limits coming from collider experiments.
In addition we also have DM which can be searched for in final states with missing energy, and current LHC data can also place limits on the mass of invisible particles.
In the past, DM model builders have used effective field theories (EFTs) to analyse NP at colliders, but in recent years it has become clear that the regions of validity of these EFTs at high energy machines such as the LHC are so small as to be almost useless \cite{Abdallah:2014hon,Abercrombie:2015wmb,Boveia:2016mrp}.
We briefly detail in the next section this point for our particular model, before moving on to a more complete analysis.

\subsection{EFT Limit}

\begin{figure}
\centering
\includegraphics[width=0.5\textwidth]{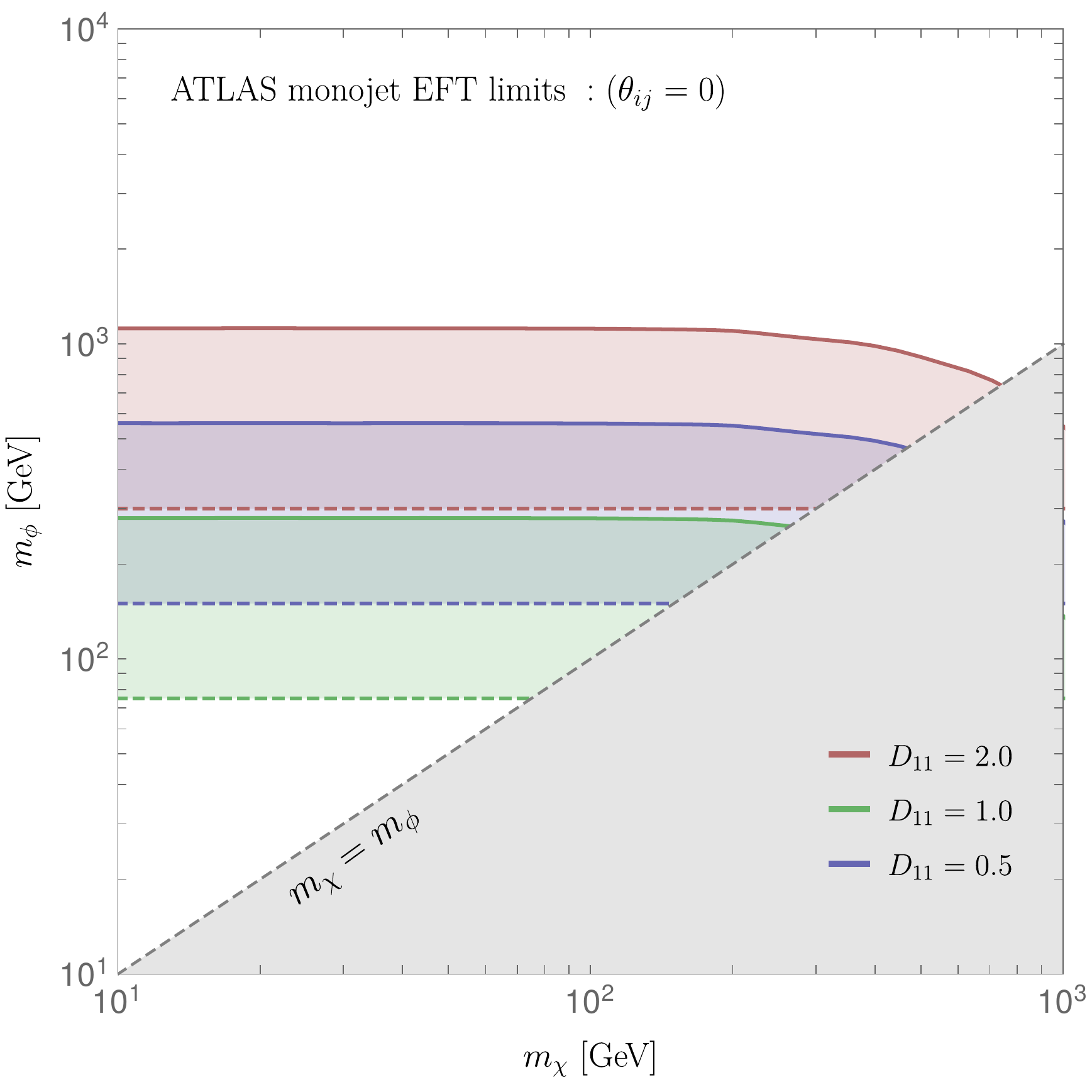}
\caption{The EFT approximation breaks down beneath the dashed lines (which are the \(R_\Lambda = 0.5\) contours with \(g \lesssim 4 \pi\)), while ATLAS excludes below the solid lines, and so only the shaded regions can robustly be excluded using the EFT.}
\label{Fig:EFTColliderBounds}
\end{figure}

In \cite{Busoni:2014haa} the validity of the EFT approximation for t-channel mediators is quantified by \(R_\Lambda\), which they define as the ratio of the cross section with the constraint \(t < \Lambda^2\) applied to the total cross section (i.e.\ the total proportion of the cross section which is valid under the EFT assumption).
The lines of \(R_\Lambda = 0.50\) are plotted alongside the EFT limits taken from ATLAS \cite{Aad:2015zva} (the \(R_\Lambda\) contour assumes \(| \eta| < 2\) and \(p_T < \SI{2}{\TeV}\), the ATLAS results assumed the same range of \(\eta\), but allow \(p_T \lesssim \SI{1.2}{\TeV}\)).
It is worth noting that the authors of \cite{Busoni:2014haa} produce results with the limit \(g \lesssim 1\), the bounds become significantly weaker by using \(g \lesssim 4 \pi\) which then permit a small region of validity as shown in \cref{Fig:EFTColliderBounds}.
The EFT breaks down entirely for \(g \lesssim 1\). Thus the EFT approximation cannot be justified in our analysis and we turn to the simulation of the full cross section.

\subsection{LHC bounds}
To try and cover a large range of constraints, we look at three different LHC processes that could place limits on our model -- monojet  with missing energy searches, where a single jet recoils off DM pair production; dijet searches with missing energy; and stop searches.
The latter are relevant to our model as we have a coloured scalar coupling to top quarks and DM, in analogy with the e.g.\ stop-top-neutralino vertex in many supersymmetric theories, and provide sensitivity to the \PphiDM-\Pqt coupling \(D_{33}\).

\begin{figure}
\centering
\includegraphics[width=0.45\textwidth]{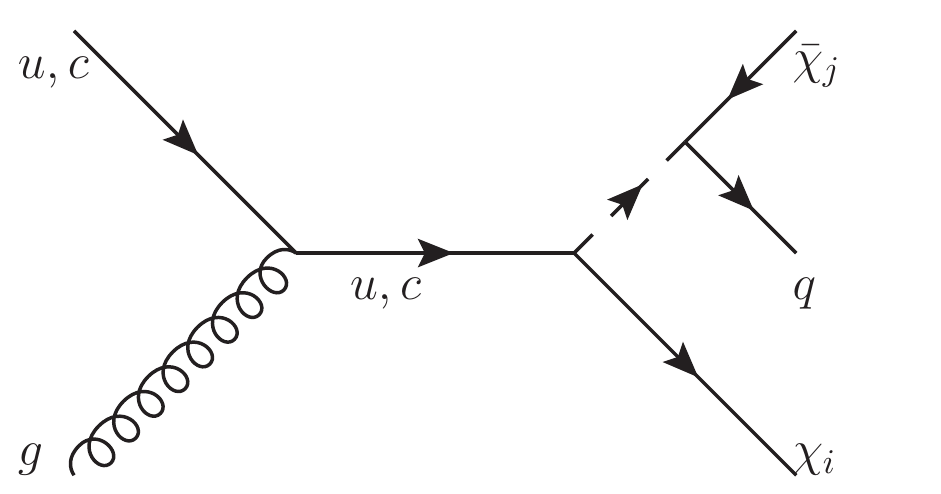}
\includegraphics[width=0.45\textwidth]{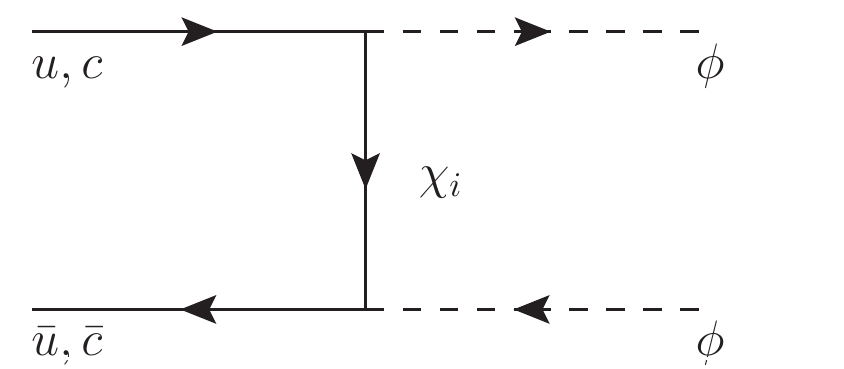}
\caption{Example Feynman diagram for the monojet (left) and dijet (right) processes.}
\label{Fig:ColliderFeynmanExample}
\end{figure}

In \cref{Fig:ColliderFeynmanExample} one example Feynman diagram that generates monojet and dijet signals is shown -- in the dijet case the decay of the mediator into quark plus DM is not shown.
Other diagrams that contribute can be seen in \cref{App:FeynmanDiagrams}.

We produce our collider constraints using MadGraph \cite{Alwall:2014hca}, replicating, except where noted below, the experimental cuts used by the experiments.

\subsubsection{Monojet searches}

\begin{figure}
\centering
\includegraphics[width=0.4\textwidth]{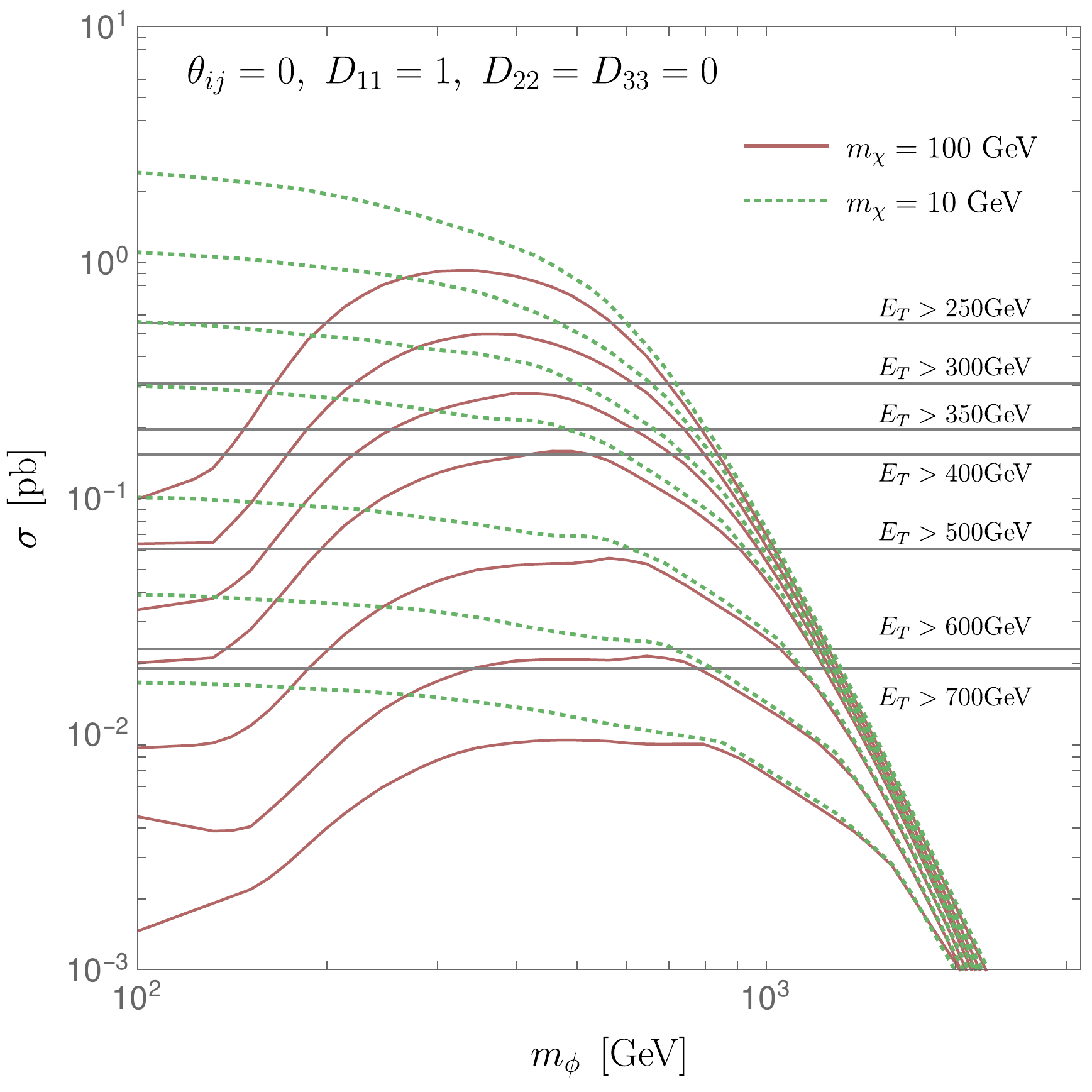}
\caption{Total cross section for the seven signal regions of the ATLAS monojet search \cite{Aaboud:2016tnv} for two DM masses.}
\label{Fig:SigmaMissingETplusJet}
\end{figure}

In our analysis, we use the most recent monojet search by ATLAS \cite{Aaboud:2016tnv} (which uses the Run 2 data (\(\sqrt{s} = \SI{13}{\TeV}\) and \(\mathcal{L} = \SI{3.2}{\per\fb}\))), along with a similar analysis performed by CMS \cite{Khachatryan:2014rra} with the Run 1 data (\(\sqrt{s} = \SI{8}{\TeV}\) and \(\mathcal{L}= \SI{19.7}{\per\fb}\)).
The total cross section as a function of \MphiDM for a benchmark scenario is shown in \cref{Fig:SigmaMissingETplusJet} with the ATLAS limits overlaid, and the constraints on our model are shown in the top of \cref{Fig:ColliderBounds}.

\subsubsection{Dijet searches}

Moving on to dijet searches, we use a Run 1 and Run 2 search by ATLAS \cite{Aad:2014wea,Aaboud:2016zdn} looking for multiple jets plus missing energy -- we restricted our comparison to the 2-jet searches which should provide the strongest constraint.
In our model, the process \HepProcess{\Pproton \Pproton \to \PphiDM \PphiDM \to \bar{\chi} \chi j j} provides the dominant contribution to this signal.

\begin{table}
\centering
\begin{tabular}{c c c c c}
\hline
SR & \(N_\text{obs}\) & \(N_\text{SM}\) & \(N_\text{n.p.}\) & \(\sigma_\text{obs}\) / \si{\fb} \\
\hline \hline
\texttt{tjl} & \num{12315} & \num{13000 \pm 1000} & \numrange{15}{704} & 60 \\
\texttt{tjm} & \num{715} & \num{760 \pm 50} & \numrange{15}{59}  & 4.3 \\
\texttt{tjt} & \num{133} & \num{125 \pm 10} & \numrange{22}{50} & 1.9 \\ 
\hline
SR & \(N_\text{obs}\) & \(N_\text{SM}\) & \(N_\text{n.p.}\) & \(\sigma_\text{obs}\) / \si{\fb} \\
\hline \hline
\texttt{tjl} & \num{263} & \num{283 \pm 24} & \numrange{12}{37} & 16 \\
\texttt{tjm} & \num{191} & \num{191 \pm 21} & \numrange{15}{58}  & 15\\
\texttt{tjt} & \num{26} & \num{23 \pm 4} & \numrange{10}{22} & 5.2
\end{tabular}
\caption{Lower limits (at \SI{95}{\percent} CL) on the visible cross section for three signal regions (SR) in the Run 1 ATLAS dijet plus missing \(E_T\) search \cite{Aad:2014wea} (top), and ATLAS dijet search from Run 2 \cite{Aaboud:2016zdn} (bottom).}
\label{Tab:ATLASdijet}
\end{table}

\begin{figure}
\centering
\includegraphics[width=0.45\textwidth]{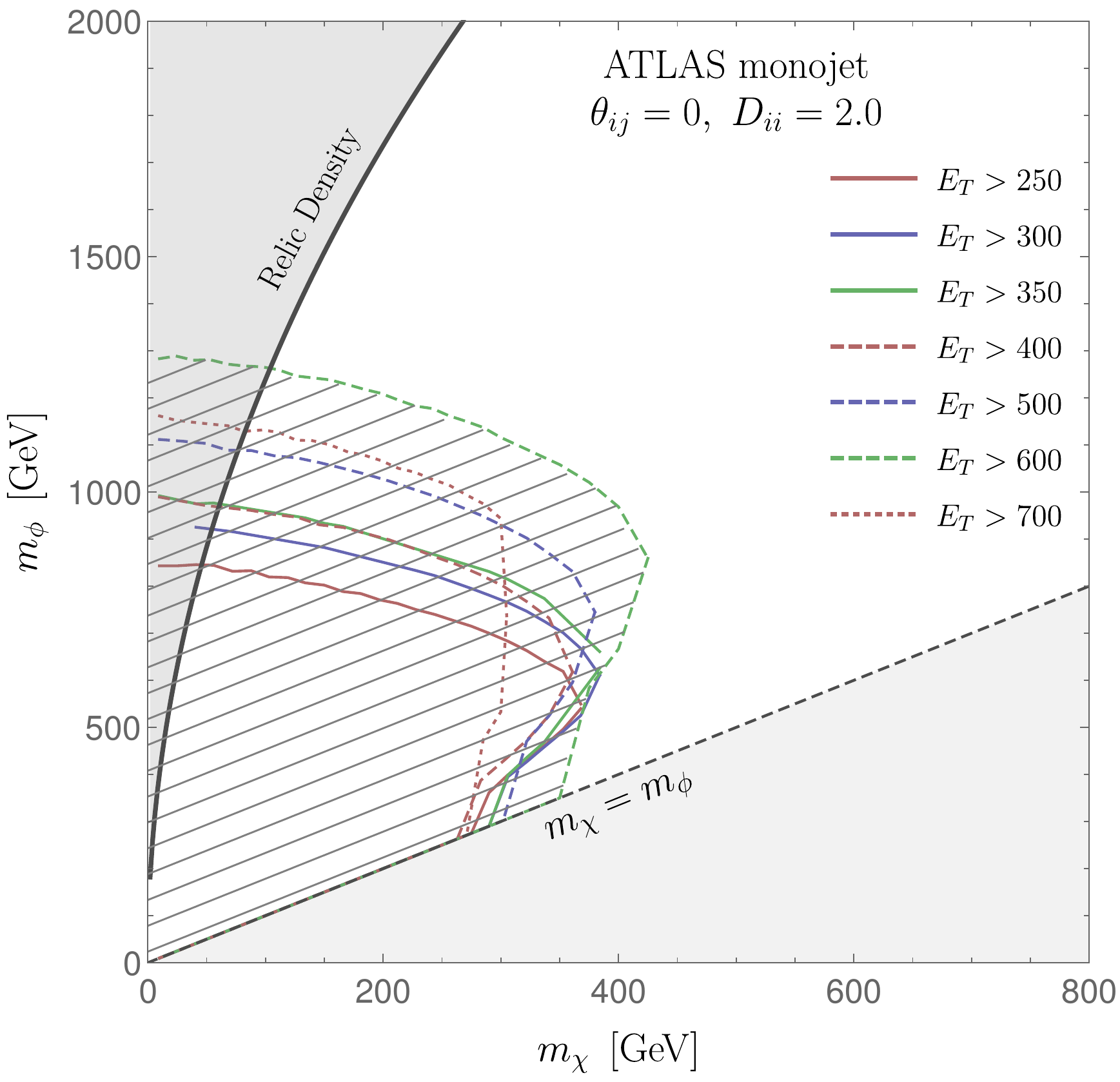}
\includegraphics[width=0.45\textwidth]{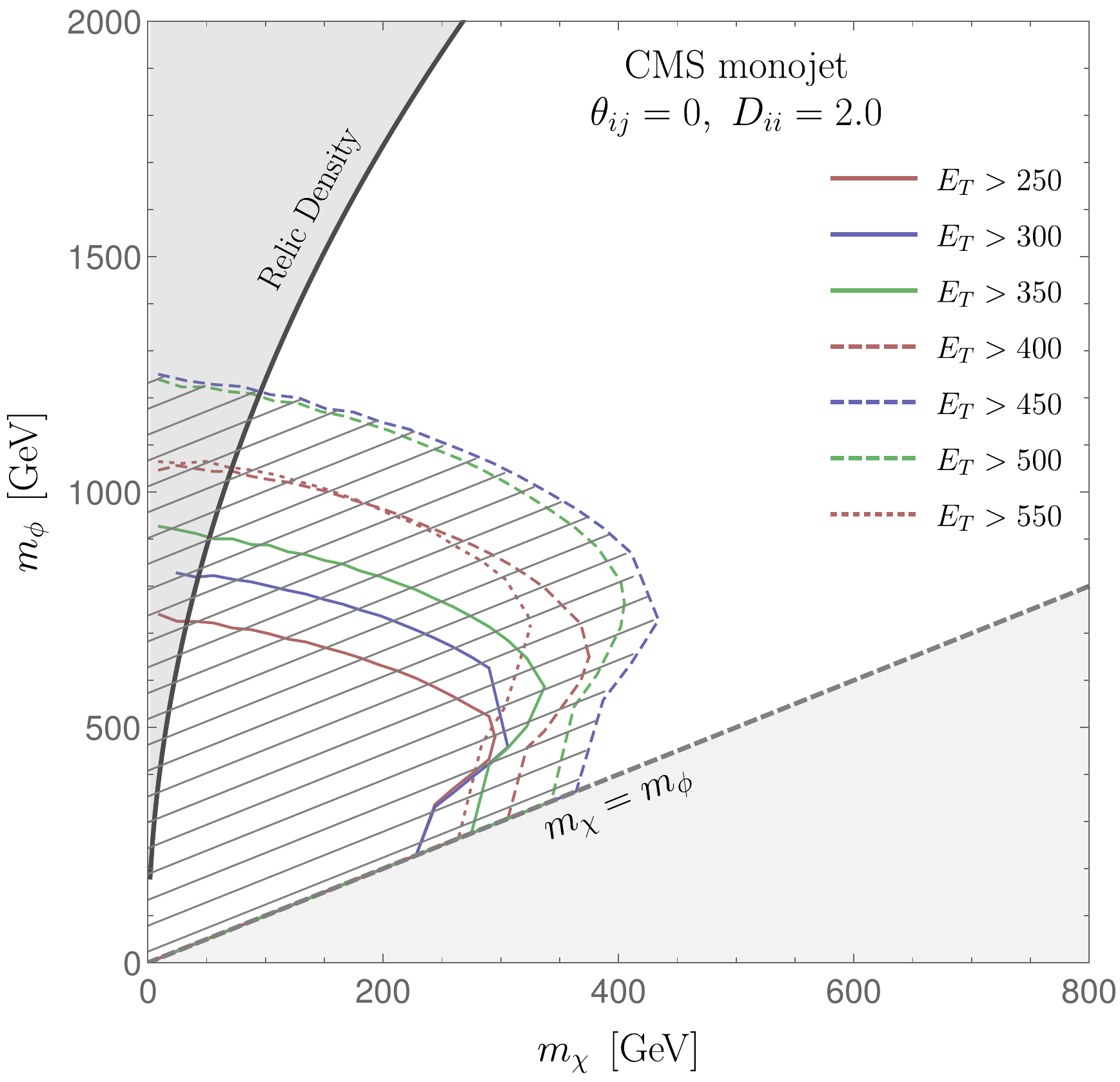}
\\
\includegraphics[width=0.45\textwidth]{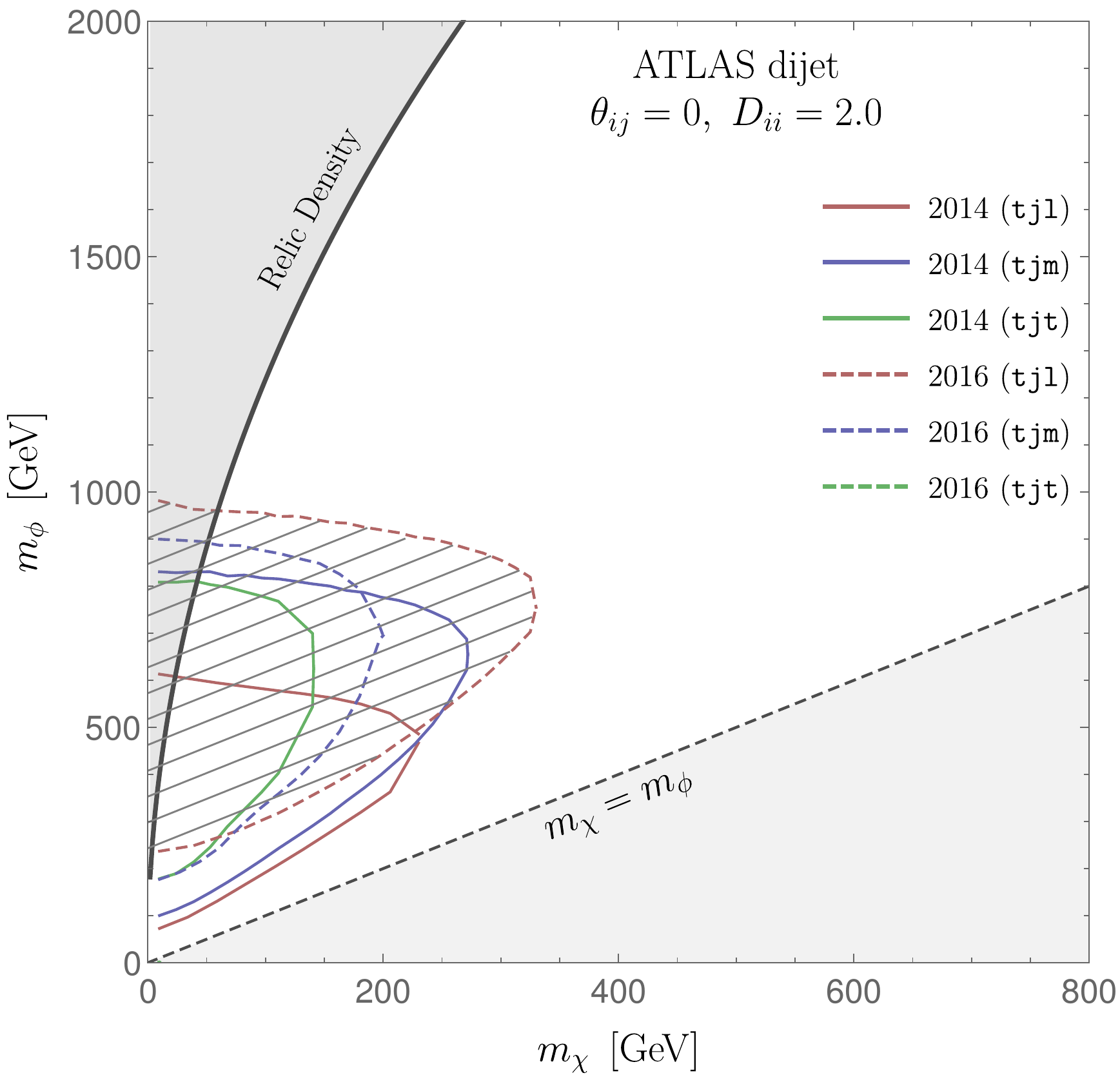}
\includegraphics[width=0.45\textwidth]{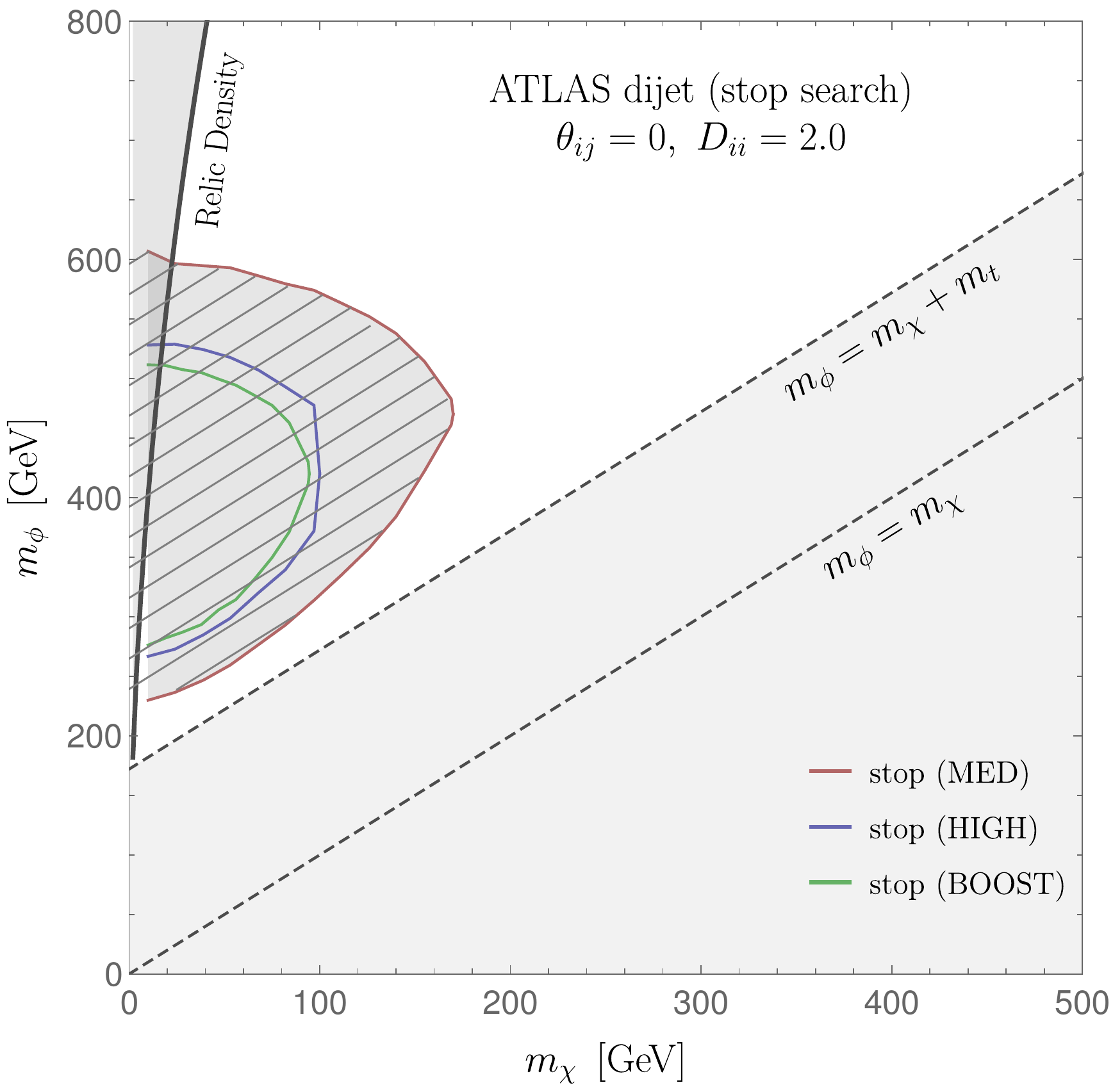}
\caption{Exclusion regions for different signal regions in the ATLAS (top left) and CMS (top right) monojet analyses, ATLAS dijet searches (bottom left), and ATLAS stop searches (bottom right).}
\label{Fig:ColliderBounds}
\end{figure}

We replicate all the main selection cuts for both analyses, in particular for the Run 1 comparison: \(E_t^\text{miss} > \SI{160}{\GeV}\), \(p_{T,(1,2)} > 130, 60\,\si{\GeV}\), \(\Delta \phi > 0.4\) (between the jets and missing momentum), and for Run 2 similar cuts are applied (full detail in Table~2 of \cite{Aaboud:2016zdn}).
The different signals regions (\texttt{tjl}, \texttt{tjm}, \texttt{tjt}) also include a minimum requirement for \(m_\text{eff}\) and \(E_{T} / \sqrt{H_T}\), which are defined as
\begin{align*}
H_T &= |p_{T,1}| + |p_{T,2}| \\
m_\text{eff} &= H_T + E_T \, ,
\end{align*}
 which we implement in MadGraph manually via Fortran code (again, see the respective papers for the cuts in each case).
The constraints this places on our model parameters are shown in the bottom left of \cref{Fig:ColliderBounds} for the case of no mixing and strong couplings for all DM particles.

\subsubsection{ATLAS 2014 Stop Search}

Lastly, a study by ATLAS \cite{Aad:2014kra} considers a set of cuts optimized for the detection of stops -- the signal consists of a lepton in the final state along with four or more jets.
There are four relevant signal regions \texttt{tN\_diag}, \texttt{tN\_med}, \texttt{tN\_high}, \texttt{tN\_boost}, each requiring a single lepton with \(p_T^l > \SI{25}{\GeV}\), and cuts in \cref{Tab:ATLASdijetStop}.\footnote{We do not include the cuts on the parameters \(am_{T2}\) and \(m^\tau_{T2}\).
From the published cut flows it can be seen that the effect of these cuts is of the order \SI{10}{\percent} and \SI{2}{\percent} respectively (although the former cut can have a more pronounced effect \(\sim \SI{30}{\percent}\) on the \texttt{tN\_med} cut choice).}

\begin{table}
\centering
\begin{tabular}{c c c c c}
Cut  & \texttt{tN\_diag} & \texttt{tN\_med} & \texttt{tN\_high} & \texttt{tN\_boost} \\
\hline \hline
\(E_{T}^\text{miss}\) / \si{\GeV} & 100 & 200 & 320 & 315  \\
\(p_{T,i}^{j}\) / \si{\GeV} & 60, 60, 40, 25 & 80, 60, 40, 25 & 100, 80, 40, 25  & 75, 65, 40, 25\\
\(m_T\) / \si{\GeV} & 60 & 140 & 200 & 175  \\
\(\Delta R(b, l)\) & 0.4 & 0.4 & 0.4 & 0.4  \\
\(\Delta \phi (j_{1,2},p_T^\text{miss})\) & 60 & 140 & 200 & 175  \\
\hline
Bound \(\sigma_\text{vis}\) / \si{\fb} &  \numrange{1.8}{2.9} & 0.4 & 0.3 & 0.3
\end{tabular}
\caption{The four relevant signal regions from \cite{Aad:2014kra} and the cuts we have implemented.}
\label{Tab:ATLASdijetStop}
\end{table}

We find that the production of the \PphiDM pair is dominated by t-channel \(\chi\) exchange and s-channel gluons; the photon and Z mediated diagrams are neglected.
We calculate in MadGraph the cross-section for a single final state (\((\APqb \Pqb) (\APqd \Pqu) + \Pelectron \)), and then multiply this by four to account for the different top quark decay options (the \(p_T\) cut means the different masses have a negligible effect).
Although the cross section is predominantly controlled by the size of \(D_{33}\), the light quark couplings \(D_{11}, D_{22}\) have a mild affect by reducing the branching ratio \HepProcess{\PphiDM \to \Pqt \bar{\chi}_i} and hence suppressing the cross section.

We also examined constraints from a similar ATLAS search for scharms \cite{Aad:2015gna} rather than stops, searching for c-tagged jets plus missing energy in the region where the branching ratio \(\PphiDM \to \Pqc \chi_i\) is large. The limits on $m_{\phi,\chi}$ are similar to the stop search, and thus do not warrant further attention when compared to the dijet searches.

\subsection{Collider Constraints within DMFV}

We have now looked at three classes of analysis: monojet searches, dijet searches, and searches optimised for a stop. Within our model we have couplings to \Pqu, \Pqc, \Pqt (which we denote here by \(\lambda_{u,c,t}\)) and the relative strengths of these dictate which signals will be dominant.

Compared to \(\lambda_u\), the monojet and dijet processes are suppressed by pure \(\lambda_c\) (due to the charm parton distribution function (PDF)), but generally are enhanced by mixtures of \(\lambda_{u,c}\).
The coupling \(\lambda_t\) reduces the signals since they dominantly come from s-channel \PphiDM resonances and thus the branching ratio to \Pqu, \Pqc jets is \(\propto (D_{33})^{-2}\) if \(\lambda_t \gg \lambda_{u,c}\).
The stop search only becomes relevant for large \(\lambda_t\) with \(\lambda_t /\lambda_{u,c} > 1\), and increasing \(\lambda_{u,c}\) suppresses the signal as the branching fraction to top quarks is reduced.

\begin{itemize}
\item \textbf{Mostly up-type}: The dominant signal will come from the monojet processes which have the least QCD suppression and which require an up quark in the initial state.
Dijet searches are also sensitive but it tends to be the monojet which sets the better constraint. 
\item \textbf{Mostly charm-type}: The monojet processes are enhanced by the presence of charm couplings, however as the up coupling is reduced the monojet processes become suppressed by the charm PDF by around a factor 10--100.
The dijet processes are very similar as for \Pqu quarks but the largest contributing diagram is again suppressed by the charm PDF.
Both searches provide constraints.
\item \textbf{Mostly top-type}: The monojet signal depends primarily on \(\lambda_{u,c}\), only indirectly on \(\lambda_t\) though the widths.
$\lambda_t$ can be probed through stop searches with jet multiplicities of \(\geq 4\).
\end{itemize}

Colliders provide very powerful exclusions (up to the \si{\TeV} scale in mediator mass), and cover the full model parameter space in coupling, although these can be significantly weakened by, for example, strong top couplings.
The DM is produced \emph{on shell}, and so the constraints are comparatively weak at high DM mass when compared with searches which depend on the cosmic abundance of DM; on the other hand the fact that the DM is produced in the collider releases any dependence on its abundance in the universe, thus allowing more powerful constraints on DM which has only a fraction of the full relic abundance (or none at all).
Similarly, low mass DM is strongly excluded, whereas the most powerful astrophysical probe (direct detection) cannot detect much below the \si{\GeV} scale due to kinematics.

When compared with the strongest direct detection limits, the collider limits are not as constraining, and this is not likely to change even with more luminosity and higher energy beams.

It is very difficult for a given parameter choice to determine the strongest bound from colliders, except in the extreme cases above, and one should therefore check all available searches as we have done.
Due to the interplay between 1 and 2 jet processes, there is no obvious scaling behaviour of the cross section with the coupling parameters, these factors make implementing collider searches in an MCMC scan difficult and slow as each cross section must be numerically computed at each point in phase space.

\section{Results}
\label{Sec:Results}

We have aimed to produce a robust statistical analysis of the eight dimensional parameter space of the DMFV model, using the Bayesian inference tool MultiNest \cite{Feroz:2007kg,Feroz:2008xx,Feroz:2013hea} and its Python interface PyMultiNest \cite{Buchner:2014nha} with 5000 live points.
The motivation for carrying out this analysis is twofold, firstly from a practical standpoint it enables very quick and efficient algorithms for scanning a large dimensional parameter space, allowing us to include all parameters in one analysis.
Secondly, a rudimentary ``hit-or-miss'' analysis leaves a large region of parameter space allowed, which is not surprising given the flexibility of 8 free parameters, with a statistical result we can quantify the regions of parameter space which are allowed but very improbable given the errors of the experimental data.
For clarity, we represent the allowed parameters as contours containing credible regions, using the method in \cite{Fowlie:2016hew}; using the posterior probability density function. The \(1,2\,\sigma\) contours give an indication of the allowed parameter range, with containment probabilities of \SIlist{68;95}{\percent} respectively.

\textbf{Regarding the use of priors:} We make one note of caution regarding the results; the credible regions depend sensitively on the choice of priors for the parameters. This is not surprising since our constraints allow large regions of parameter space to be equally well allowed, and so the use of priors which bias the parameters to lower values (i.e. log-uniform compared with linearly uniform) is reflected in the final result.
Nonetheless, we are careful to limit the statements made in the text to those which are independent of the choice of priors. In all figures the log-uniform priors have been used for the masses and for $D_{ii}$, as this represents the more conservative choice. The ranges and priors for the parameters of the scan are summarized in \cref{Tab:Priors}.
\begin{table*}
\begin{center}
\begin{tabular}{| c | c | c | }
\hline
Parameter & Range & Prior \\
\hline
\hline
\(m_\chi\ /\ \si{\GeV}\) & \numrange{1}{e5} & Log-Uniform \\
\(m_\phi\ /\ \si{\GeV}\) & \numrange{1}{e5} & Log-Uniform \\
\(\theta_{ij}\) &  \numrange[quotient-mode=fraction,fraction-function=\tfrac]{0}{\pi/4} & Uniform \\
\(D_{ii}\) &  \numrange{e-2}{4\pi} & Log-Uniform \\
\hline
\end{tabular}
\end{center}
\caption{Allowed ranges for the parameters used in the MCMC scan, along with the assumed prior likelihood, which is uniform on either a linear or logarithmic scale.}
\label{Tab:Priors}
\end{table*}

\begin{figure}
\centering
\includegraphics[width=0.45\textwidth]{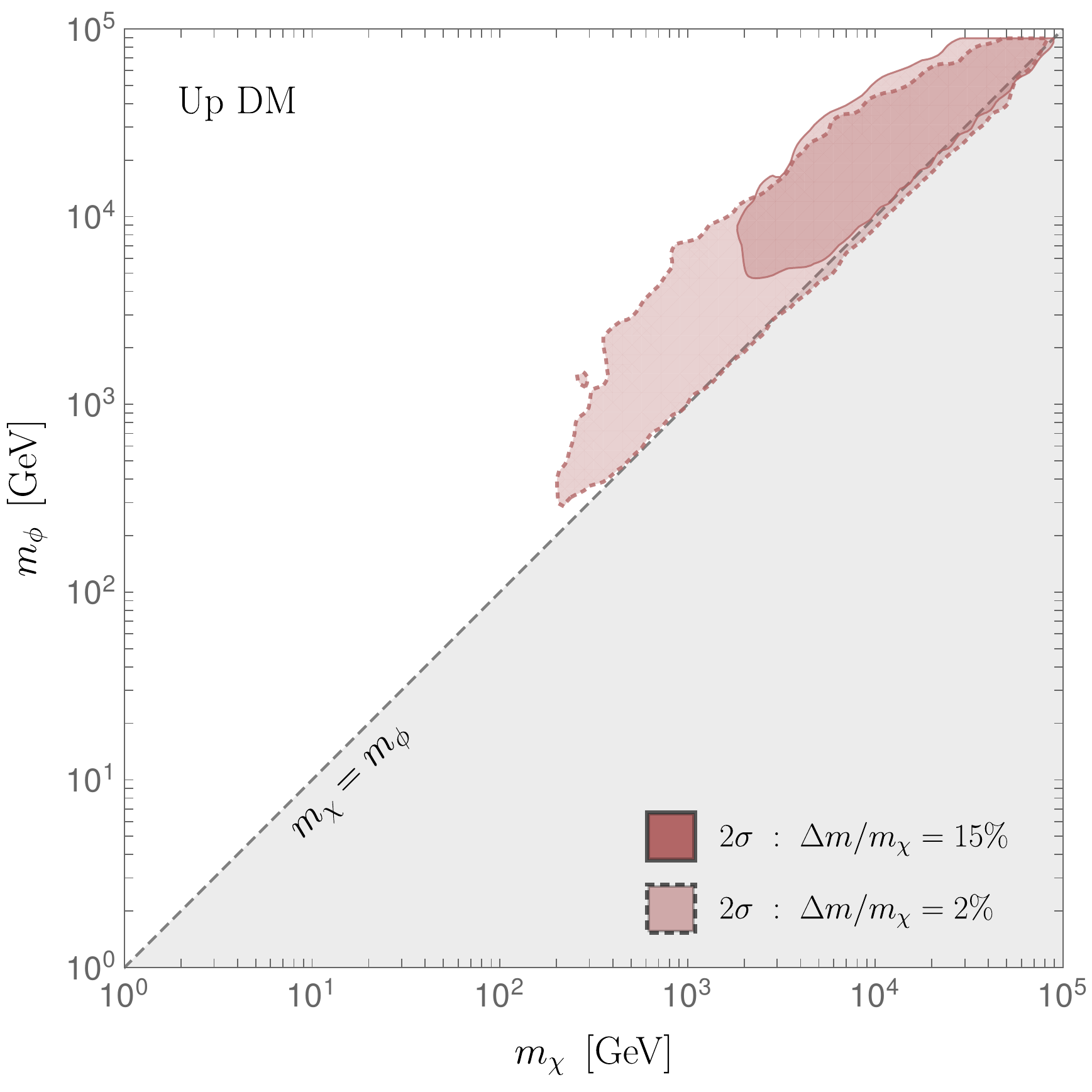}
\includegraphics[width=0.45\textwidth]{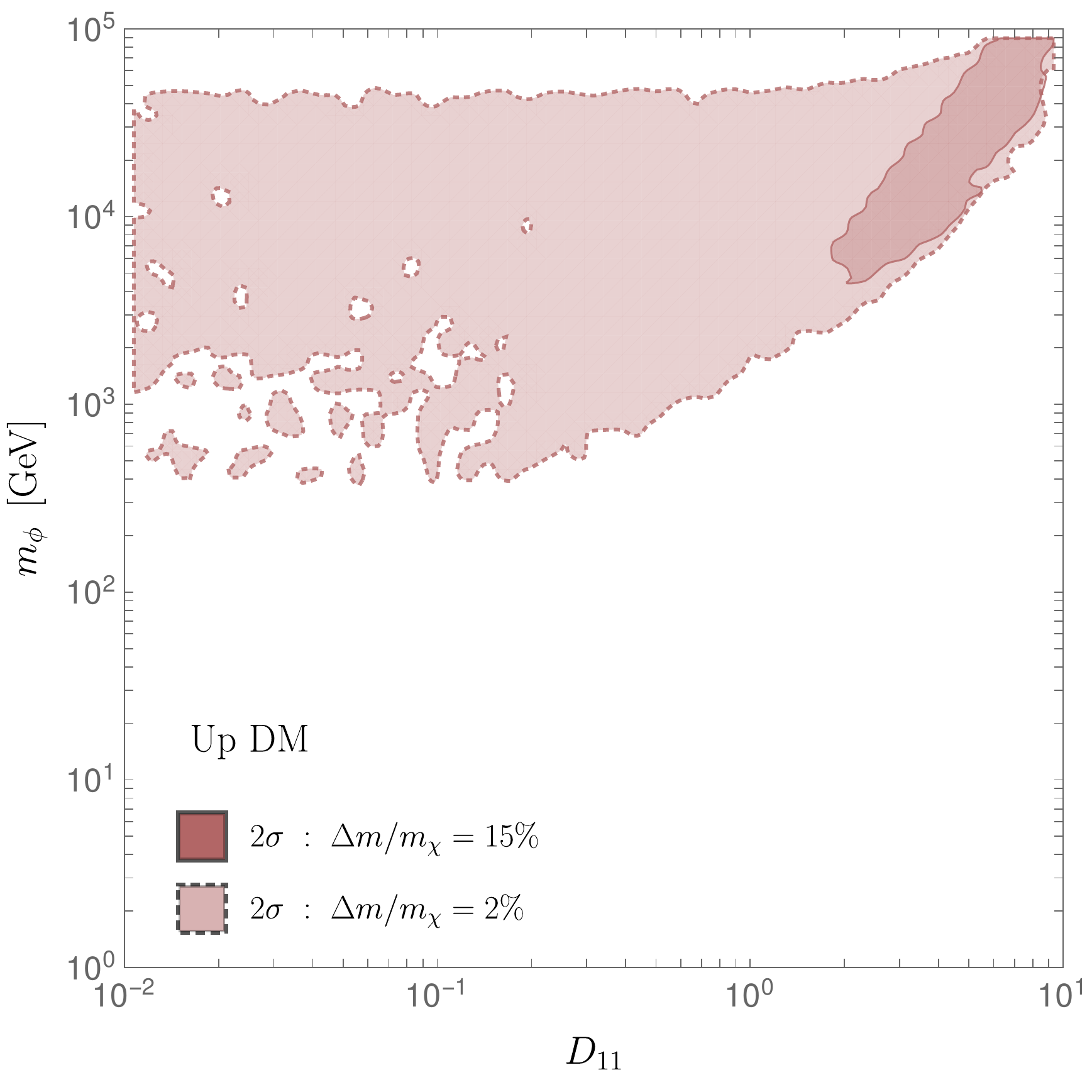}
\includegraphics[width=0.45\textwidth]{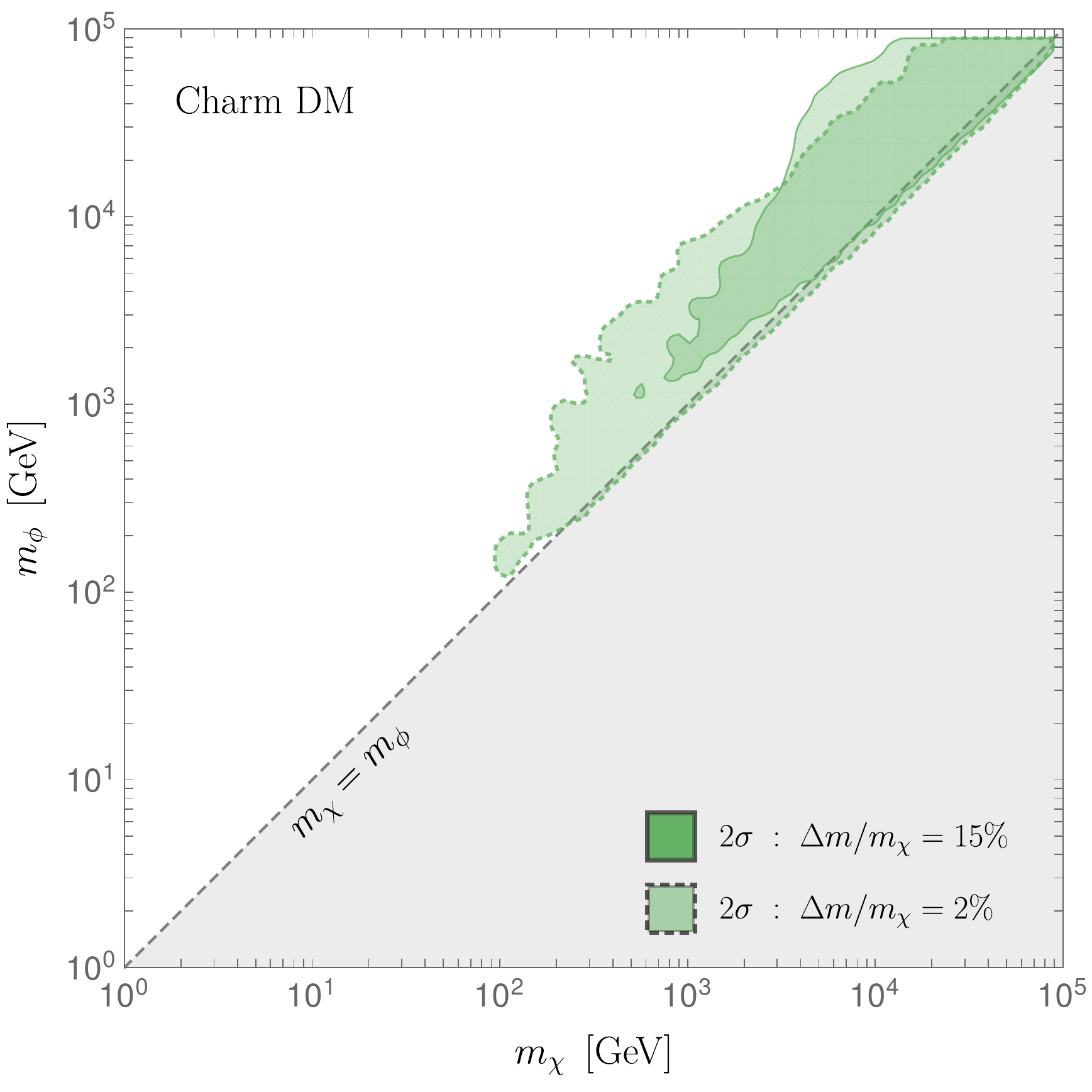}
\includegraphics[width=0.45\textwidth]{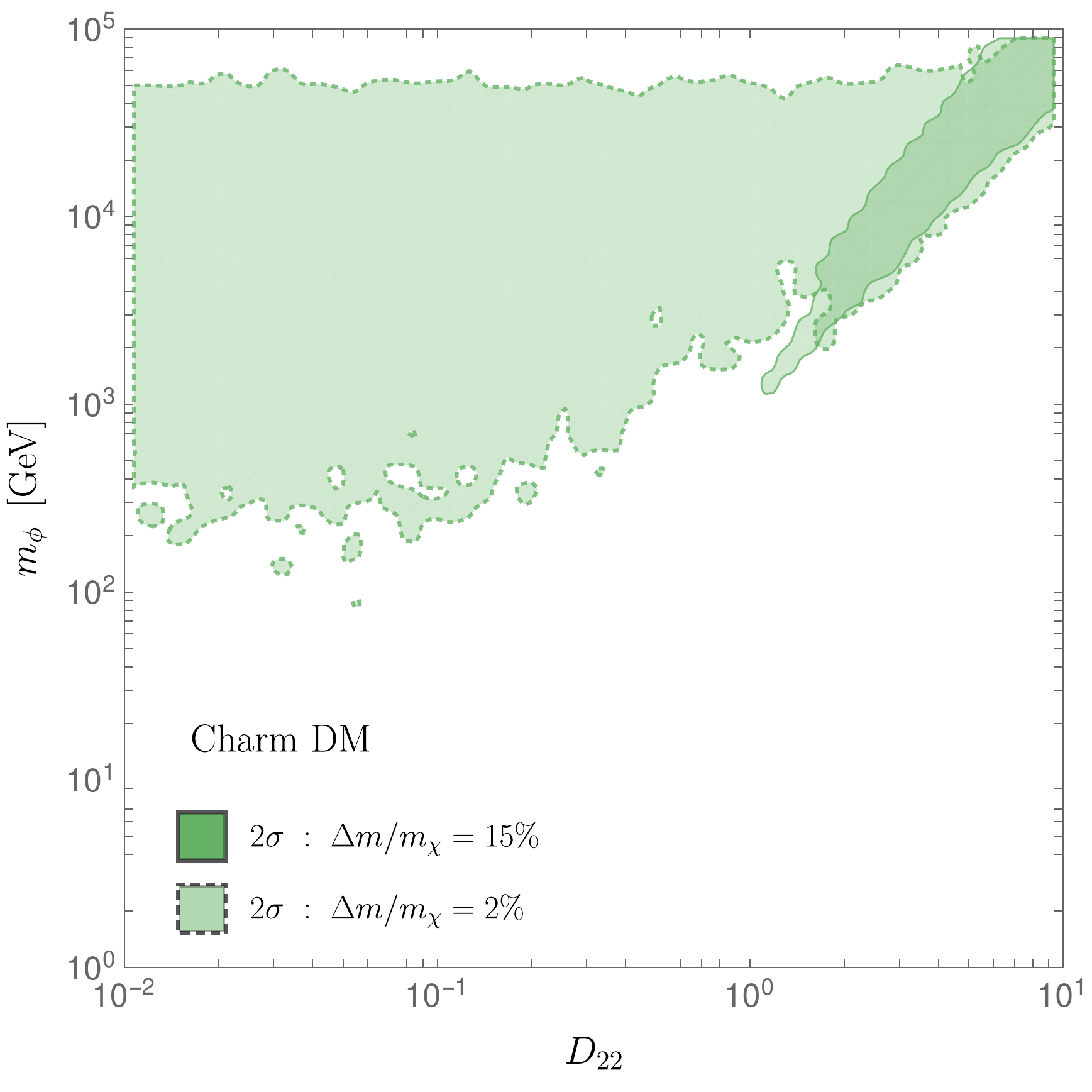}
\includegraphics[width=0.45\textwidth]{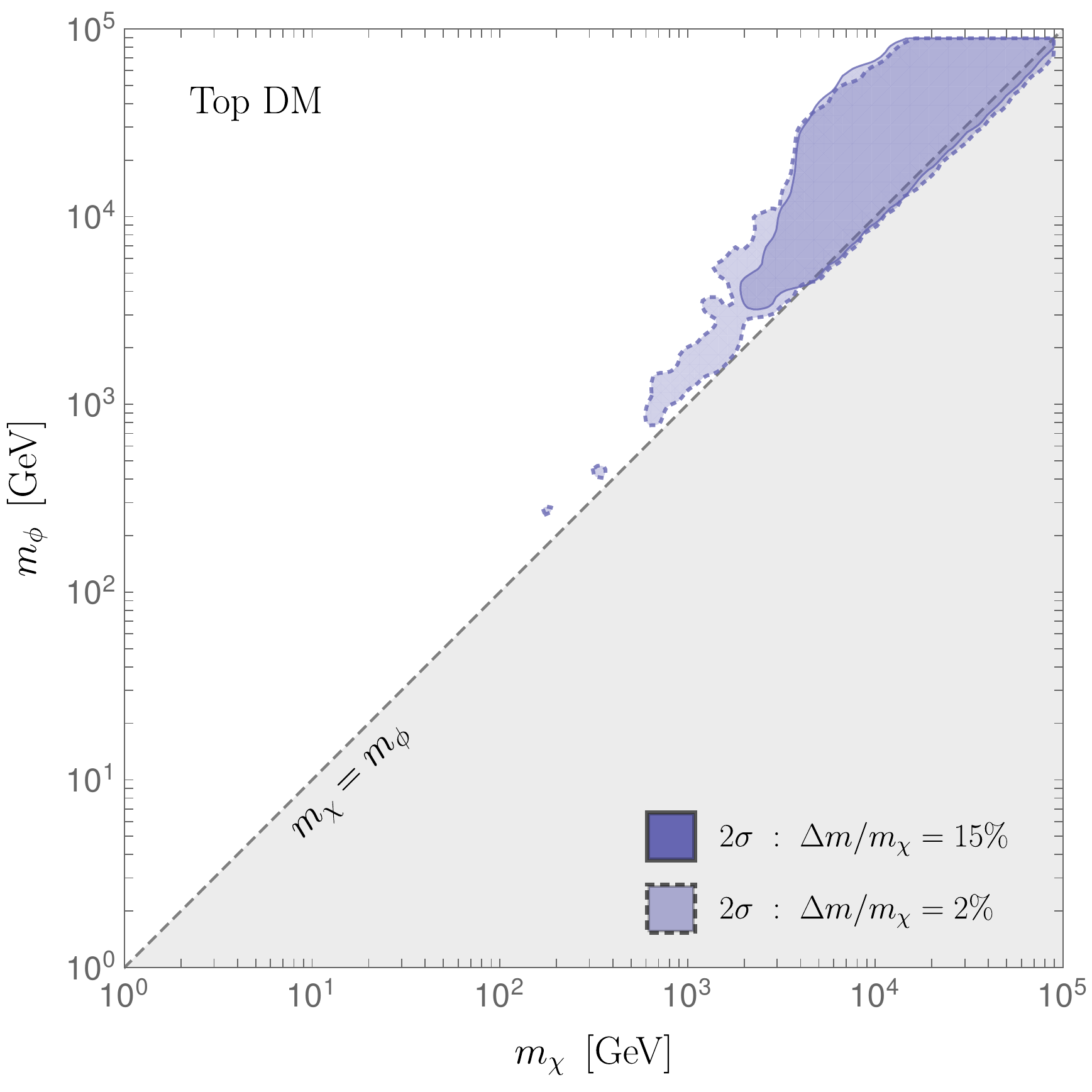}
\includegraphics[width=0.45\textwidth]{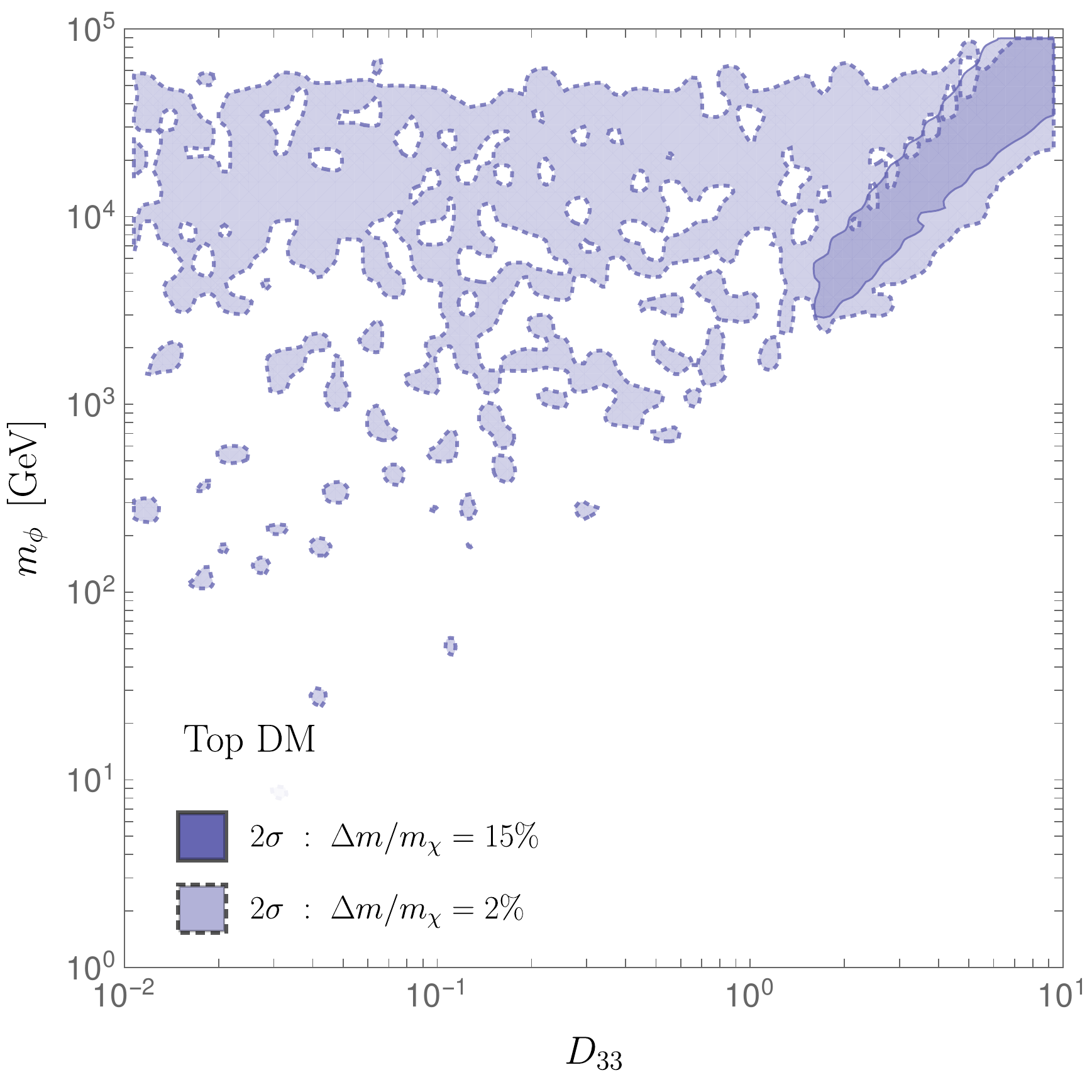}
\caption{Credible regions (\(2\,\sigma\) contours) in the \(m_\chi-\MphiDM\) plane (left) and \(D_{ii}-\MphiDM\) (right) where the DM is \(\chi_1\) (top), \(\chi_2\) (middle) or \(\chi_3\) (bottom). Two values of a mass splitting are chosen, shown with solid and dashed contours respectively.}
\label{Fig:ResultsMass}
\end{figure}

\begin{figure}
\centering
\includegraphics[width=0.45\textwidth]{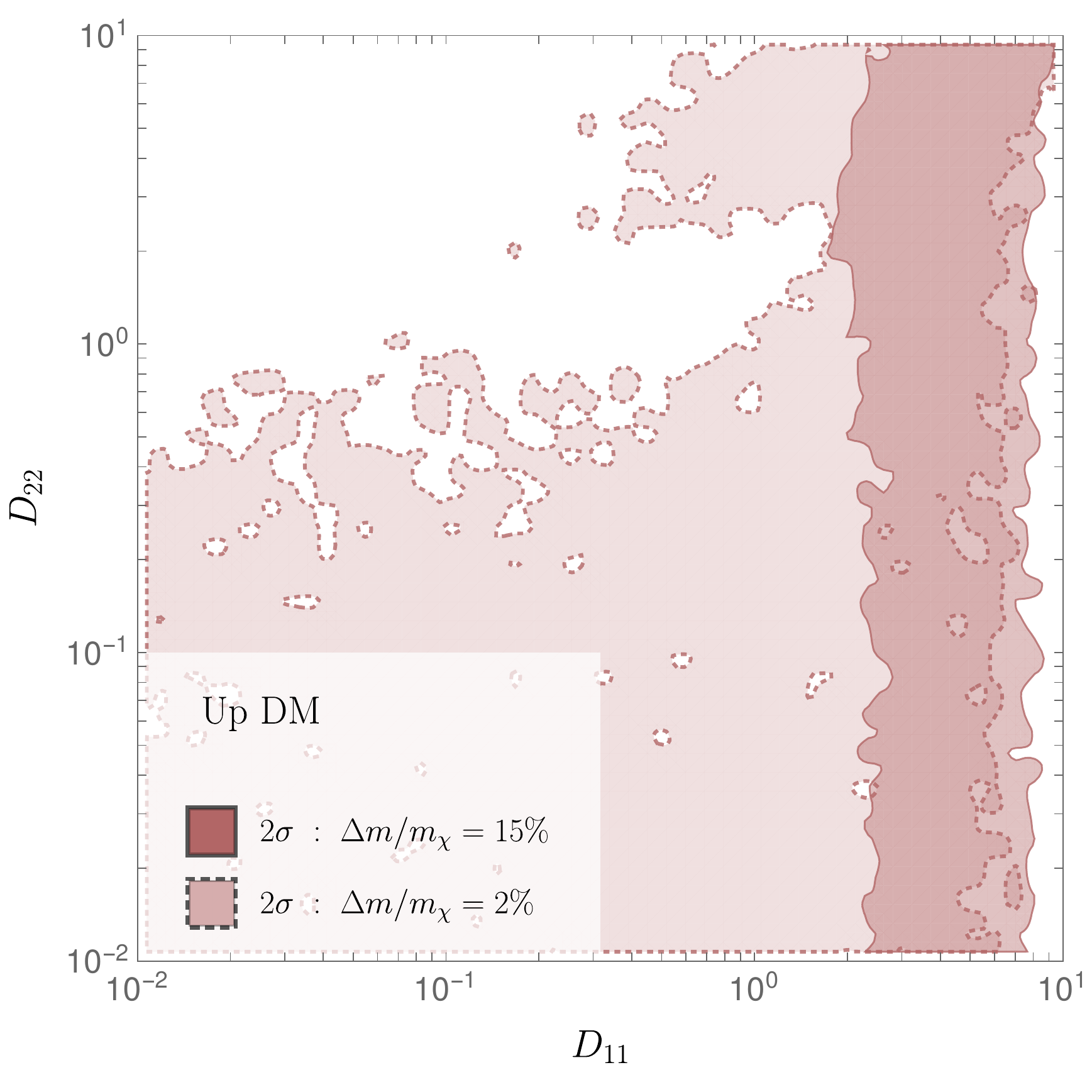}
\includegraphics[width=0.45\textwidth]{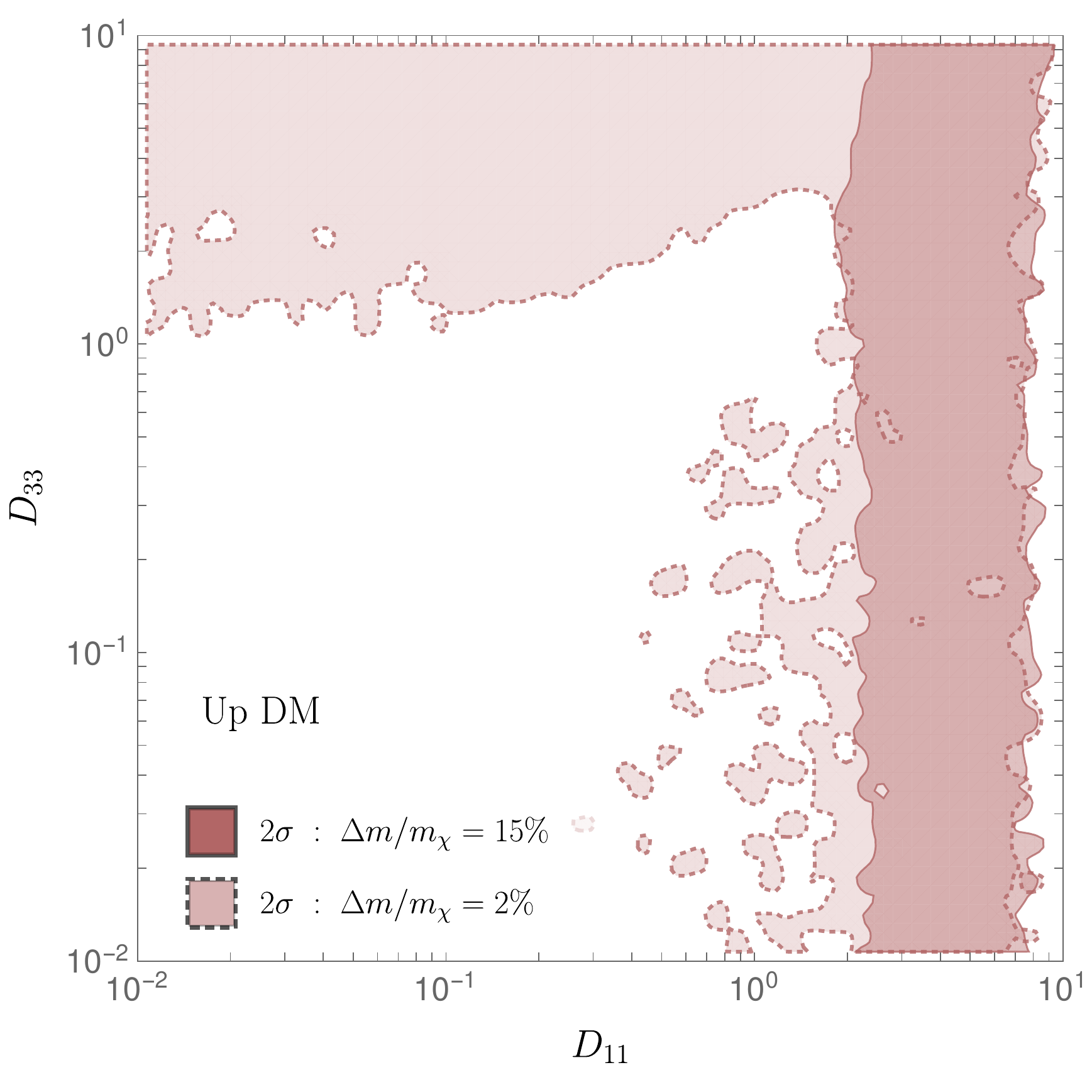}
\includegraphics[width=0.45\textwidth]{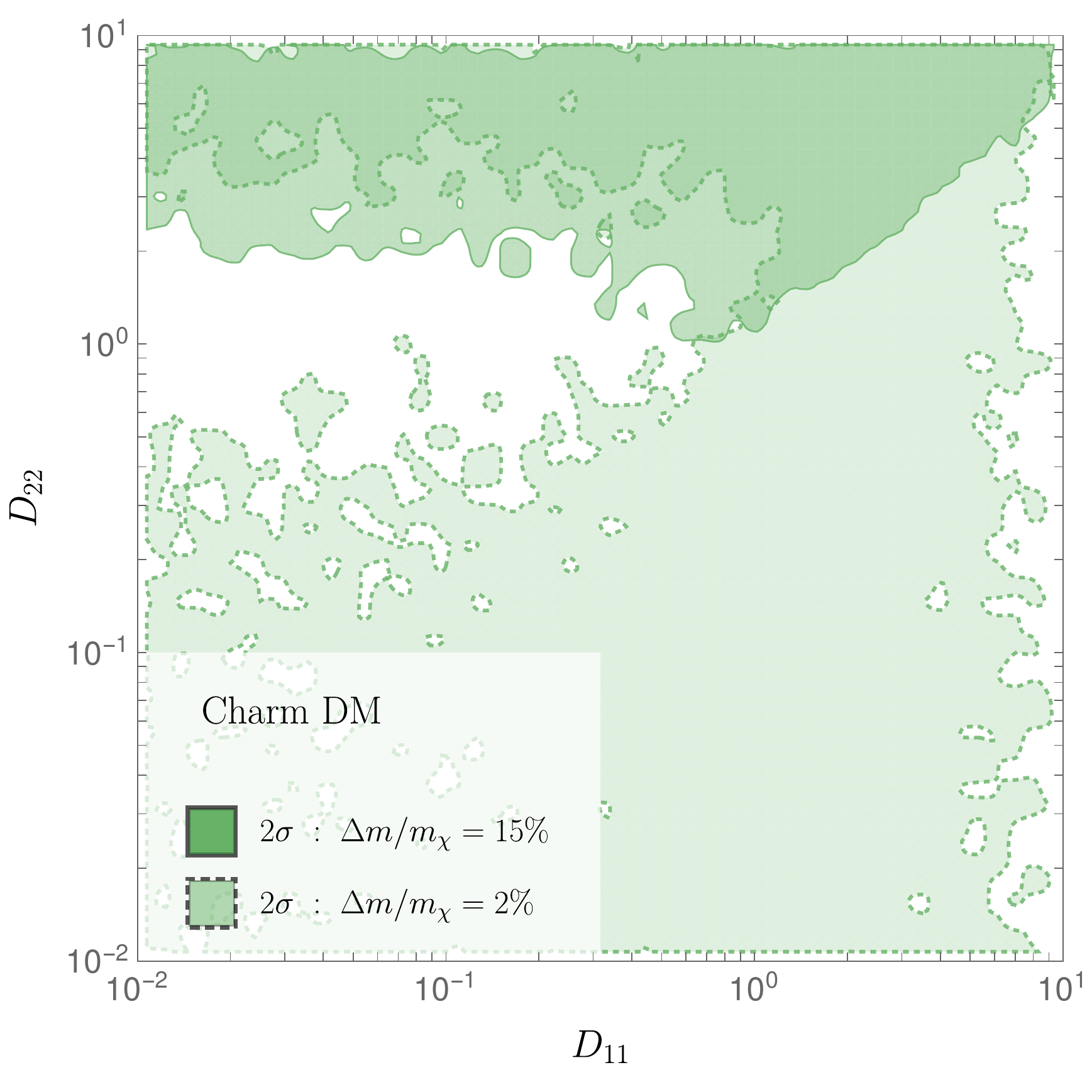}
\includegraphics[width=0.45\textwidth]{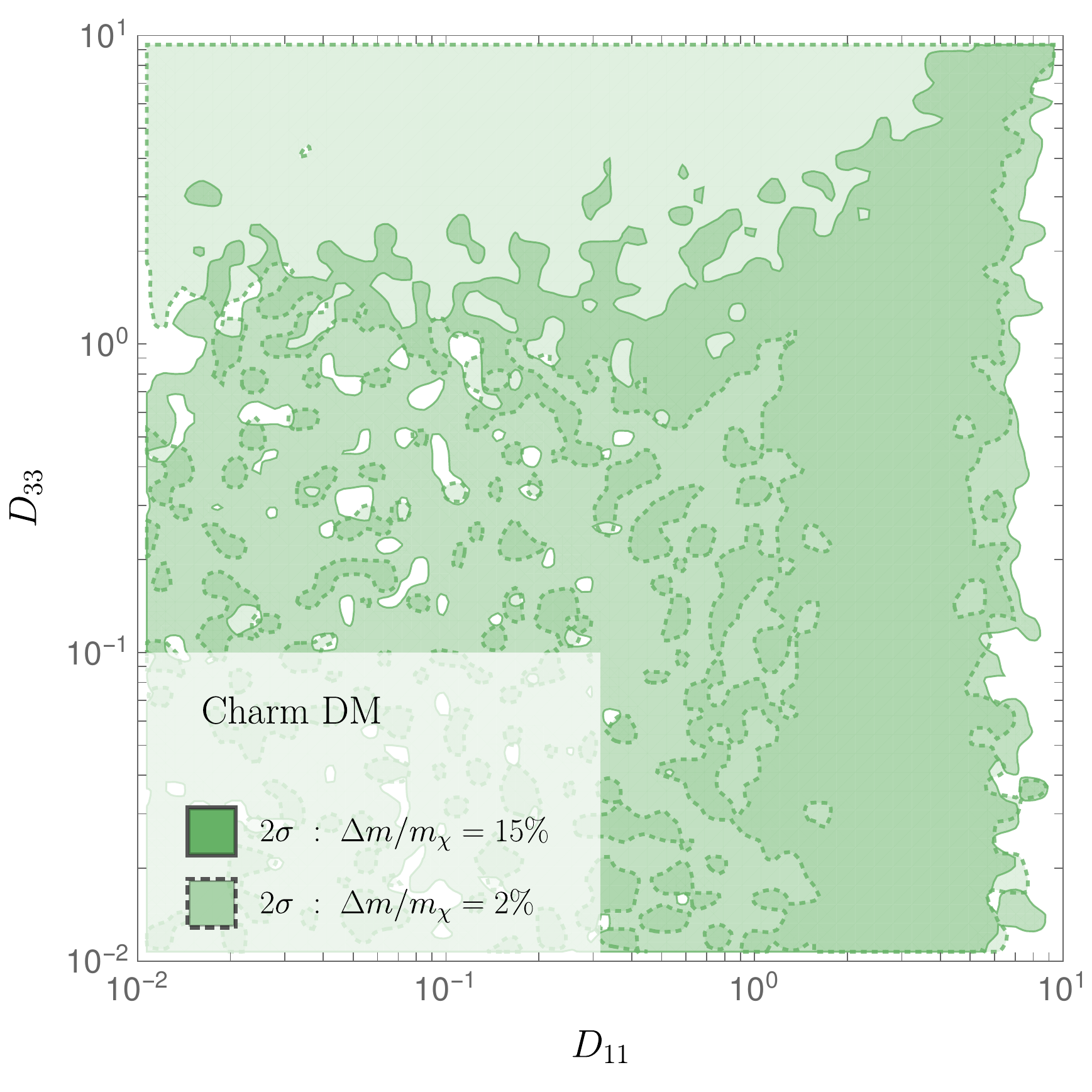}
\includegraphics[width=0.45\textwidth]{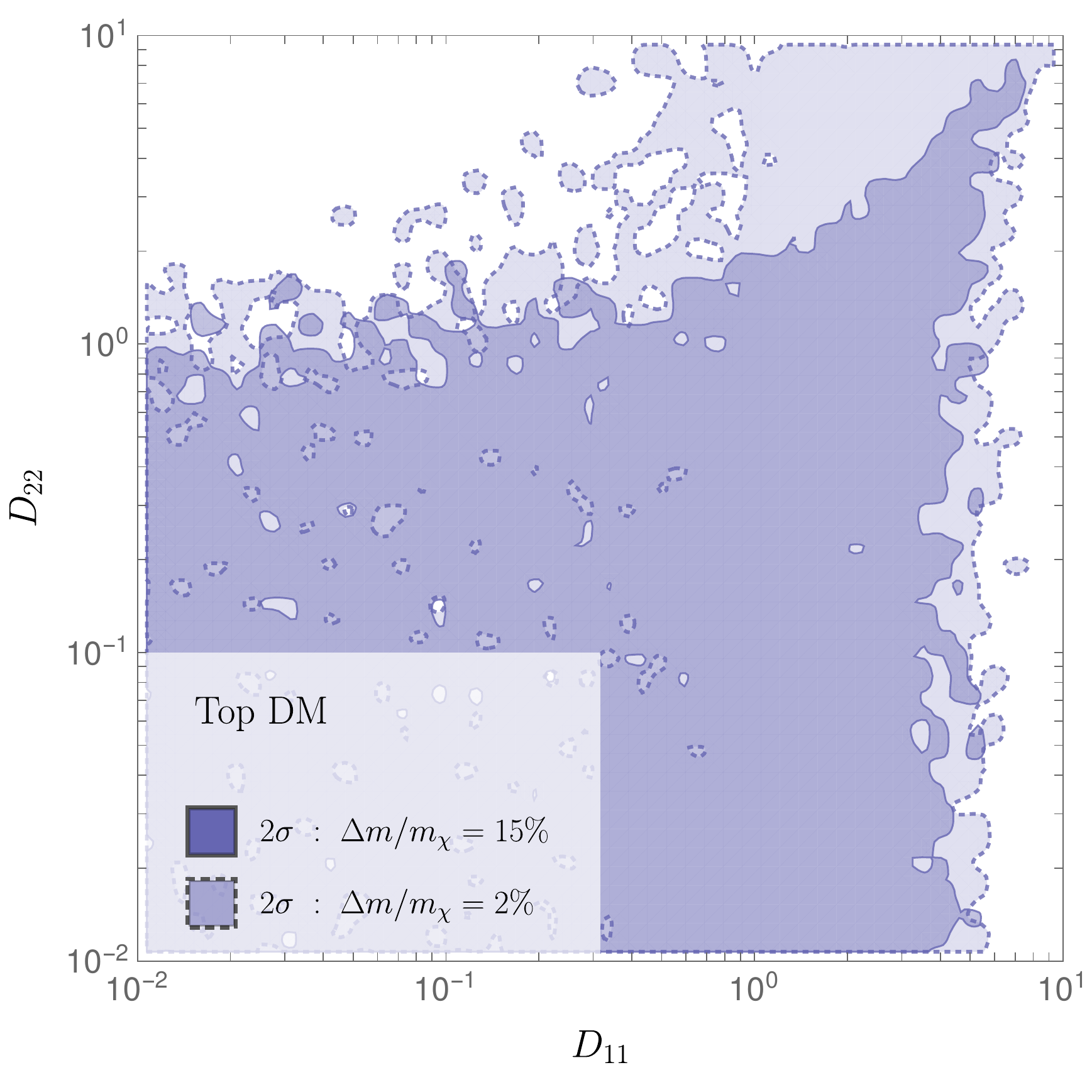}
\includegraphics[width=0.45\textwidth]{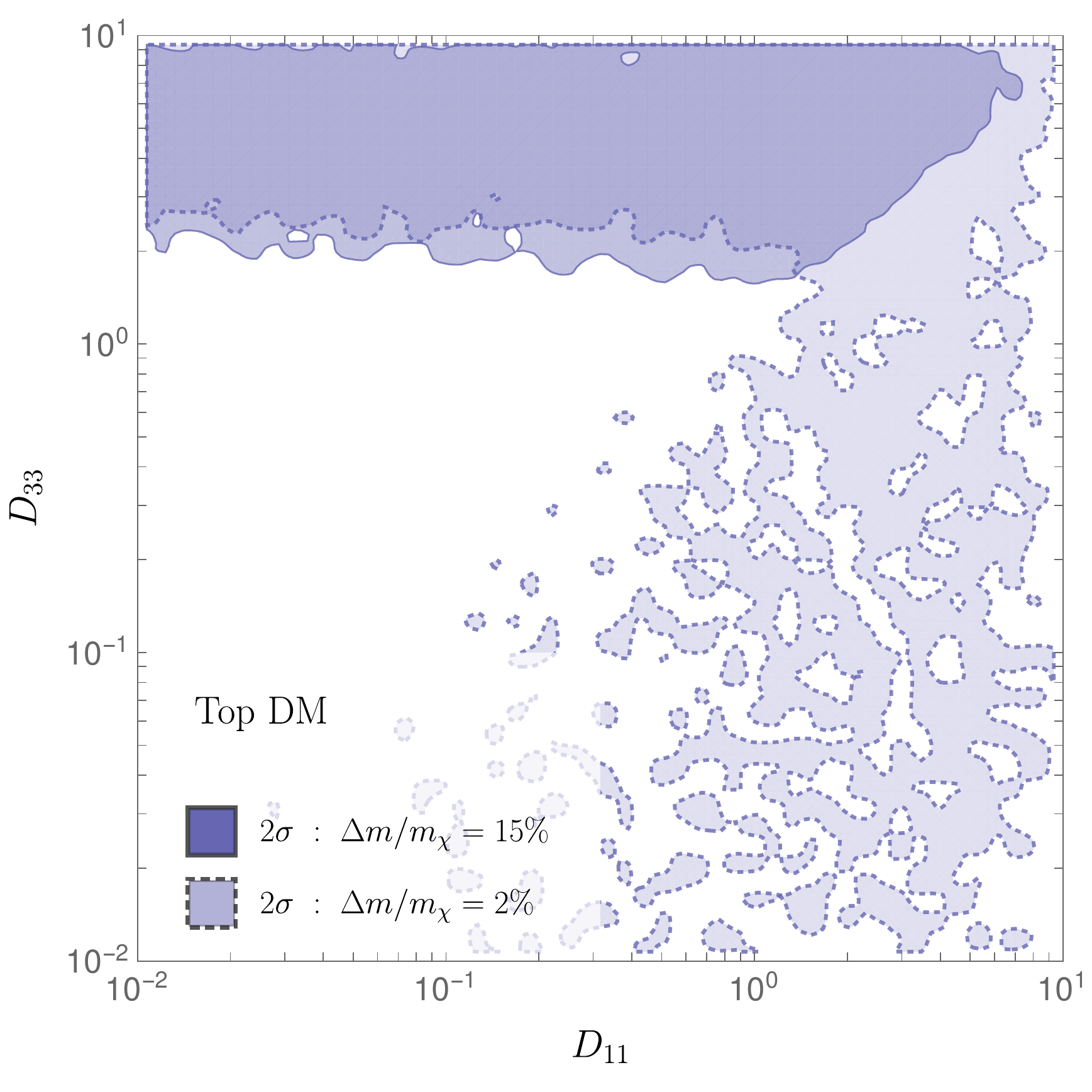}
\caption{As for \cref{Fig:ResultsMass} but for the $D_{11}-D_{22}$ plane (left) and $D_{11}-D_{33}$ (right), for two values of mass splitting (dashed shaded, and solid darker shaded respectively).}
\label{Fig:ResultsD11vsD22}
\end{figure}

\begin{figure}
\centering
\includegraphics[width=0.45\textwidth]{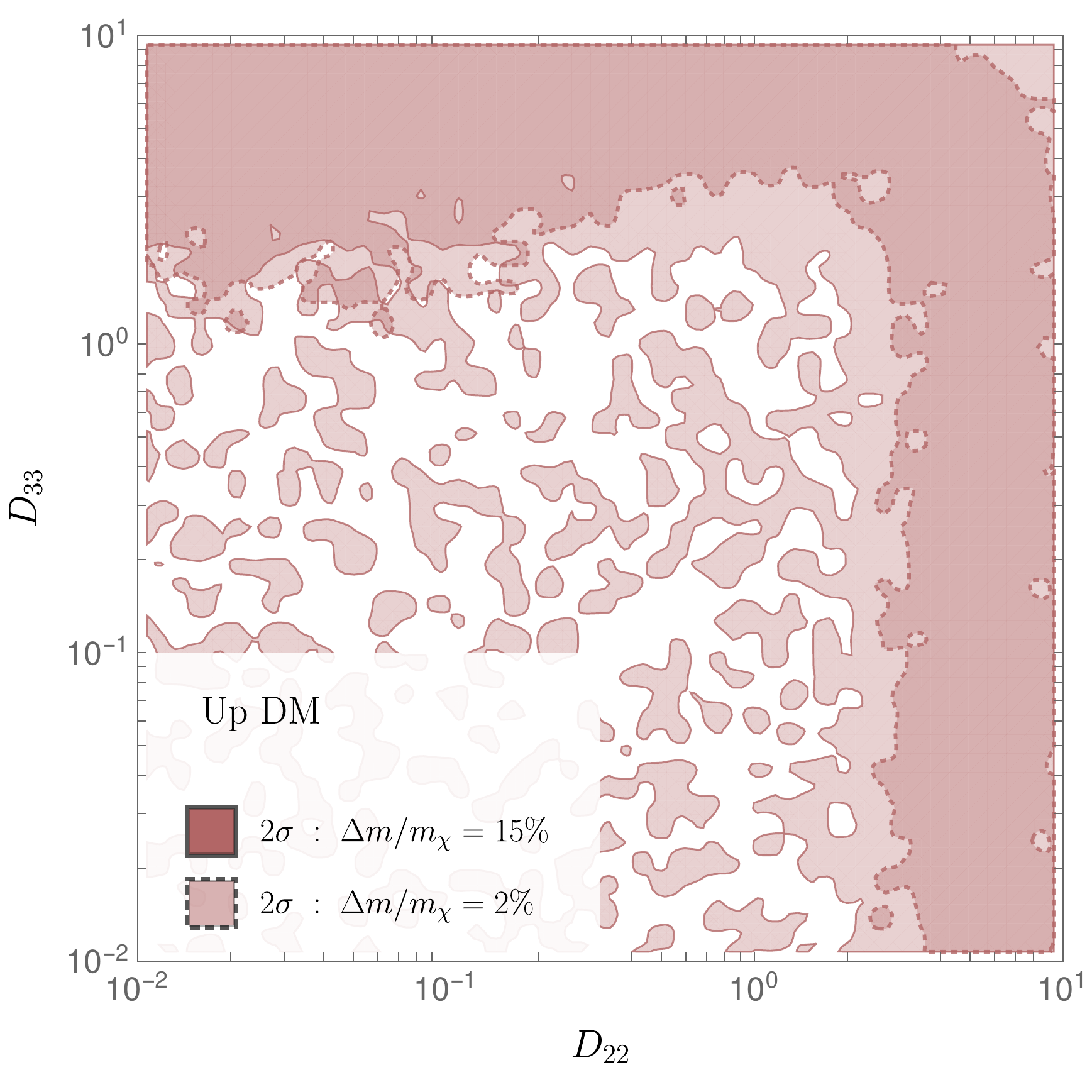}
\includegraphics[width=0.45\textwidth]{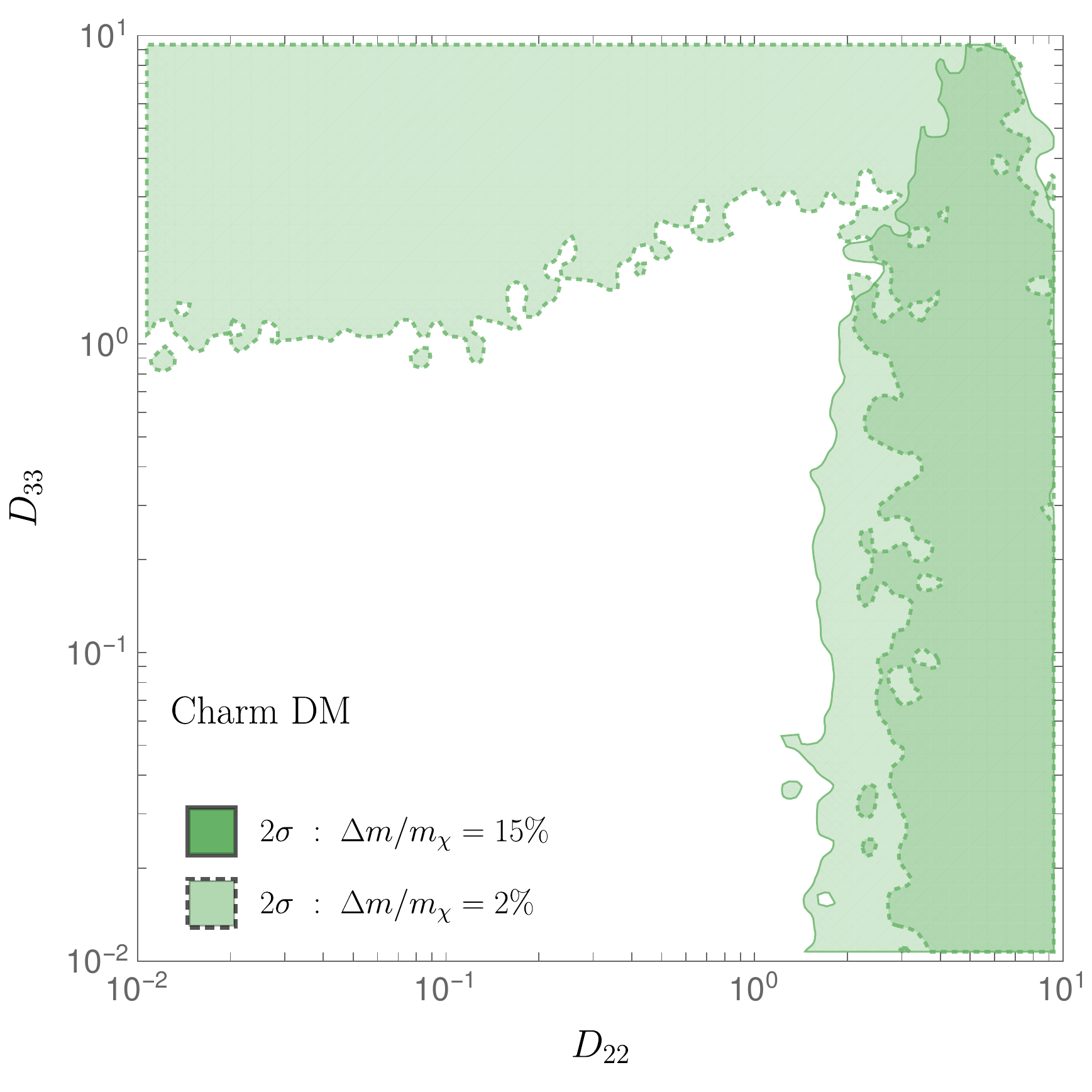}
\includegraphics[width=0.45\textwidth]{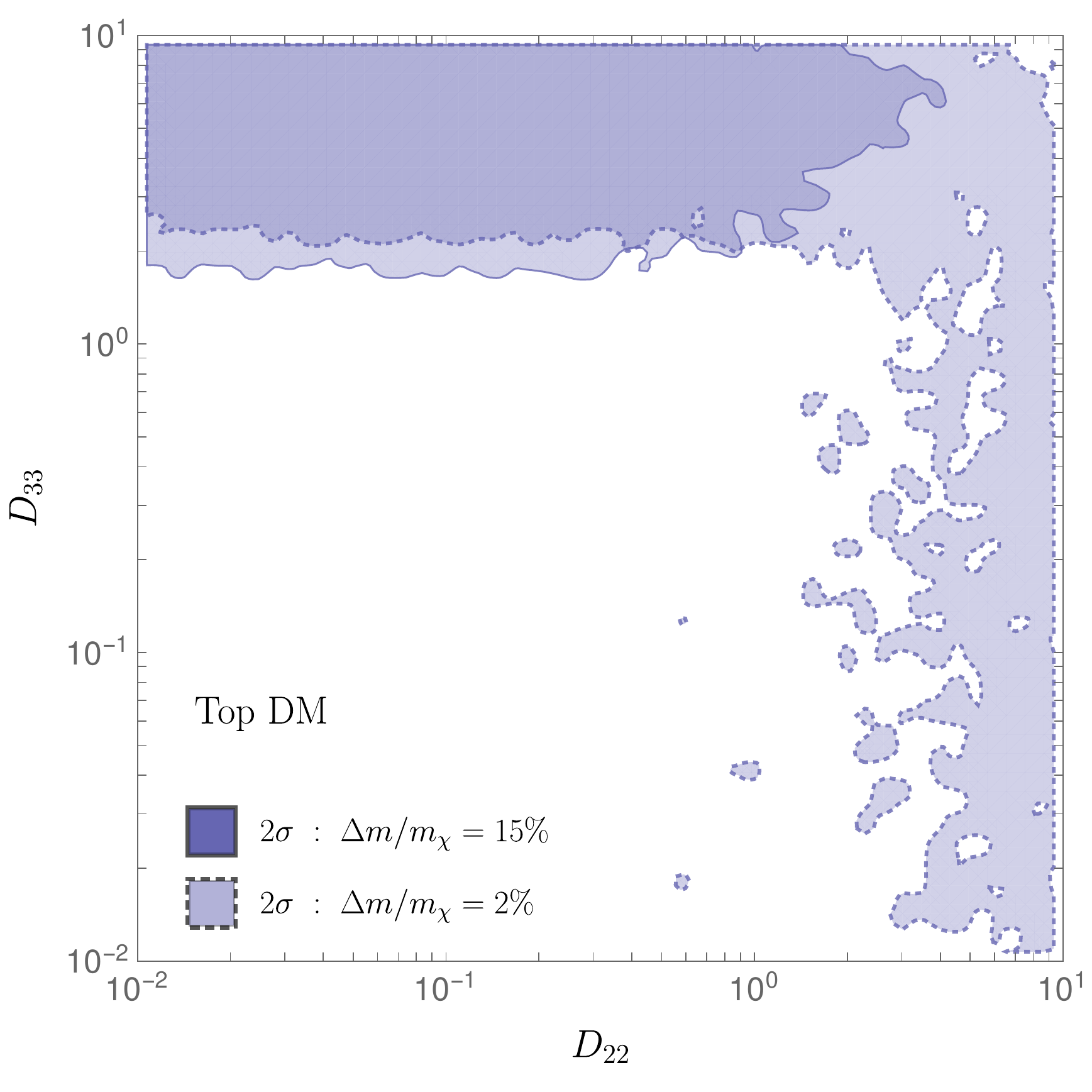}
\caption{As for \cref{Fig:ResultsMass} but for the $D_{22}-D_{33}$ plane.}
\label{Fig:ResultsD22vsD33}
\end{figure}

Our results are summarized in \cref{Fig:ResultsMass,Fig:ResultsD11vsD22,Fig:ResultsD22vsD33} as \(2\,\sigma\) contours, and in \cref{Fig:1Dcredibleregions} as one-dimensional \(1\,\sigma\) intervals.
We consider three separate samples in which the DM (the lightest $\chi$) is the first, second and third member of the triplet (denoted `up', `charm' and `top' DM).
Within each sample we present a low and high mass splitting (\SIlist{2;15}{\percent}), which primarily distinguish the effects caused by coannihilation in the calculation of relic density, but affect all other bounds to some extent as we have explicitly included the masses in each.

As we see from see \cref{Fig:ResultsMass}, the masses of the DM and mediator are both required to be in the TeV range, with upper limits in the tens of TeV,
The DM and mediator masses are strongly correlated with the \(D_{ii}\), as in \cref{Fig:ResultsMass}, due to the relic density and mixing bounds which both scale approximately as \((D / m)^4\) in the high mass limit.
Masses in the TeV range favour the \(D_{ii}\) to be \(\gtrsim\mathcal{O}(1)\).
The mixing angles are not well constrained in general; $\theta_{ij}=0$ is favoured, but the full range of angles are usually allowed with \(2 \sigma\) credibility.

The $D_{ii}$ themselves are highly correlated from the mixing constraints (see \cref{Fig:ResultsD11vsD22,Fig:ResultsD22vsD33}) which depend on $(\lambda \lambda^\dag)_{12}$ which is approximately
\begin{equation}
(\lambda \lambda^\dag)_{12} \approx  \left( s_{13} s_{23} (D_{22}^2 -D_{11}^2) + s_{12} (D_{33}^2 - D_{11}^2)  \right) \ ,
\end{equation}
where \(s_{ij} = \sin \theta_{ij}\) and so we see $D_{11} \sim D_{33}$ (and less strongly $D_{11} \sim D_{22}$). Because the correlation between $D_{22}, D_{33}$ is less pronounced, the RD bound controls the behaviour and produces an anti-correlation, since the annihilation cross section scales like 
\begin{equation}
\braket{\sigma v}_\text{eff} \propto (D_{11}^2 + D_{22}^2 + D_{33}^2)^2 \sim \SI{3e-26}{\cm\cubed\per\s}
\end{equation}
due to coannihilations, as such the trend is most pronounced for small mass splitting. This is seen in the range of $D_{22}$ for the small splitting data, \cref{Fig:ResultsD22vsD33}.

\clearpage

In all cases, increasing the mass splitting reduces the available parameter space of the masses and couplings of the DM since the coannihilations and annihilations of the heavy particles have a reduced effect on the relic density (scaling with a Boltzmann factor \(\exp(-\Delta m)\)).
This allows less flexibility in the DM parameters whilst potentially opening up the allowed parameters of the heavy particles, since their couplings are out of reach of the astronomical constraints (indirect and direct searches) which are proportional to the relic density of the lightest $\chi$ (scaling as $\Omega^2$ and $\Omega$ respectively).
This effect can be clearly seen in the right panels of \cref{Fig:ResultsMass}, where the \SI{2}{\percent} splitting allows much smaller DM couplings compared with the \SI{15}{\percent} splitting, contrastingly in \cref{Fig:ResultsD11vsD22} (middle right panel) the non DM coupling space opens up with a larger splitting.
Of course, since we have fixed the mass splitting by hand, the heavy particle parameters are not totally free, and so the parameter space is still reduced by the constraints we consider.

Top quark threshold effects are absent in the MCMC scan, due to the high masses (\(m_\chi \gtrsim m_{\Pqt}\)). Since \(m_\chi, m_\phi \gg m_{\Pqt}\) the three quarks are kinematically equivalent, and so the bounds are not strongly dependent on the flavour of DM.
The main differences arise due to the quarks SM interactions which impact the DD and ID limits.

As described in \cref{Sec:Collider}, we have studied collider bounds on our model, but these were not directly incorporated into our MultiNest routine as these bounds are much more computationally intensive than the others.
However, as we see from \cref{Fig:ColliderBounds}, the collider bounds only rule out sub-\si{\TeV} scale masses, even at large couplings and so we do not expect that a full likelihood function incorporating the LHC constraints would give significantly different results.
As a test, we checked a sample of the points inside the \SI{68}{\percent} (\(1\,\sigma\)) credible regions and found only a small minority (of order \SI{1}{\percent}) that would be excluded by collider data.
We produce, for each parameter, a marginalized posterior integrated over the remaining 7 parameters. From this distribution we find the \(1\,\sigma\) credible interval.
The results are shown in \cref{Fig:1Dcredibleregions}.
This contains results for both uniform and log-uniform  priors on $D_{ii}$, $m_{\chi}$ and  $m_\phi$; when the two cases are discrepant by \(> 1\,\sigma\) this is due to a flat posterior, and using the \(2\,\sigma\) band instead the two agree.

\begin{figure}
\centering
\includegraphics[width=0.45\textwidth]{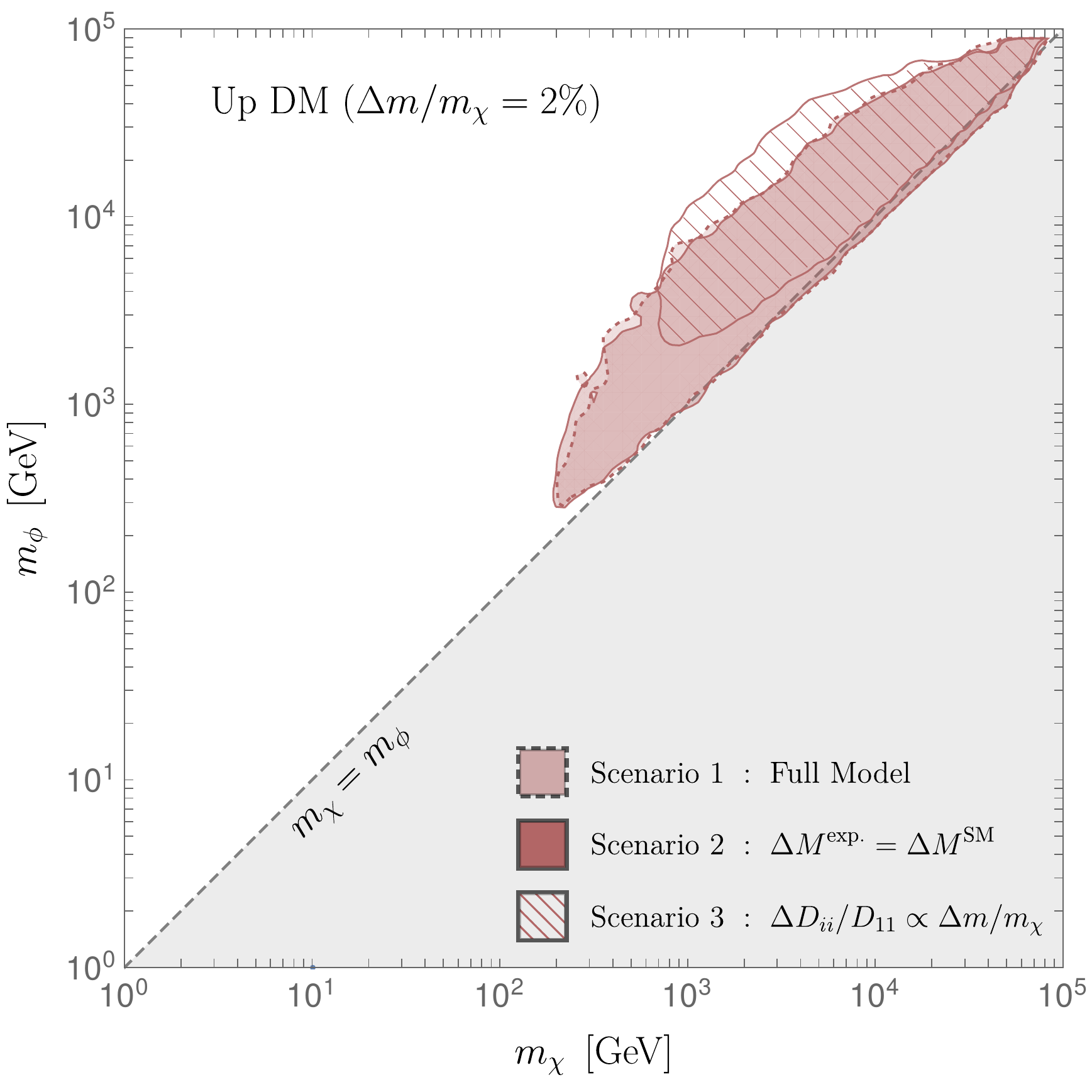}
\includegraphics[width=0.45\textwidth]{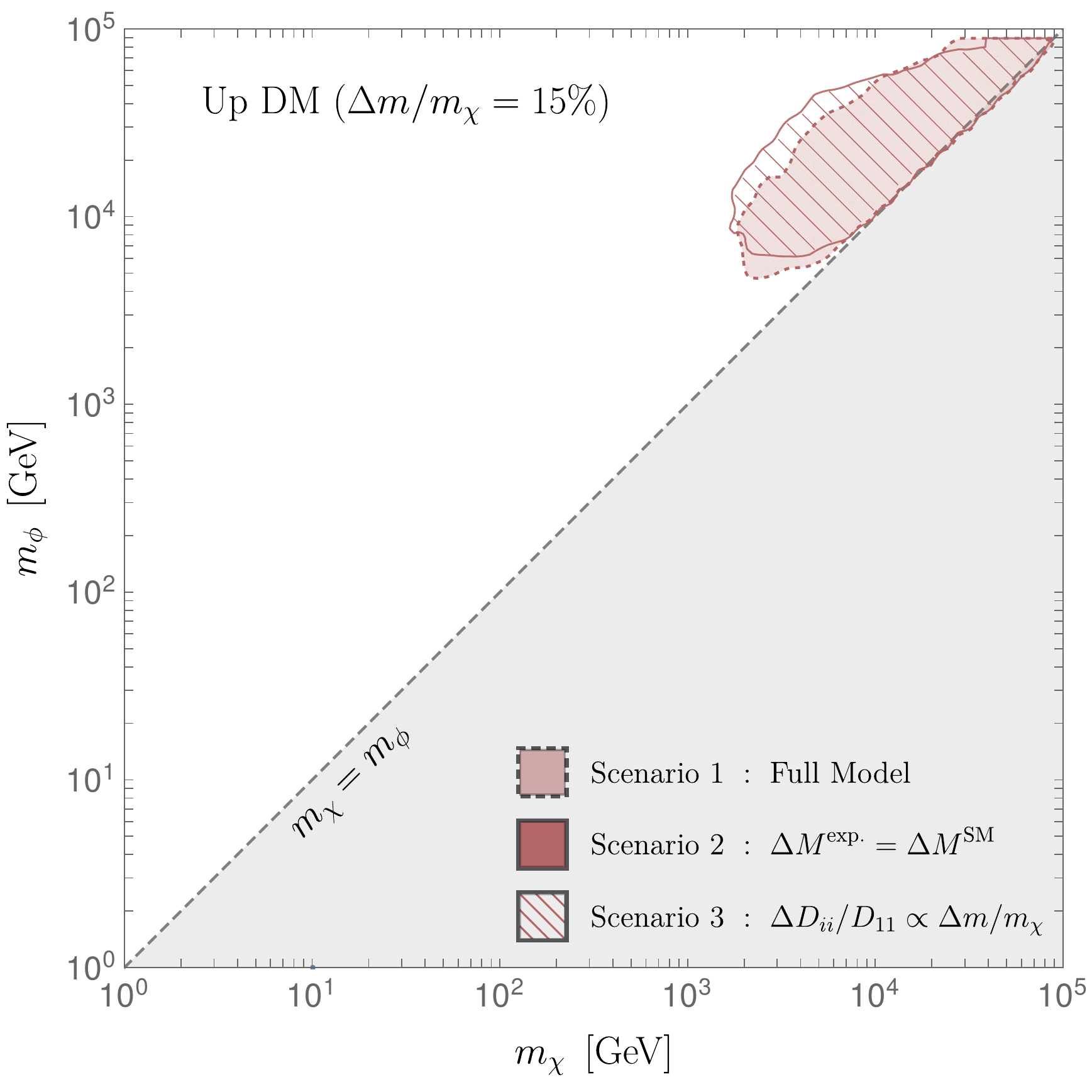}
\includegraphics[width=0.45\textwidth]{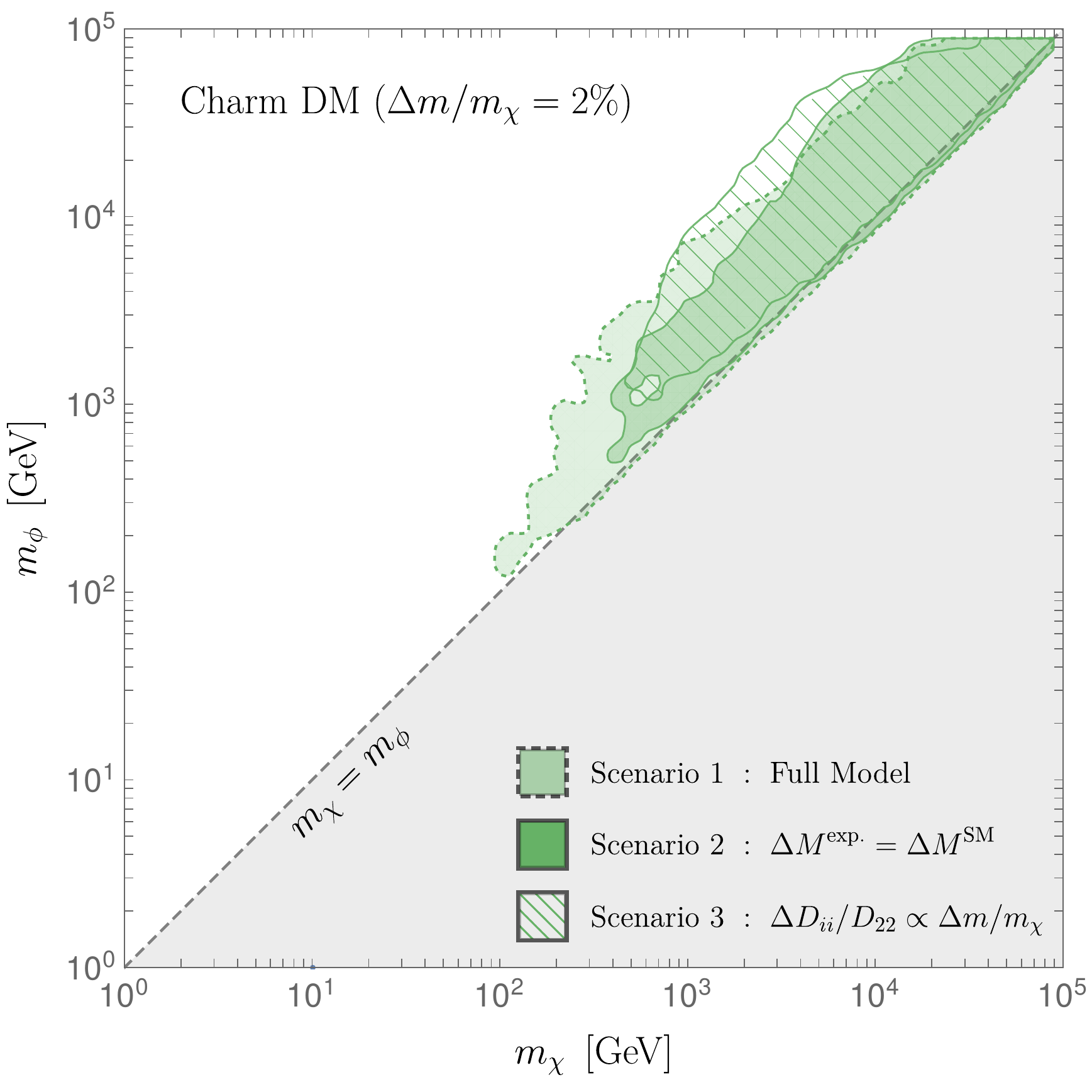}
\includegraphics[width=0.45\textwidth]{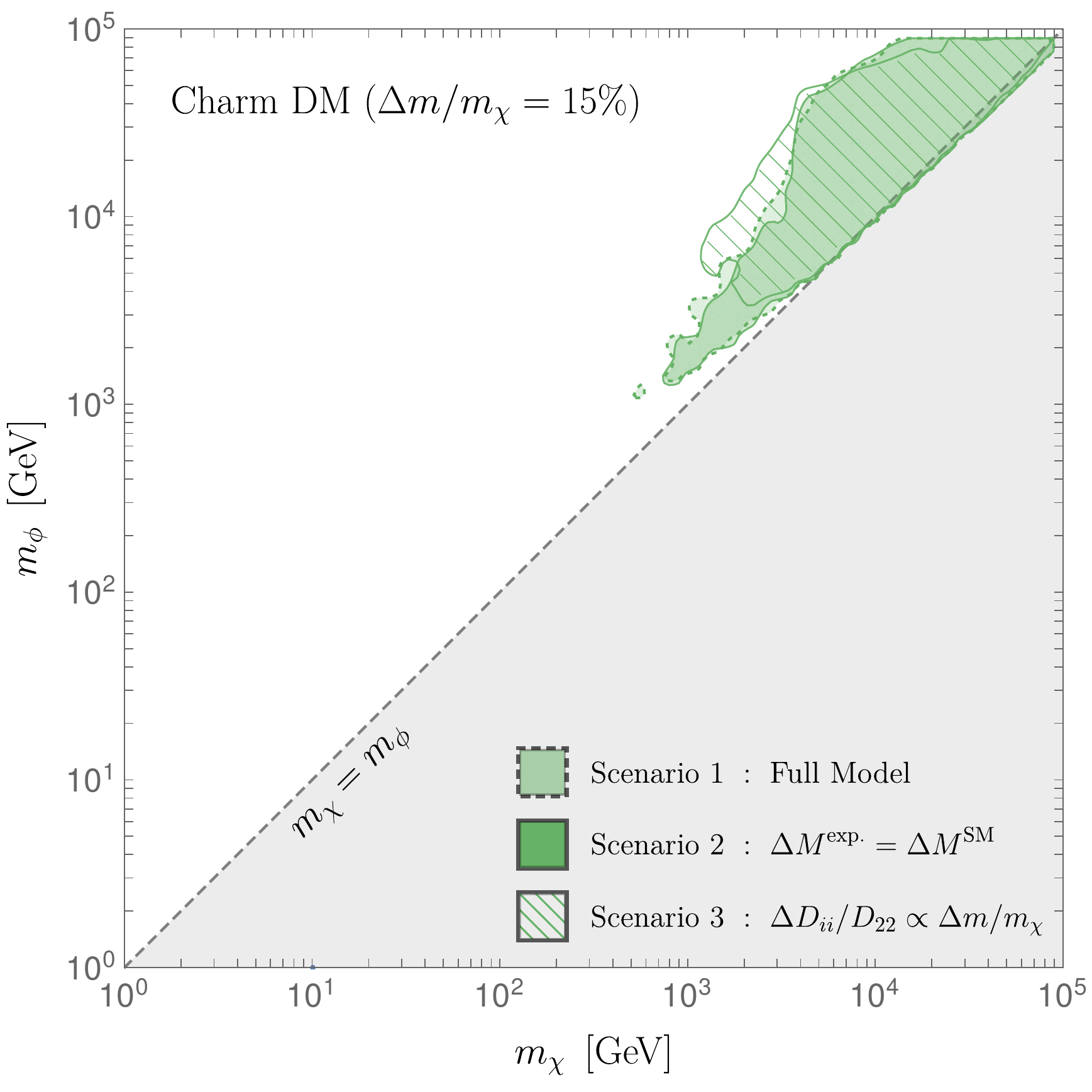}
\includegraphics[width=0.45\textwidth]{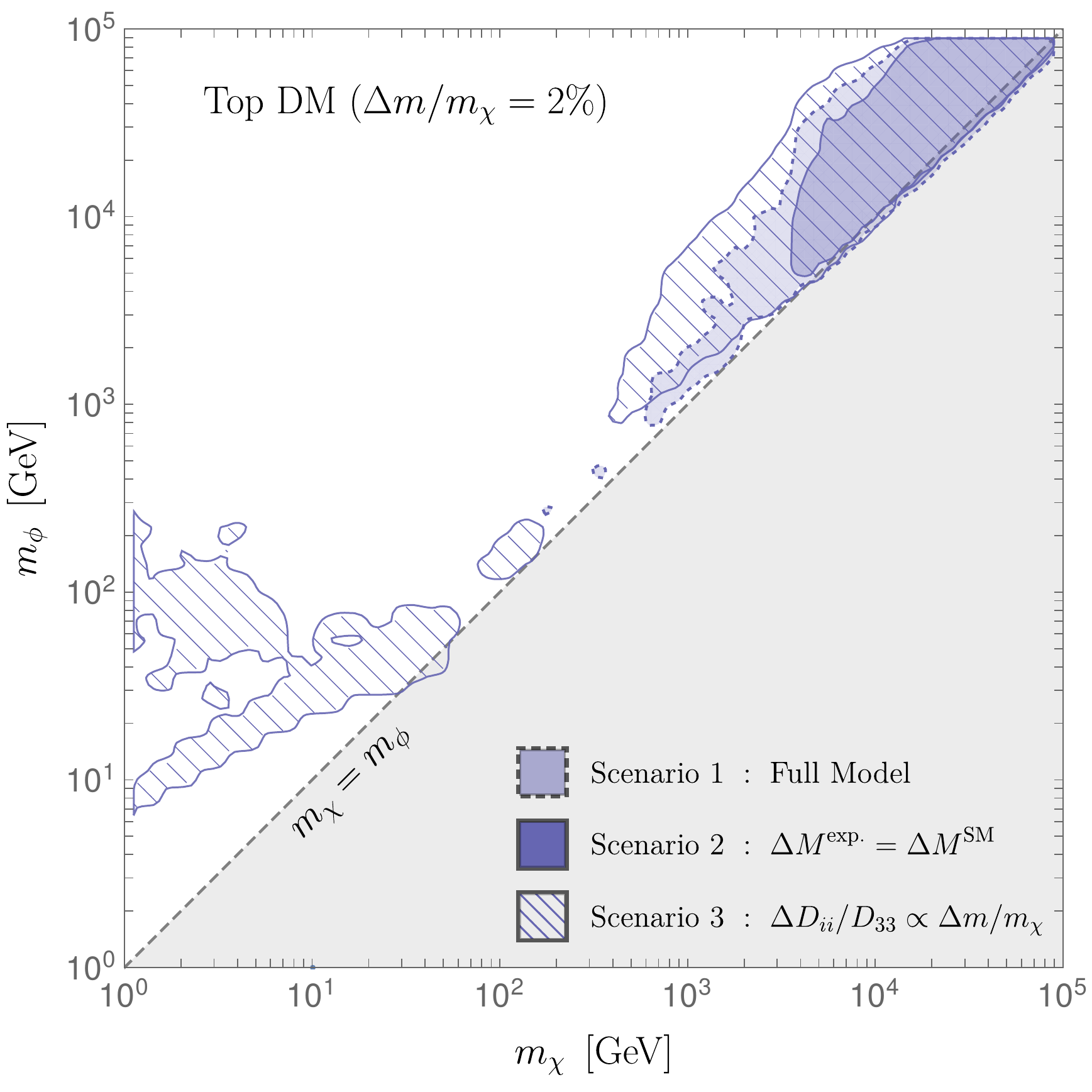}
\includegraphics[width=0.45\textwidth]{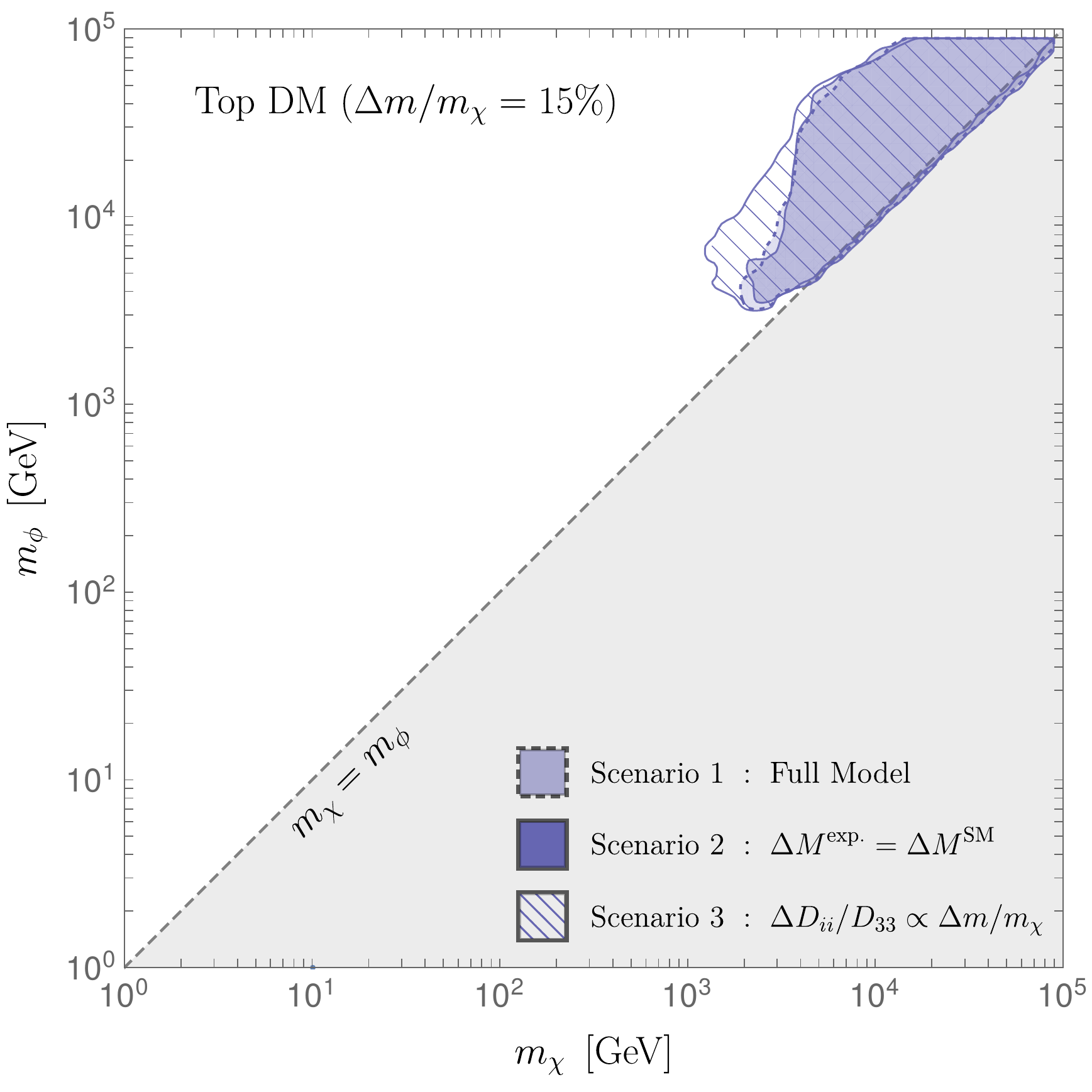}
\caption{Comparison between \(2\,\sigma\) contours of the full MCMC scan and two extensions discussed in the text, for a mass splitting of \SI{2}{\percent} (left) or \SI{15}{\percent} (right).}
\label{Fig:ResultsConstrainedScenarios}
\end{figure}

\subsection{Constrained Scenarios}
\label{sub:ConsScen}

We consider two extensions to the previous results:
\begin{enumerate}
\item In \cref{sub:RenormalizationMassSplitting} we found that the mass splitting which is generated through RG running of the DM self-energy is approximately proportional to \(D_{ii}^2\), this motivates us to consider a scenario in which the couplings \(D_{ii}\) are correlated with the masses (thus introducing a coupling splitting \(\Delta D_{ii} / D_\chi \propto \Delta m_{ij} / m_\chi\)).
The reduced parameter space enforces almost degenerate couplings which leads to two important effects; firstly, it subjects all three \(\chi\) to the astrophysical constraints of indirect and direct detection, despite the heavier particles having no relic density.
By this we mean that, upon fixing the mass splitting, any limits on the coupling strength of the relic particle are translated to restrict the non-relic particles.
Secondly, because the \(D_{ii}\) are equal the mixing effects are naturally small and as a result the mixing angles are much less constrained as they do not need to be small to counteract flavour effects.

This scenario is representative of a model in which MFV is broken only slightly, since the couplings to quark flavours are roughly equal, differing due to the mixing angles and the small differences in the $D_{ii}$.
It is actually only slightly less constrained in both mass and couplings than models in which flavour violation is allowed, which counteracts the naive assumption that without MFV, flavour observables restrict NP very high scales (\(\mathcal{O}(\SI{100}{\TeV})\)).
\item When compared with the down-type quark sector, flavour bounds are weaker due to \PDzero being less well measured and our conservative treatment in which we assume the SM contribution to \PDzero mixing is zero and the experimental value comes entirely from the new physics.
This is not entirely unreasonable, since short distance calculations of the observable are known to be very discrepant, nor is it completely reasonable, since long distance calculations are able to bring the SM into agreement.

To cover this caveat we consider a future scenario in which the SM calculation reproduces the experimental number (but the precision of the measurement stays at its current value).
This is also conservative, since any interference terms between the SM and DMFV amplitude are likely to be large. The constraints on the mixing angles are more pronounced
\end{enumerate}

Results for these two further scenarios are shown in \cref{Fig:ResultsConstrainedScenarios}, and the \(1\,\sigma\) intervals in \cref{Fig:1DcredibleregionsSplitting,Fig:1DcredibleregionsSMMixing}.

\begin{sidewaystable*}
\begin{tabular*}{0.75 \textwidth}{ c  c  c c c } 
 &  & $ \theta_{12} $ & $ \theta_{13}$ & $\theta_{23}$ \vspace{2mm} \\ 
& \includegraphics[trim={0 0 2cm 0}, width=2cm]{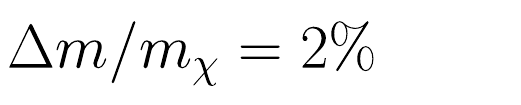}  &  
\includegraphics[width=0.28\textwidth]{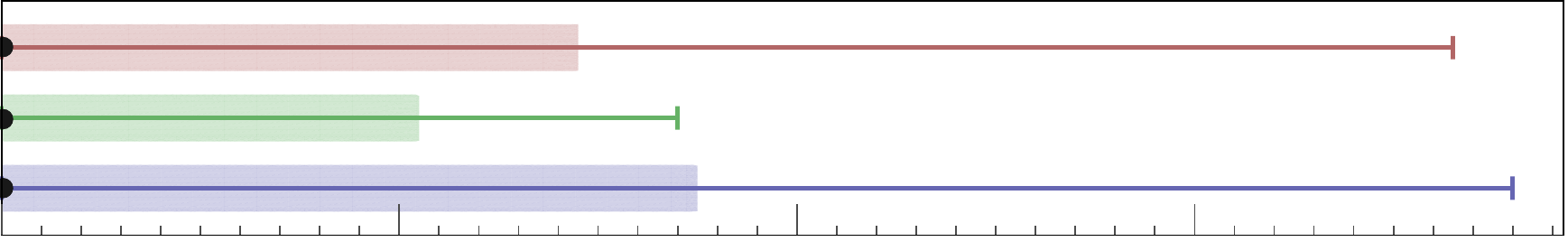} & \includegraphics[width=0.28\textwidth]{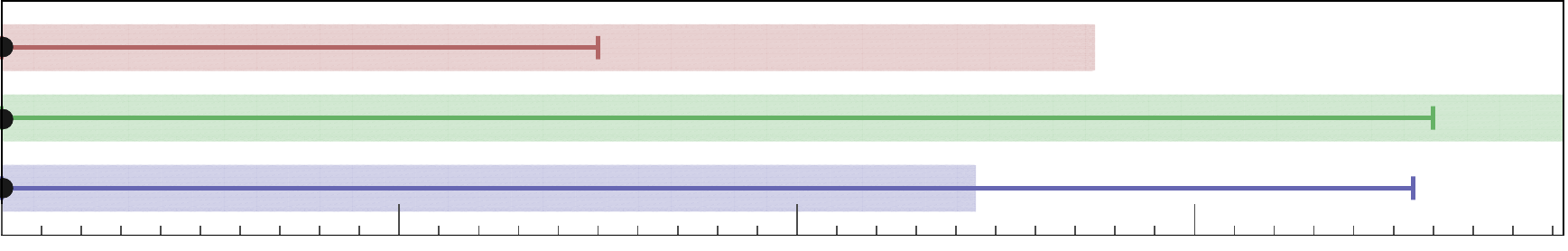}& \includegraphics[width=0.28\textwidth]{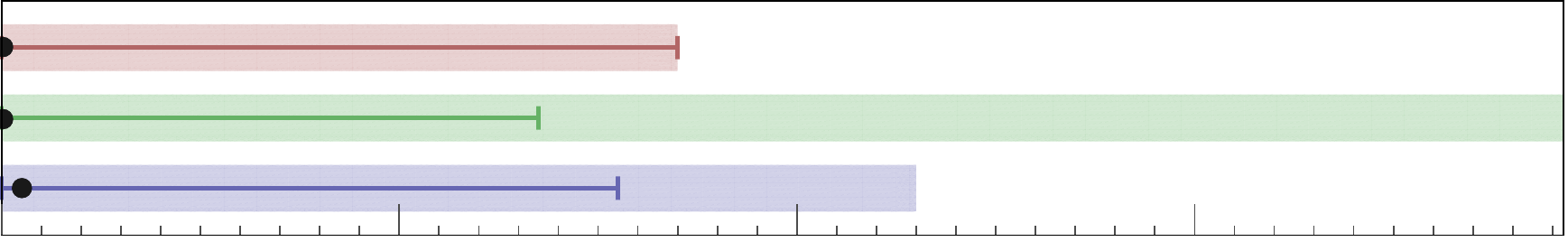}   \\
& \includegraphics[trim={0 0 1.8cm 0}, width=2cm]{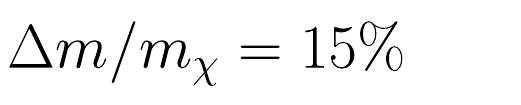}  &  
\includegraphics[width=0.28\textwidth]{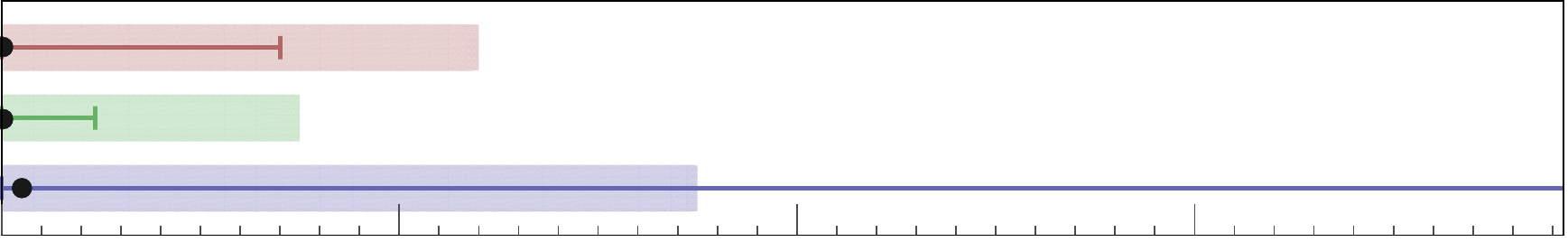} & \includegraphics[width=0.28\textwidth]{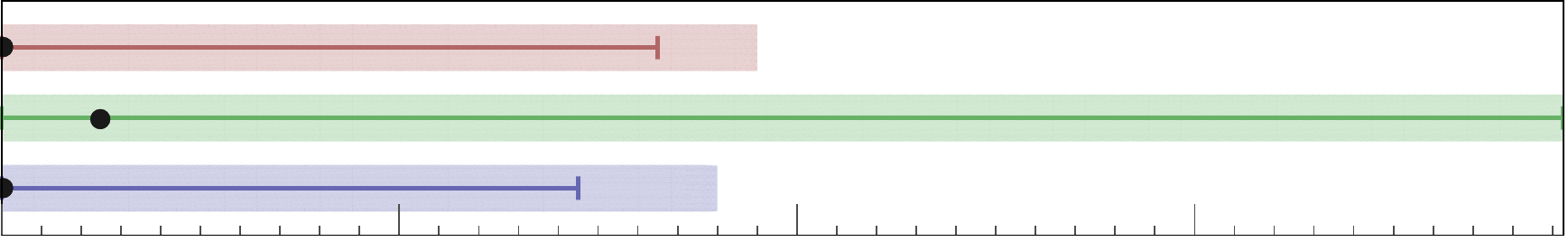}& \includegraphics[width=0.28\textwidth]{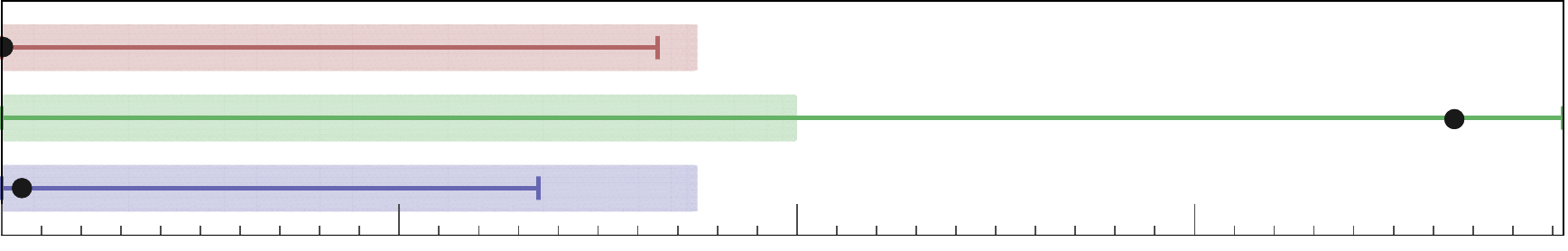} \\
&  & 
\includegraphics[width=0.29\textwidth ,trim={0 0 0 2cm}]{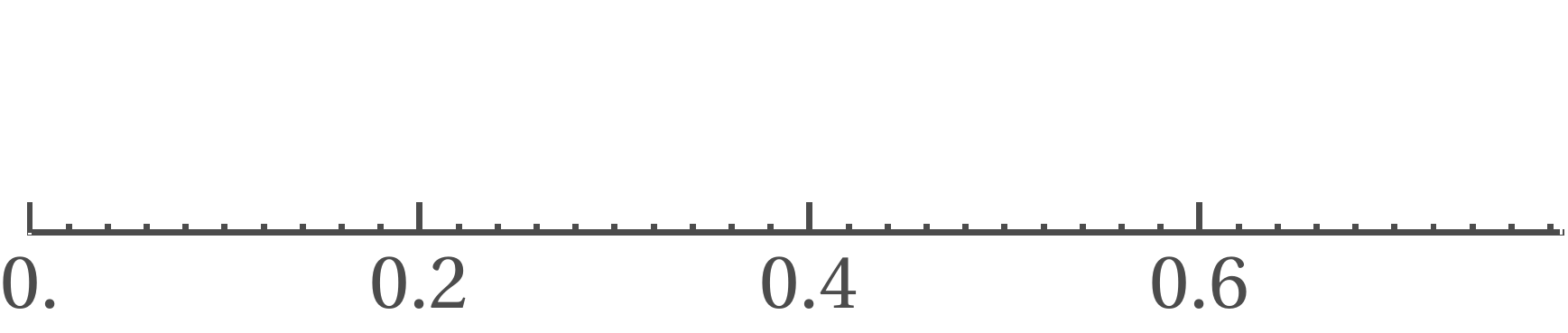} & \includegraphics[width=0.29\textwidth ,trim={0 0 0 2cm}]{1D_axis_theta} & \includegraphics[width=0.29\textwidth ,trim={0 0 0 2cm}]{1D_axis_theta}  \\
 &  & $ D_{11} $ & $ D_{22} $ & $ D_{33}$ \vspace{2mm} \\
& \includegraphics[trim={0 0 2cm 0}, width=2cm]{1D_sidelabel1}  &  
\includegraphics[width=0.28\textwidth]{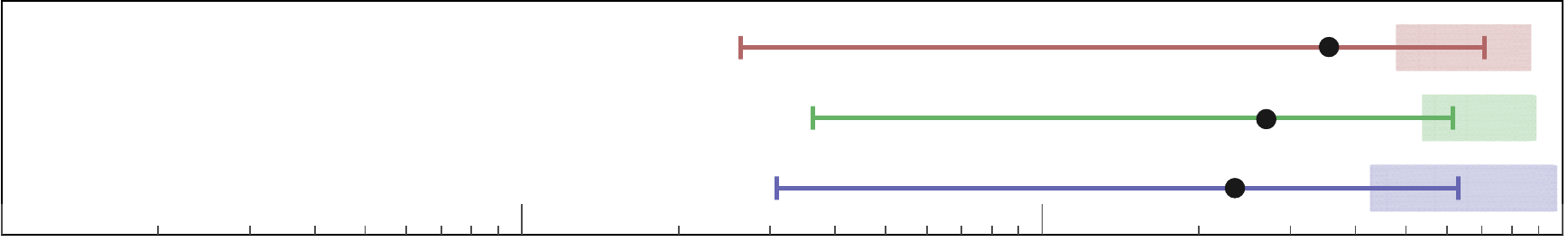} & \includegraphics[width=0.28\textwidth]{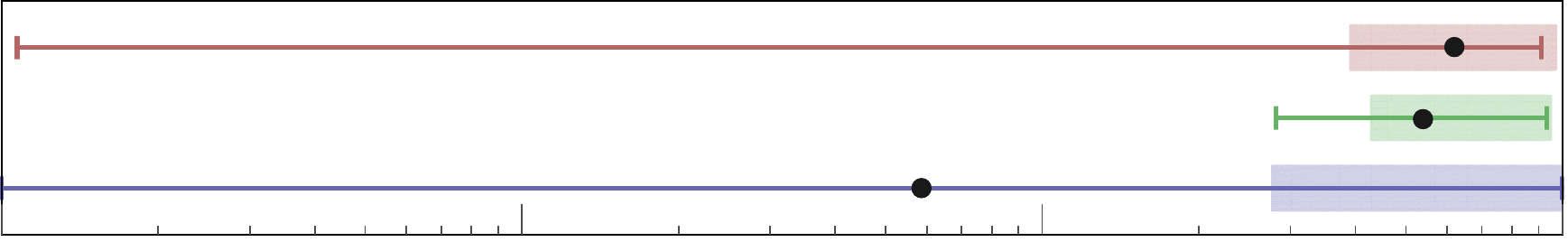}& \includegraphics[width=0.28\textwidth]{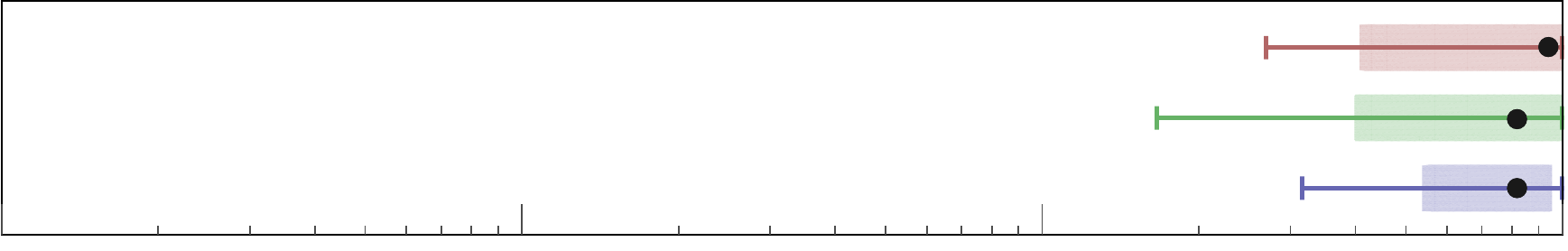}   \\
& \includegraphics[trim={0 0 1.8cm 0}, width=2cm]{1D_sidelabel2}  &   
\includegraphics[width=0.28\textwidth]{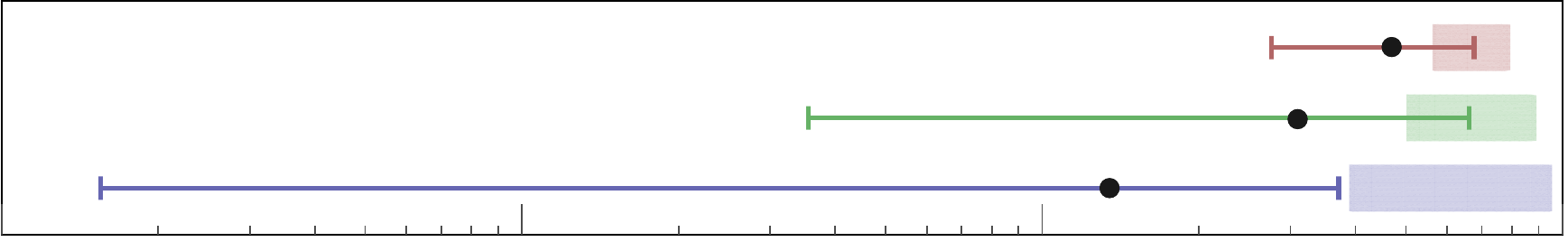} & \includegraphics[width=0.28\textwidth]{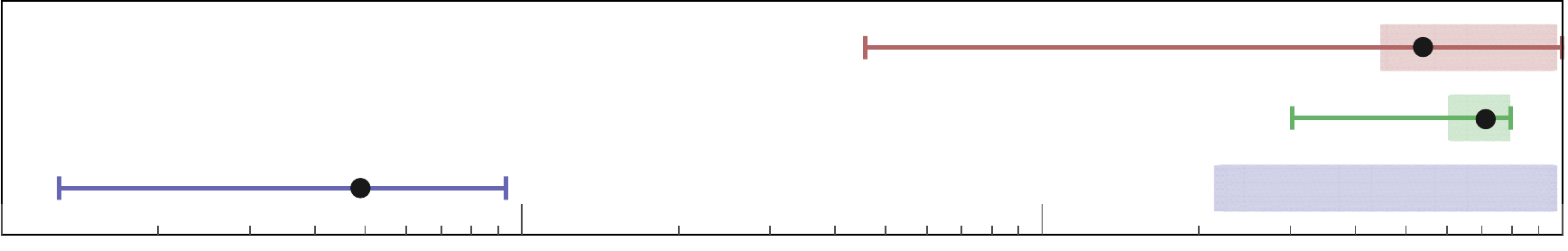}& \includegraphics[width=0.28\textwidth]{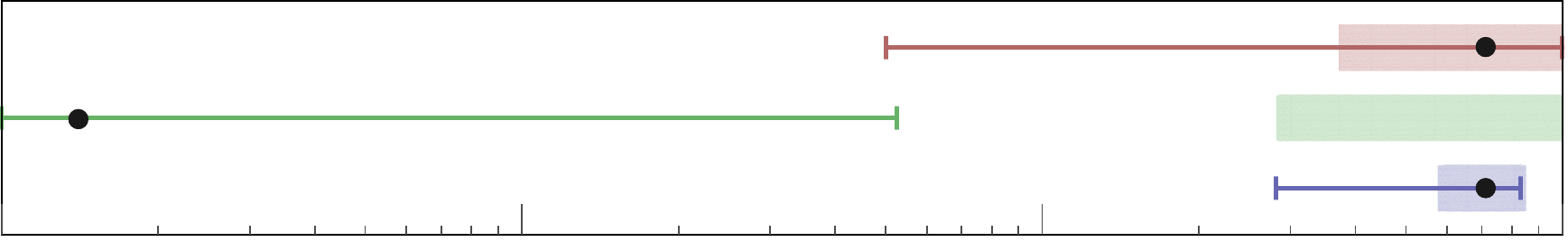} \\
&  & 
\includegraphics[width=0.29\textwidth ,trim={0 0 0 2cm}]{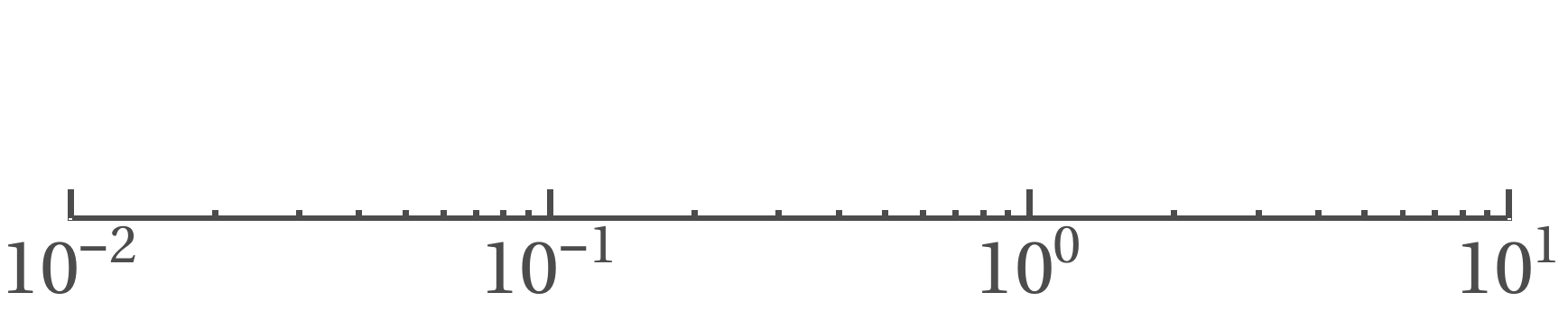} & \includegraphics[width=0.29\textwidth ,trim={0 0 0 2cm}]{1D_axis_Dii} & \includegraphics[width=0.29\textwidth ,trim={0 0 0 2cm}]{1D_axis_Dii}  \\
 &  & \(m_\chi\ [\si{\GeV}]\) & \(m_\phi\ [\si{\GeV}]\) &  \vspace{2mm} \\
& \includegraphics[trim={0 0 2cm 0}, width=2cm]{1D_sidelabel1}  &  
\includegraphics[width=0.28\textwidth]{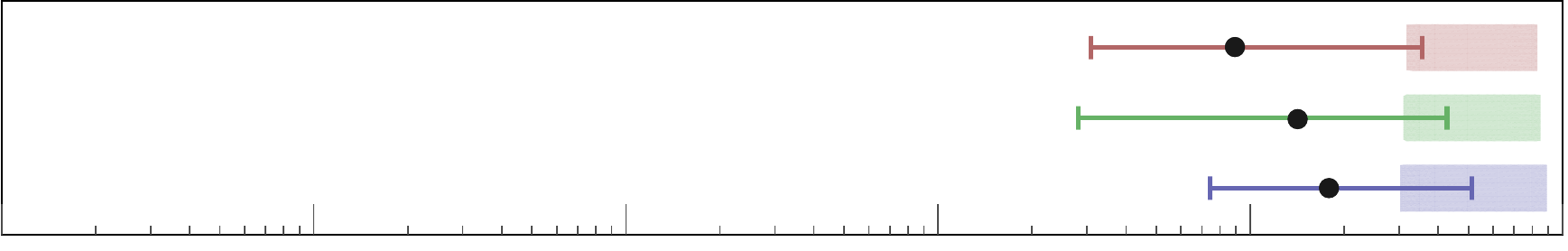} & \includegraphics[width=0.28\textwidth]{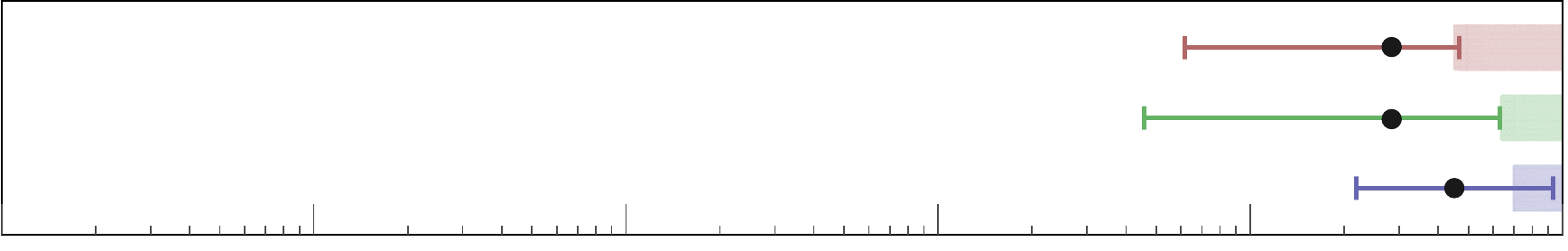}&
\includegraphics[width=0.28\textwidth]{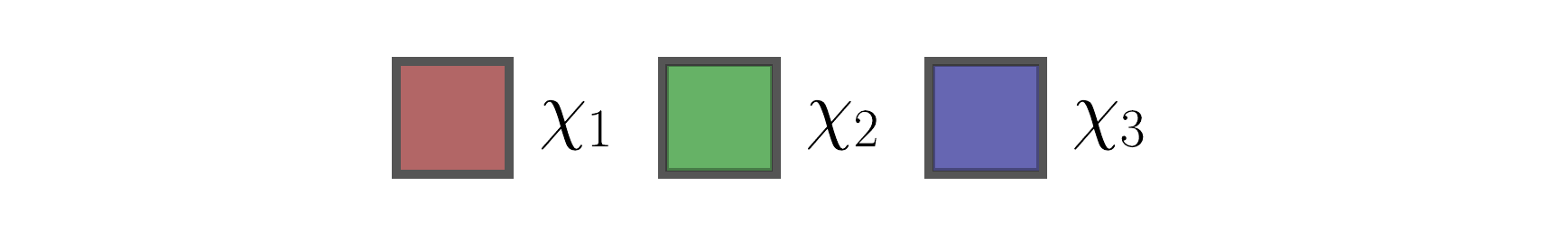}   \\
& \includegraphics[trim={0 0 1.8cm 0}, width=2cm]{1D_sidelabel2}  &  
\includegraphics[width=0.28\textwidth]{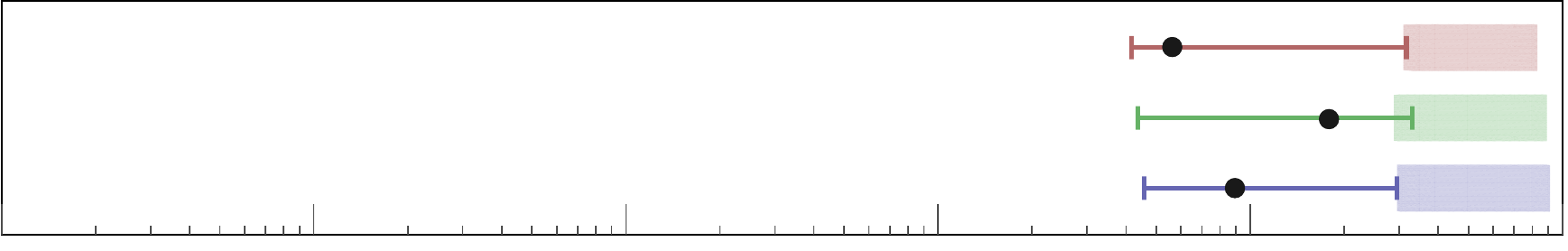} & \includegraphics[width=0.28\textwidth]{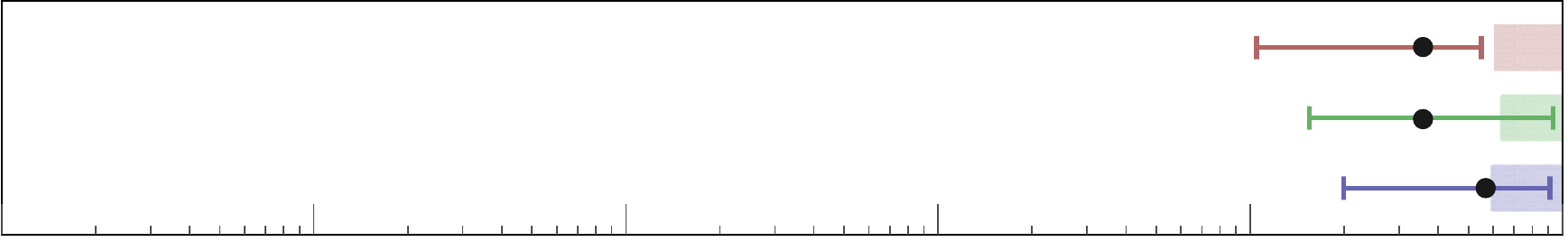}& 
\includegraphics[width=0.28\textwidth]{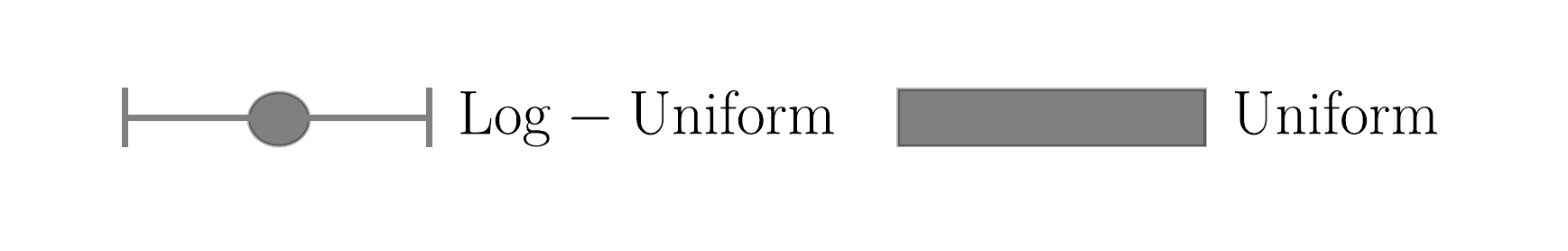}   \\
&  & 
\includegraphics[width=0.29\textwidth ,trim={0 0 0 2cm}]{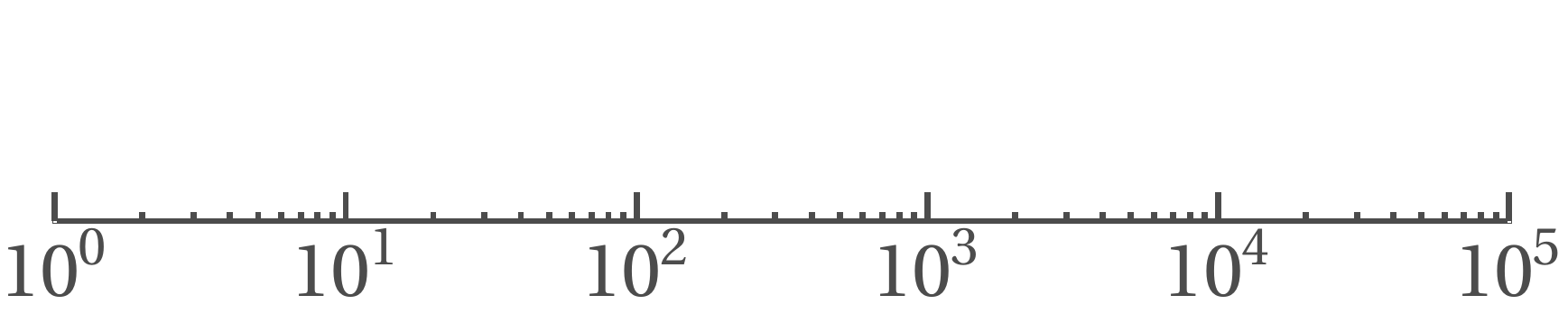} & \includegraphics[width=0.29\textwidth ,trim={0 0 0 2cm}]{1D_axis_m} &  \\
\end{tabular*}
\caption{The 1D \(1\,\sigma\) credible intervals for all 8 parameters of the MCMC scan. Colours denote the DM flavour (red, green and blue for up, charm and top DM respectively). The two prior choices on $D_{ii}$ and the masses (log-uniform vs uniform) are shown as lines and shaded regions respectively, the modal average of the log-uniform prior choice is shown as a dot. The two mass splitting cases are contained in different panels.}
\label{Fig:1Dcredibleregions}
\end{sidewaystable*}

\begin{sidewaystable*}
\begin{tabular*}{0.75 \textwidth}{ c  c  c c c } 
 &  & $ \theta_{12} $ & $ \theta_{13}$ & $\theta_{23}$ \vspace{2mm} \\
& \includegraphics[trim={0 0 2cm 0}, width=2cm]{1D_sidelabel1}  &  
\includegraphics[width=0.28\textwidth]{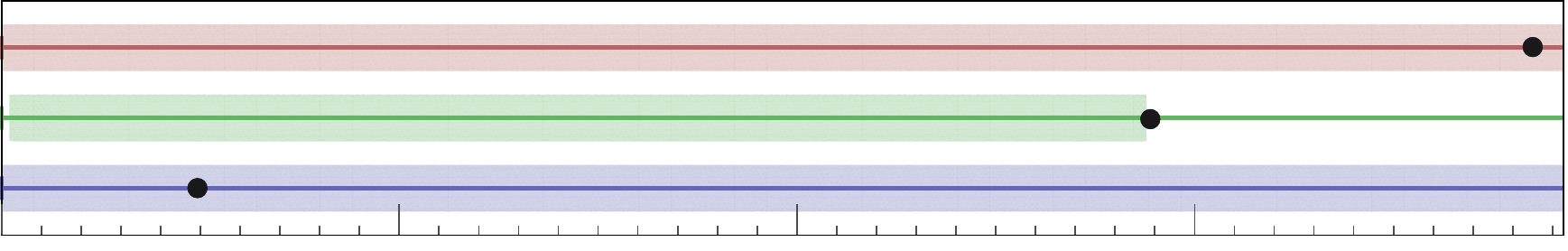} & \includegraphics[width=0.28\textwidth]{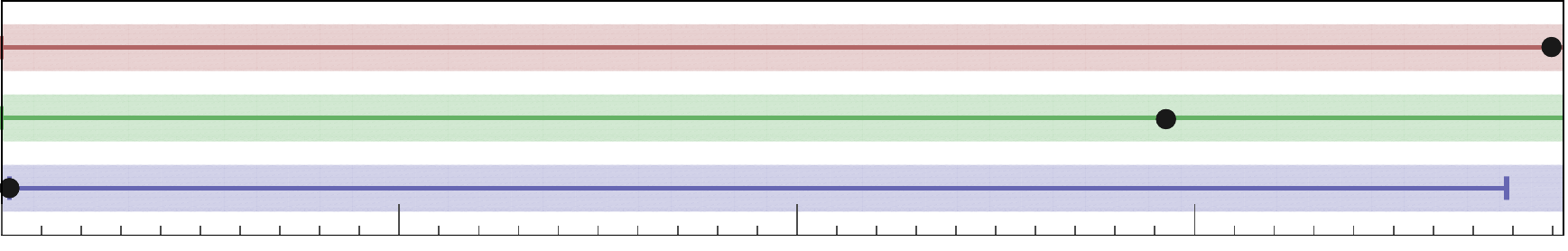}& \includegraphics[width=0.28\textwidth]{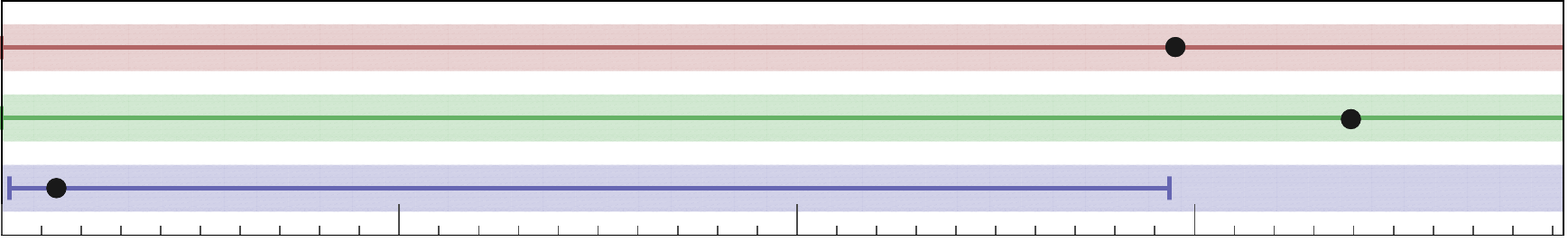}   \\
& \includegraphics[trim={0 0 1.8cm 0}, width=2cm]{1D_sidelabel2}  &  
\includegraphics[width=0.28\textwidth]{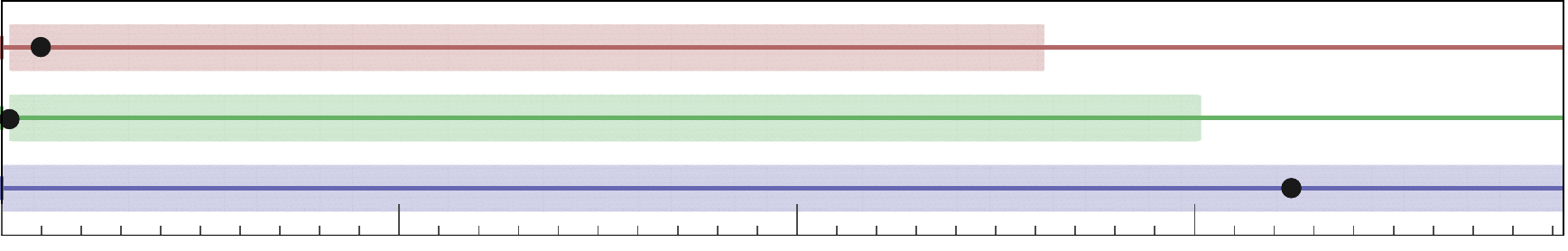} & \includegraphics[width=0.28\textwidth]{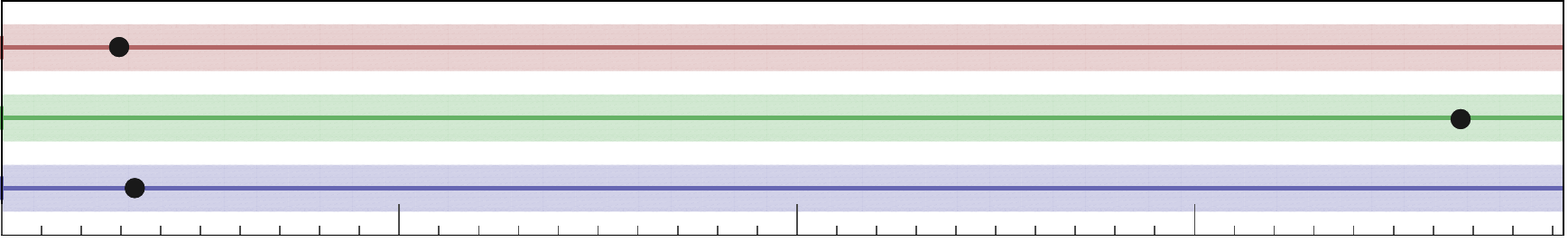}& \includegraphics[width=0.28\textwidth]{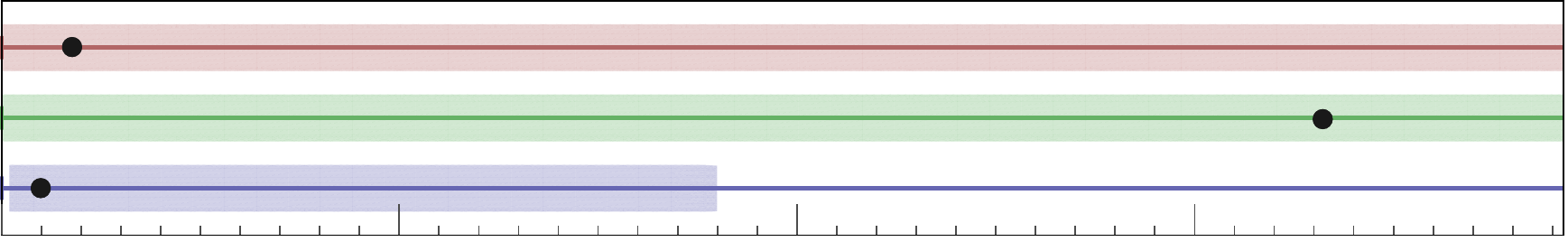} \\
&  & 
\includegraphics[width=0.29\textwidth ,trim={0 0 0 2cm}]{1D_axis_theta} & \includegraphics[width=0.29\textwidth ,trim={0 0 0 2cm}]{1D_axis_theta} & \includegraphics[width=0.29\textwidth ,trim={0 0 0 2cm}]{1D_axis_theta}  \\
& & $ D_{ii} $ & \(m_\chi\ [\si{\GeV}]\) & \(m_\phi\ [\si{\GeV}]\) \vspace{2mm} \\
& \includegraphics[trim={0 0 2cm 0}, width=2cm]{1D_sidelabel1}  &  
\includegraphics[width=0.28\textwidth]{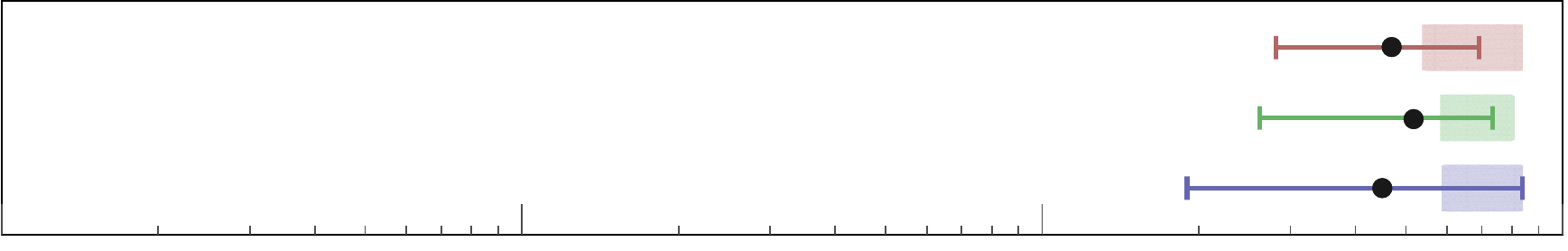} & \includegraphics[width=0.28\textwidth]{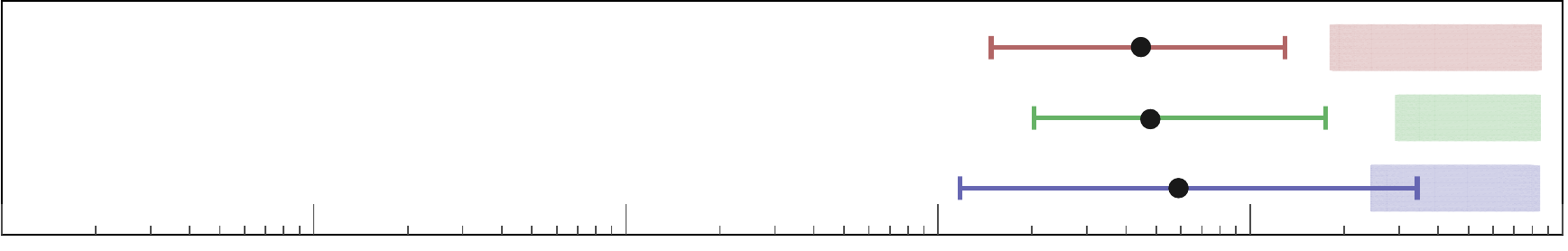}& \includegraphics[width=0.28\textwidth]{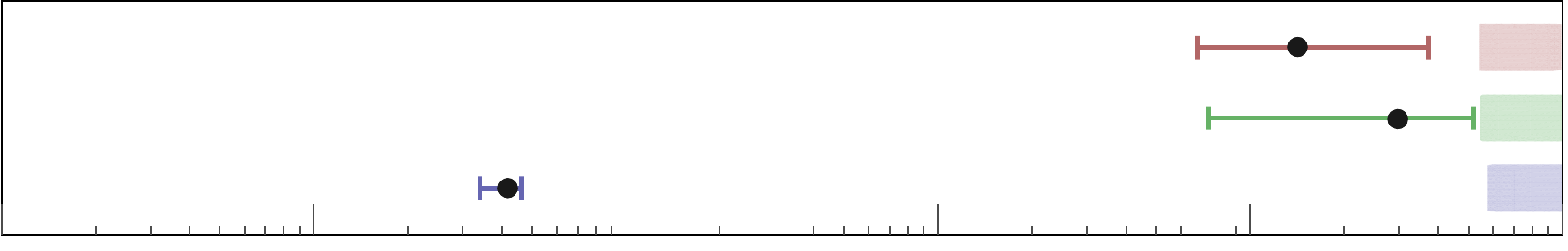}   \\
& \includegraphics[trim={0 0 1.8cm 0}, width=2cm]{1D_sidelabel2}  &   
\includegraphics[width=0.28\textwidth]{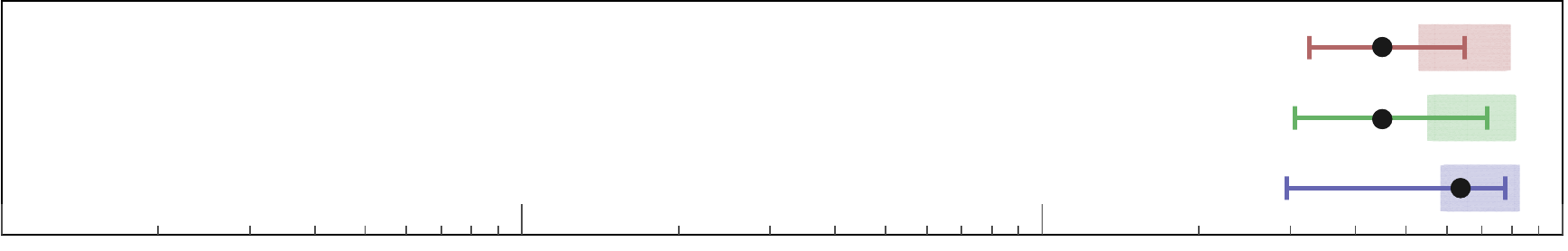} & \includegraphics[width=0.28\textwidth]{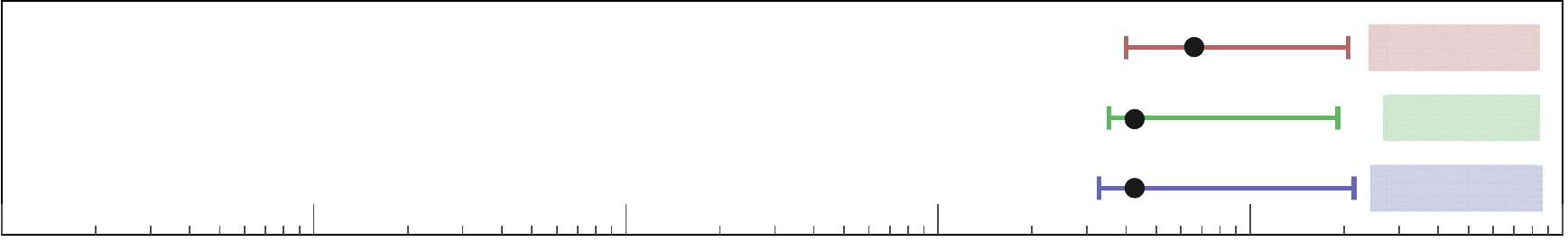}& \includegraphics[width=0.28\textwidth]{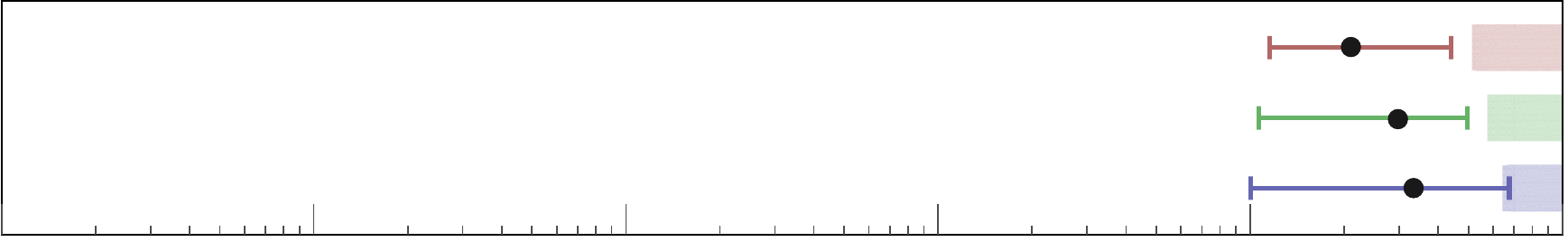} \\
&  & 
\includegraphics[width=0.29\textwidth ,trim={0 0 0 2cm}]{1D_axis_Dii} & \includegraphics[width=0.29\textwidth ,trim={0 0 0 2cm}]{1D_axis_m} & \includegraphics[width=0.29\textwidth ,trim={0 0 0 2cm}]{1D_axis_m}  \\
 &  &  &  &  \vspace{2mm} \\
& \includegraphics[trim={0 0 2cm 0}, width=2cm]{1D_sidelabel1}  &  
\includegraphics[width=0.28\textwidth]{1D_Legend1}  & &  \\
& \includegraphics[trim={0 0 1.8cm 0}, width=2cm]{1D_sidelabel2}  &  
\includegraphics[width=0.28\textwidth]{1D_Legend2} & &  \\
&  &  &  &  \\
\end{tabular*}
\caption{The 1D \(1\,\sigma\) credible intervals for the 6 parameters of constrained scenario in which the \(D_{ii}\) splitting are proportional to the \(\Delta m / m_\chi\).
Colours denote the DM flavour (red, green and blue for up, charm and top DM respectively).
The two prior choices on $D_{ii}$ and the masses (log-uniform vs uniform) are shown as lines and shaded regions respectively, the modal average of the log-uniform prior choice is shown as a dot. The two mass splitting cases are contained in different panels.}
\label{Fig:1DcredibleregionsSplitting}
\end{sidewaystable*}

\begin{sidewaystable*}
\begin{tabular*}{0.75 \textwidth}{ c  c  c c c } 
 &  & $ \theta_{12} $ & $ \theta_{13}$ & $\theta_{23}$ \vspace{2mm} \\
& \includegraphics[trim={0 0 2cm 0}, width=2cm]{1D_sidelabel1}  &  
\includegraphics[width=0.28\textwidth]{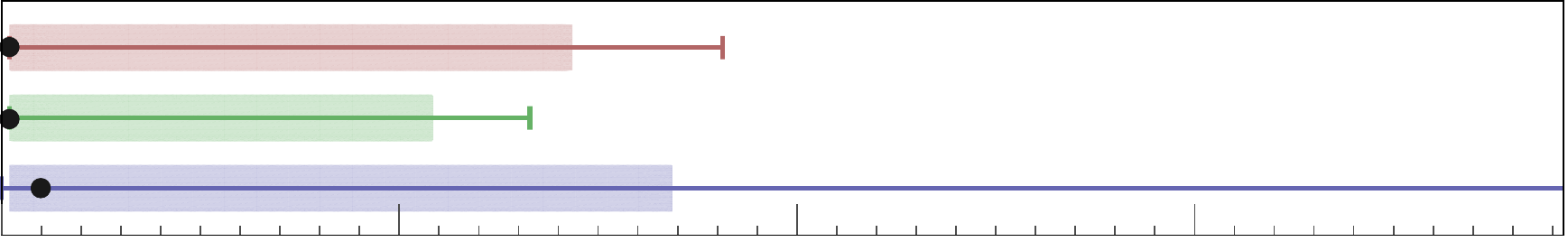} & \includegraphics[width=0.28\textwidth]{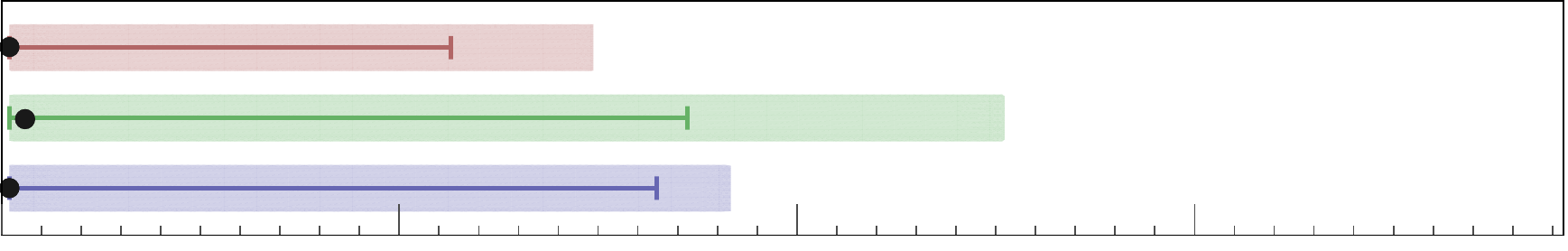}& \includegraphics[width=0.28\textwidth]{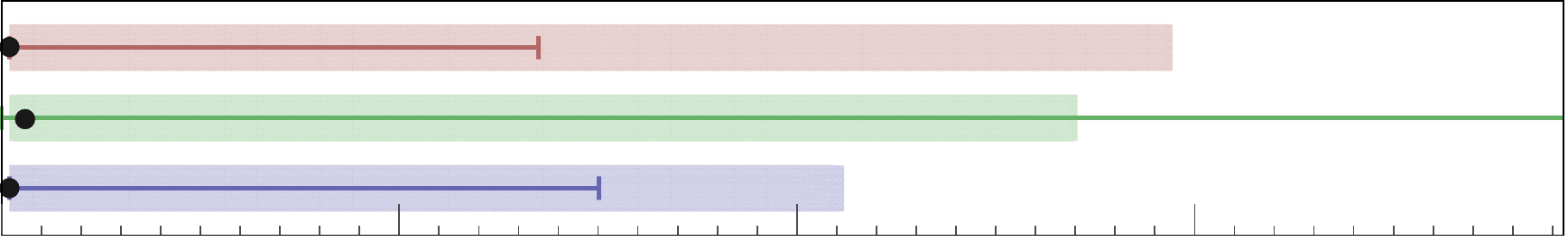}   \\
& \includegraphics[trim={0 0 1.8cm 0}, width=2cm]{1D_sidelabel2}  &  
\includegraphics[width=0.28\textwidth]{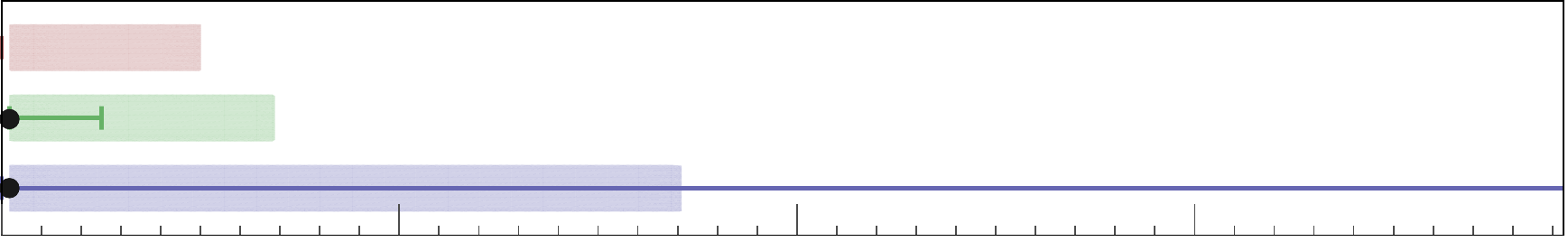} & \includegraphics[width=0.28\textwidth]{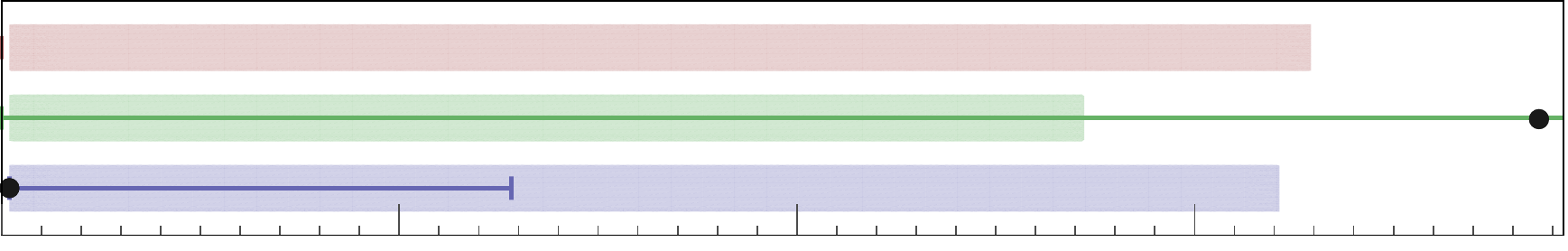}& \includegraphics[width=0.28\textwidth]{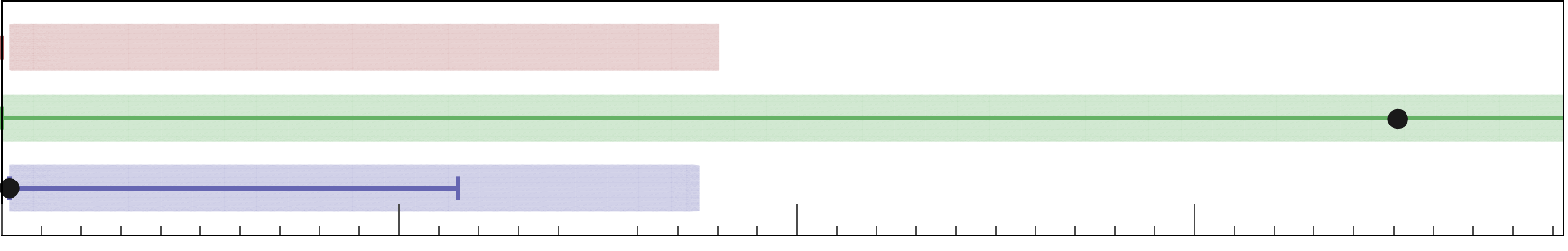} \\
&  & 
\includegraphics[width=0.29\textwidth ,trim={0 0 0 2cm}]{1D_axis_theta} & \includegraphics[width=0.29\textwidth ,trim={0 0 0 2cm}]{1D_axis_theta} & \includegraphics[width=0.29\textwidth ,trim={0 0 0 2cm}]{1D_axis_theta}  \\
 &  & $ D_{11} $ & $ D_{22} $ & $ D_{33}$ \vspace{2mm} \\
& \includegraphics[trim={0 0 2cm 0}, width=2cm]{1D_sidelabel1}  &  
\includegraphics[width=0.28\textwidth]{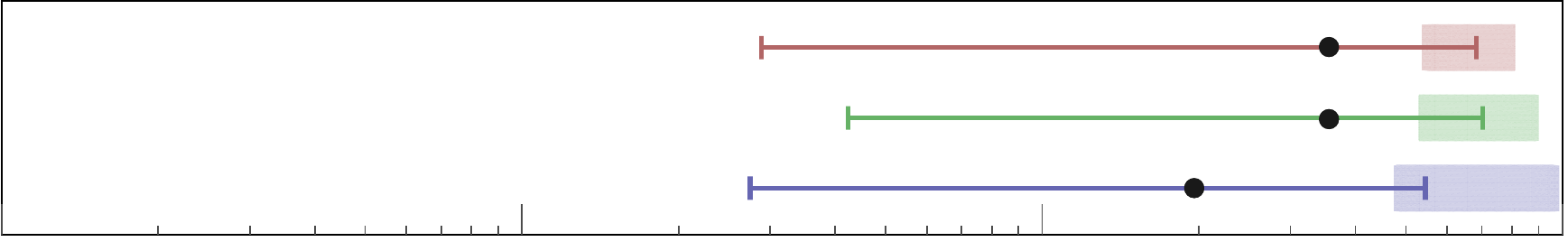} & \includegraphics[width=0.28\textwidth]{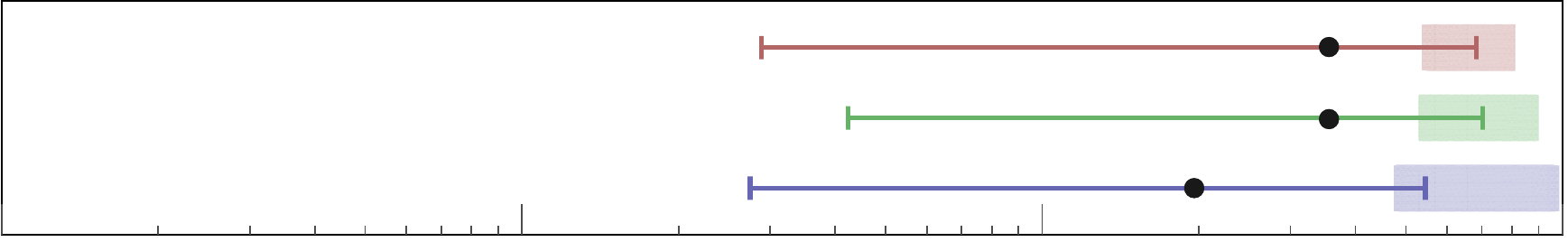}& \includegraphics[width=0.28\textwidth]{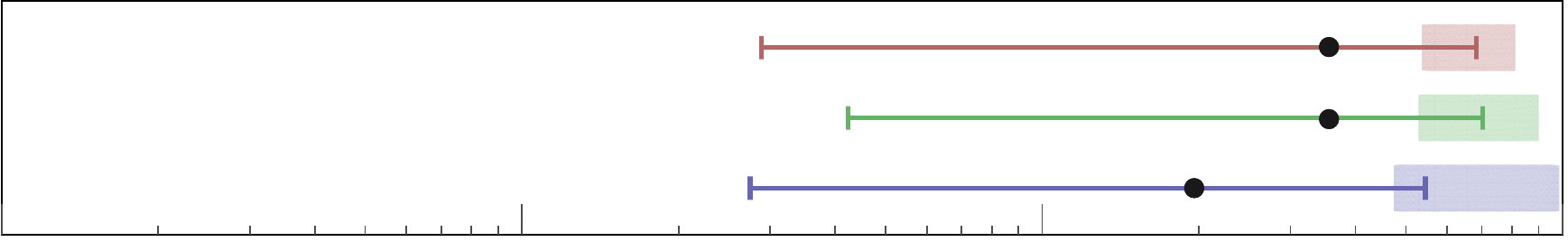}   \\
& \includegraphics[trim={0 0 1.8cm 0}, width=2cm]{1D_sidelabel2}  &   
\includegraphics[width=0.28\textwidth]{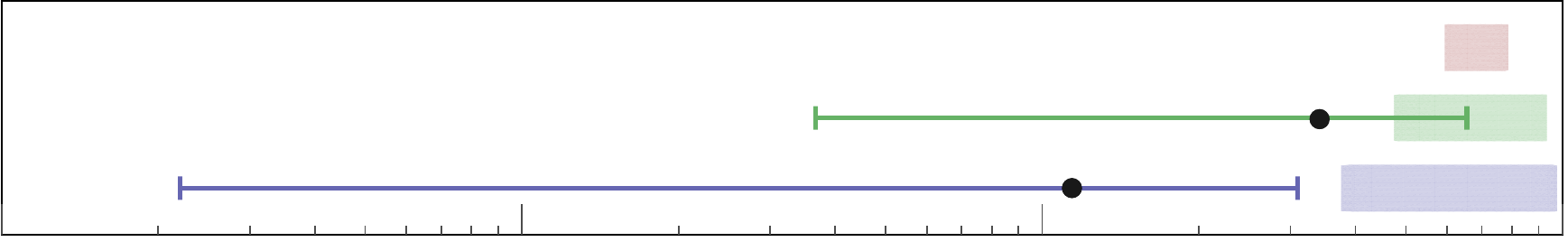} & \includegraphics[width=0.28\textwidth]{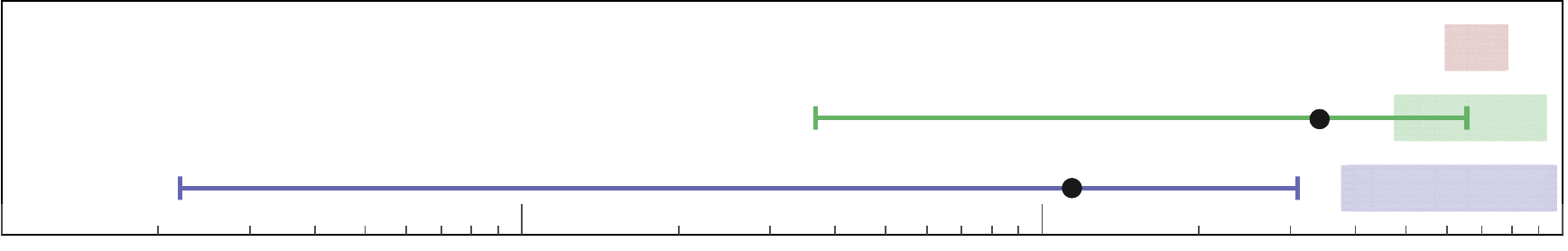}& \includegraphics[width=0.28\textwidth]{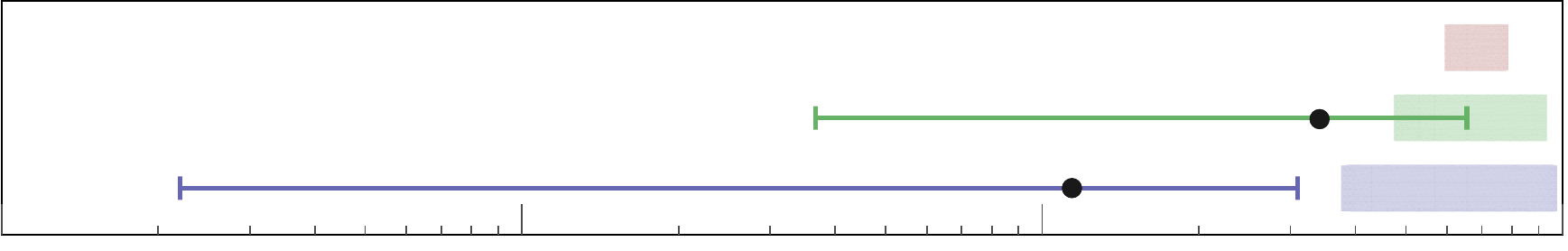} \\
&  & 
\includegraphics[width=0.29\textwidth ,trim={0 0 0 2cm}]{1D_axis_Dii} & \includegraphics[width=0.29\textwidth ,trim={0 0 0 2cm}]{1D_axis_Dii} & \includegraphics[width=0.29\textwidth ,trim={0 0 0 2cm}]{1D_axis_Dii}  \\
& & \(m_\chi\ [\si{\GeV}]\) & \(m_\phi\ [\si{\GeV}]\) &  \vspace{2mm} \\
& \includegraphics[trim={0 0 2cm 0}, width=2cm]{1D_sidelabel1}  &  
\includegraphics[width=0.28\textwidth]{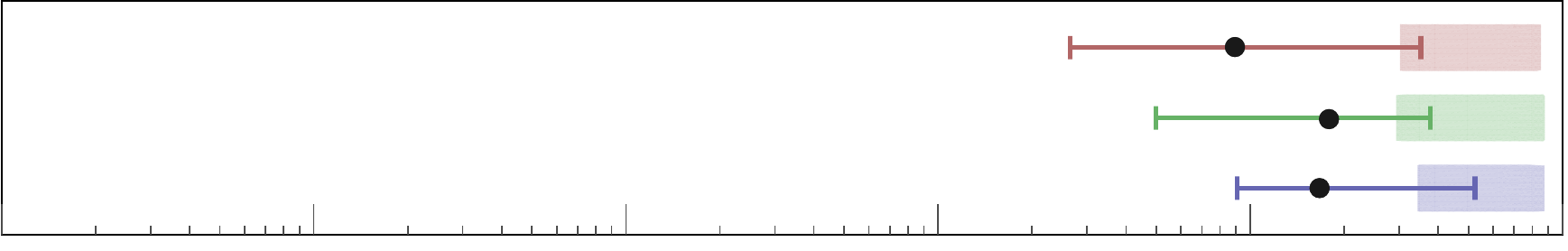} & \includegraphics[width=0.28\textwidth]{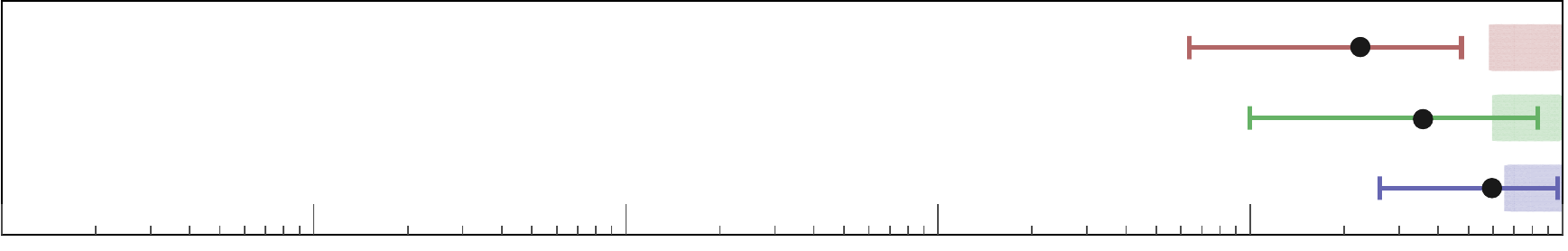}&
\includegraphics[width=0.28\textwidth]{1D_Legend1}   \\
& \includegraphics[trim={0 0 1.8cm 0}, width=2cm]{1D_sidelabel2}  &  
\includegraphics[width=0.28\textwidth]{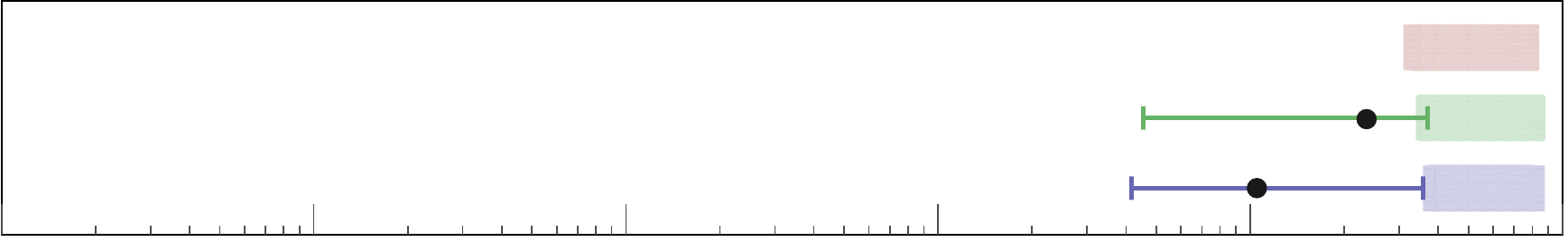} & \includegraphics[width=0.28\textwidth]{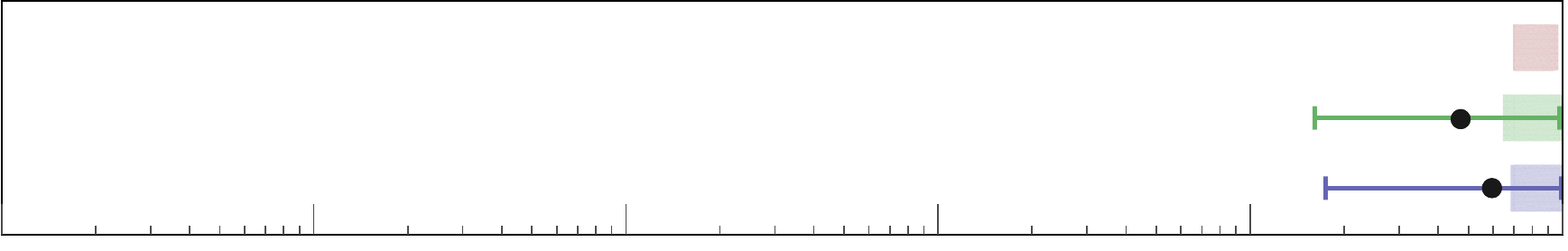}& 
\includegraphics[width=0.28\textwidth]{1D_Legend2}   \\
&  & 
\includegraphics[width=0.29\textwidth ,trim={0 0 0 2cm}]{1D_axis_m} & \includegraphics[width=0.29\textwidth ,trim={0 0 0 2cm}]{1D_axis_m} &  \\
\end{tabular*}
\caption{The 1D \(1\,\sigma\) credible intervals for a future scenario in which the SM prediction for the mixing matches the experimental value. Colours denote the DM flavour (red, green and blue for up, charm and top DM respectively). The two prior choices on $D_{ii}$ and the masses (log-uniform vs uniform) are shown as lines and shaded regions respectively, the modal average of the log-uniform prior choice is shown as a dot. The two mass splitting cases are contained in different panels.}
\label{Fig:1DcredibleregionsSMMixing}
\end{sidewaystable*}

\clearpage

\section{Conclusions}
\label{sec:Conclusions}

In this work, we have analysed a model of dark matter, based on \cite{Agrawal:2014aoa} but coupling to up type quarks, that goes beyond MFV in order to allow potentially large new effects in the flavour sector, and have seen how the combination of a wide range of constraints can be used to place limits on models of this type.
We approached this task of combining many different constraints using the MCMC tool Multinest, which allowed us to place limits on the high dimensional parameter space of our particular model.

As we can see from \cref{Fig:ResultsMass}, the MCMC places lower bounds on the new particle masses of at least \SI{1}{\TeV} for Top DM, and a few hundred \si{\GeV} for Up and Charm DM in certain cases.
Our collider bounds (\cref{Fig:ColliderBounds}) cannot further exclude Top DM, even in the case of strong couplings, but could remove a small area of allowed parameter space from the bottom end of the mass range in the case of Up/Charm DM.

Ref.~\cite{Blanke:2017tnb} considers this model, but examined the region of parameter space with dominant top quark couplings.
Our results in general agree with their conclusions if we look at their more focused parameter space.
For example, they find strong constraints on \(\theta_{12}\) except in the case of some degeneracy in the \(D_{ii}\), which we replicate.
Similarly the strong constraints on DM mass from relic density and direct detection are reproduced.
In their work, they explain how loop-level diagrams contributing to direct detection favour the dominant top coupling -- however as we explain in \cref{sec:dd}, RG effects mean even when DM doesn't couple to up quarks directly, the mixing is substantial enough to weaken this conclusion (as long as the mediator mass is large enough).

Given the current level of data, the model we examine of flavoured DM coupling to up-type quarks has large sections of its parameter space still allowed, so long as one considers large mass new particles.
However, even without the complimentary collider results, the lower mass, phenomenologically interesting, regions of parameter space are disfavoured by flavour, relic density, and direct detection considerations.

The MFV assumption is frequently invoked in simplified models in order to evade potentially large flavour-violating effects.
The level of robustness of this assumption varies considerably between up-type and down-type quark couplings in the DMFV model; for RH down-type quarks strong flavour bounds do ensure that the assumption is a good one.
However for couplings to RH up-type quarks we have seen that in fact the flavour bounds are avoided in a large region of MFV-breaking parameter space.

One particular future development could alter this picture however -- if a precise theoretical prediction of \PDzero mixing observables could be obtained then either (a) a significant discrepancy requiring new physics is present, or (b) the SM predictions are reproduced with a high precision.
The former would motivate the exploration of models which go beyond MFV, and the latter would make the MFV assumption a necessary assumption of the DMFV simplified model if one wants to avoid some fine-tuning.

\section*{Acknowledgements}
The authors would like to thank Richard Ruiz, Thomas Rauh, Olivier Mattelaer, Ulrich Haisch, and Thomas Hahn for useful discussions.
TJ and MK are supported by Durham University and STFC respectively.

\appendix

\section{Rare decays}
\label{app:rareDecays}
\begin{figure}[h]
\centering
\includegraphics[scale=0.5,trim={0 5cm 0 0},clip]{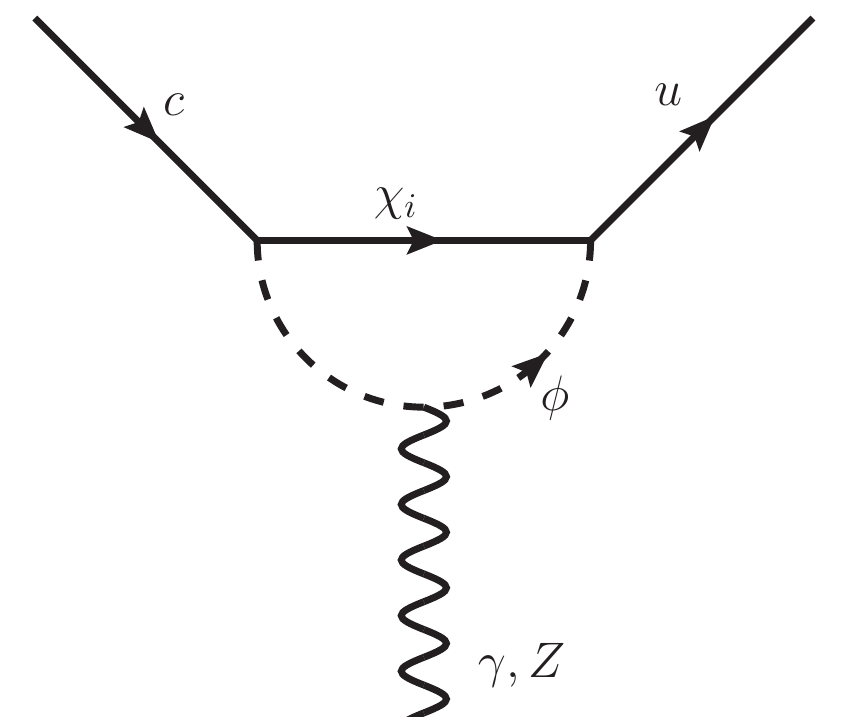}
\includegraphics[scale=0.5,trim={0 5cm 0 0},clip]{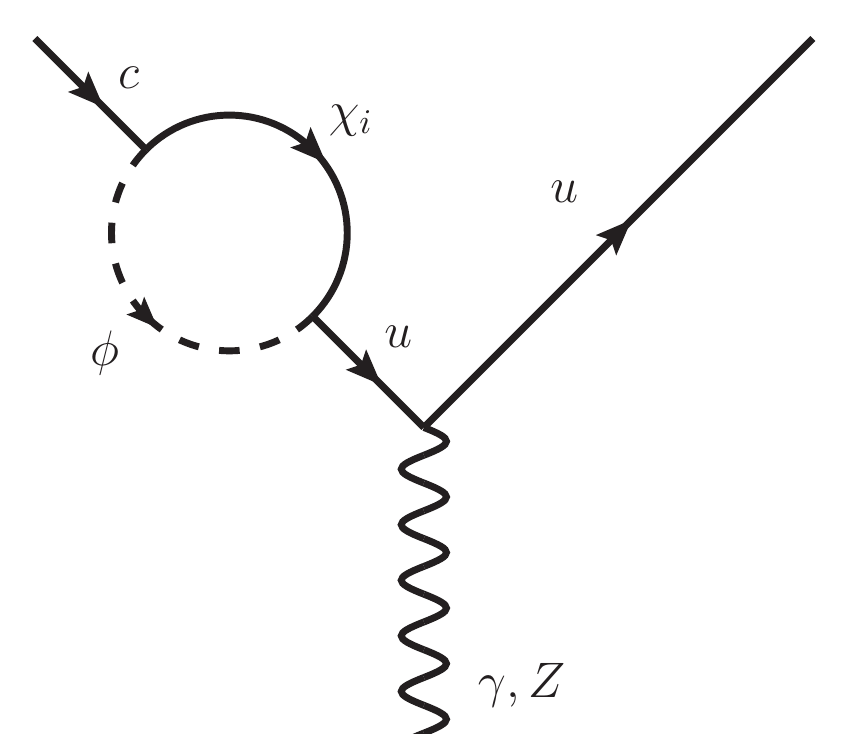}
\includegraphics[scale=0.5,trim={0 5cm 0 0},clip]{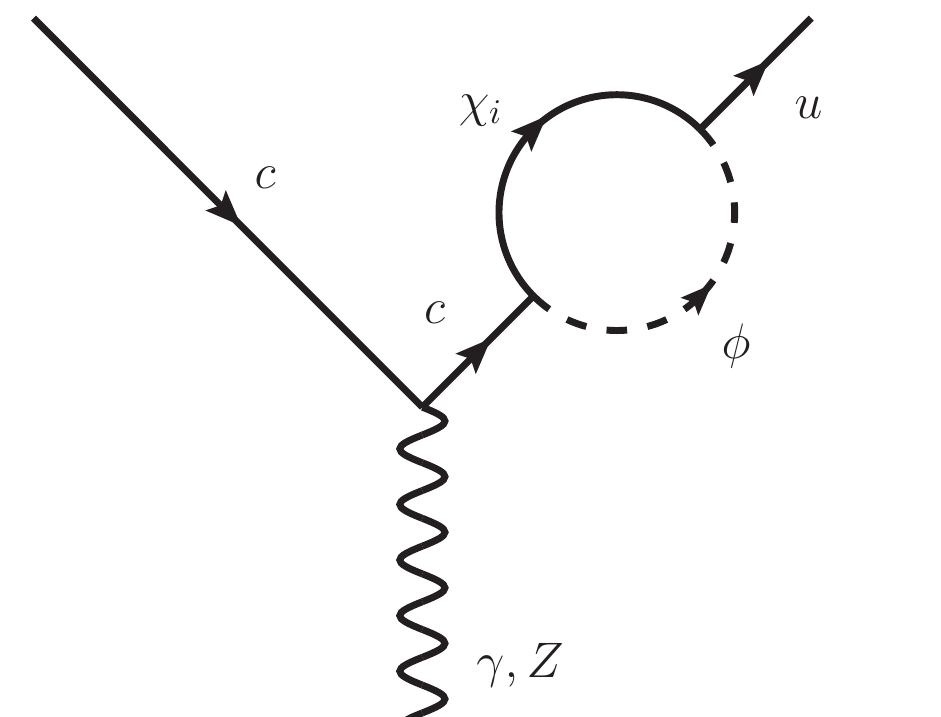}
\caption{The DMFV model contribution to the effective operators governing rare decays of charm mesons, including explicit self-energy corrections to the external quark legs as explained in the text. The $\gamma,Z$ couple to a lepton pair.}
\label{Fig:RareDecayDiagram}
\end{figure}

The non-zero Wilson coefficients arise from electroweak penguins (shown in \cref{Fig:RareDecayDiagram}), and neglecting \PZ penguins since the small momentum transfer means they amount to an \(\mathcal{O}(1\%)\) correction, we find
\begin{align}
C_7^\prime &= \sum_{i} \frac{\lambda_{1i} \lambda^*_{2i}}{6 \sqrt{2} G_F} \bigg[
	\begin{aligned}[t]
		&C_1(m_c^2, q^2, 0, m_{\chi_i}^2, \MphiDM^2, \MphiDM^2) + C_{11}(m_c^2, q^2, 0, m_{\chi_i}^2, \MphiDM^2, \MphiDM^2)
		\\
		+ &C_{12}(m_c^2, q^2, 0, m_{\chi_i}^2, \MphiDM^2, \MphiDM^2) \bigg] \ ,
	\end{aligned}
\\
C_9^\prime &= \sum_{i} \frac{\lambda_{1i} \lambda^*_{2i}}{3 \sqrt{2} G_F q^2} \bigg[
	\begin{aligned}[t]
		&B_1(m_c^2, m_{\chi_i}^2, \MphiDM^2) + 2 C_{00}(m_c^2, q^2, 0, m_{\chi_i}^2, \MphiDM^2, \MphiDM^2)
		\\
		+ &m_c^2 \Big\{
			\begin{aligned}[t]
				&C_1(m_c^2, q^2, 0, m_{\chi_i}^2, \MphiDM^2, \MphiDM^2) + C_{11}(m_c^2, q^2, 0, m_{\chi_i}^2, \MphiDM^2, \MphiDM^2)
				\\
				+ &C_{12}(m_c^2, q^2, 0, m_{\chi_i}^2, \MphiDM^2, \MphiDM^2) \Big\} \bigg] \ ,
			\end{aligned}
	\end{aligned}
\end{align}
where \(B\) and \(C\) are loop functions using LoopTools \cite{Hahn:1998yk} notation.

\section{Direct Detection}
\label{app:dd}

\subsection{LUX}
For situations where we have both a measured event count, \(N^\text{obs}_k\) (binned into energy bins labelled by \(k\)) and theoretical background \(N^\text{bck}_k\), we can use the \emph{likelihood ratio test}, a method based on a hypothesis test between a background only, and background+signal model, with likelihoods \(\mathcal{L} , \mathcal{L}_\text{bck}\) respectively \cite{DelNobile:2013sia}.

The likelihood of observing the data, \(D\), assuming a particular set of parameters \(\{ \lambda \}\), is denoted \(\mathcal{L}(D | \{ \lambda \})\).
The likelihood of each bin is a Poisson distribution $\text{Poiss}(N^\text{obs} , N^\text{th} (\lambda) )$ where $N^\text{th}_k$ are the predicted number of signal events (including background),
\begin{align}
\mathcal{L}(N^\text{obs} | \{ \lambda \}) = \prod_k \frac{\left( N^\text{th}_k \right)^{N^\text{obs}_k}}{N^\text{obs}_k !} \exp \left[ - N^\text{th}_k  \right]
\end{align}
where \(N^\text{th}(\lambda) = N^\text{DM}(\lambda) + N^\text{bck}\).
 The background only model is identical but with \(N^\text{th} = N^\text{bck}\).
 Then the test statistic,
\begin{align}
\text{TS}(\lambda) = -2 \log{\left( \frac{\mathcal{L}}{\mathcal{L}_\text{bck}}  \right)} \approx 2 \sum_k \left( N_k^\text{th} - N_k^\text{obs} \log{\left[ \frac{N_k^\text{th} + N_k^\text{bck}}{N_k^\text{bck}}  \right]}   \right) \ ,
\label{Eq:TestStatistic1}
\end{align}
follows a \(\chi^2\) distribution -- the cumulative probability density function of \(\chi^2(x)\) represents the probability that we observe the data given the model parameters \(\lambda\).
The value of \(x\) such that \(\chi^2(x) = C\) (i.e. the \(C\,\%\) confidence limit) depends on the number of parameters \(\{\lambda\}\) -- for only one parameter for example one can look up that \(\chi^2(2.71) = 0.9\), which means that the \SI{90}{\percent} confidence bounds on \(\lambda\) are given by \(\text{TS}(\lambda) = 2.71\).

\subsection{CDMSlite}
For CDMSlite, we use a conservative method based on the statement that the \SI{90}{\percent} confidence limit is such that \emph{there is a probability of 0.9 that if the model were true, then the experiment would have measured more events (\(n\)) than have been measured (\(n_\text{obs}\))}.
Using the Poisson distribution this probability is,
\begin{align}
P(n > n_\text{obs} | \mu ) = \sum_{n= n_\text{obs}}^\infty \frac{\mu^n}{n !} \exp(- \mu) \approx  \int_{n_\text{obs}}^\infty \frac{1}{\sqrt{2 \pi \mu}} \exp \left( - \frac{(t - \mu )^2}{2 \mu}  \right) dt = 0.9
\end{align}
and in the limit \(n_\text{obs} \gg 1\), this can be approximated by
\begin{align}
P(n > n_\text{obs} | \mu ) = \frac{1}{2} \left( \text{Erfc} \left( \frac{n_\text{obs} - \mu }{\sqrt{2 \mu}}  \right) \right) = 0.9 \ .
\end{align}
This equation is numerically solvable for \(\mu\) giving a required signal \(\mu = 109^{+51}_{-50} , 88 \pm 14, 635 \pm 37 \text{ and } 207 \pm 20\) events for energy bins 1 to 4 respectively.
This is conservative since a large portion of the measured events are background, and the resulting limits are slightly weaker than those given by the CDMSlite collaboration.

\newpage

\section{Feynman Diagrams for collider searches}
\label{App:FeynmanDiagrams}

\subsection{Monojet processes}

The dominant diagrams contributing to the pure monojet process. Each processes
scales as \(\sigma \propto (\lambda \lambda^\dagger) \alpha_s\) and can become extremely large for large \(\lambda \). The cross section is dominated by the diagrams containing a heavy \(\phi\) resonance.

\begin{figure}[hb]
\centering
\includegraphics[width=0.45\textwidth]{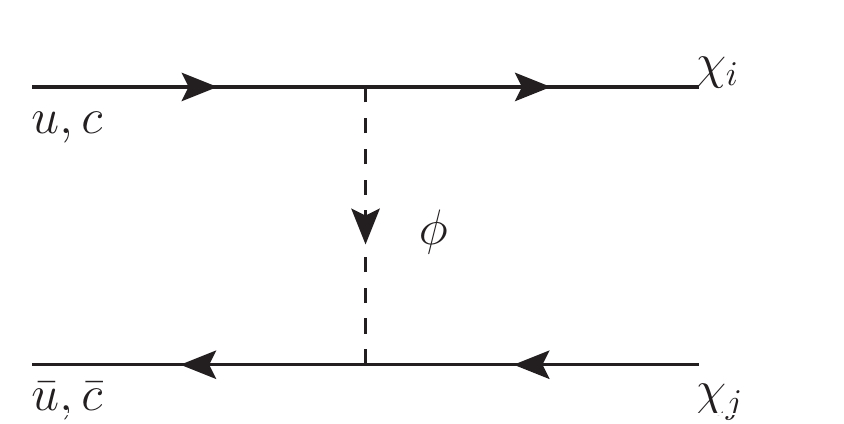}
\caption{The above diagram must include initial/final state radiation from external legs or
internal bremsstrahlung from the mediator. The contribution is roughly equal amongst these
emissions.}
\label{Fig:Monojet_diag1}
\end{figure}

\begin{figure}[hb]
\centering
\includegraphics[width=0.45\textwidth]{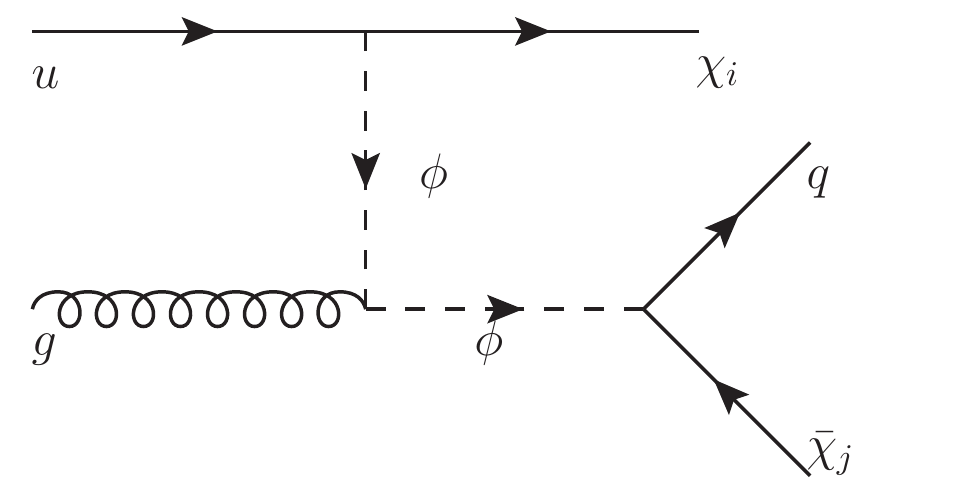}
\includegraphics[width=0.45\textwidth]{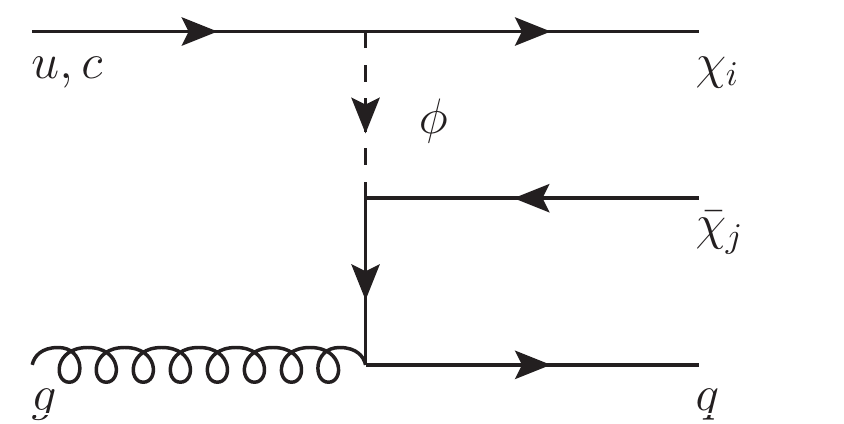}
\includegraphics[width=0.45\textwidth]{monojet_diagram_4}
\caption{The s-channel \(\phi\) resonance is responsible for (top left) and (bottom) dominating over (top right), and the additional enhancement due to the gluon pdf over \cref{Fig:Monojet_diag1} makes these the overall dominant monojet contribution. For very heavy mediators (top left) is suppressed due to the two propagators.}
\label{Fig:Monojet_diag234}
\end{figure}

\clearpage

\subsection{Dijet processes}

\begin{figure}[hb]
\centering
\includegraphics[width=0.35\textwidth]{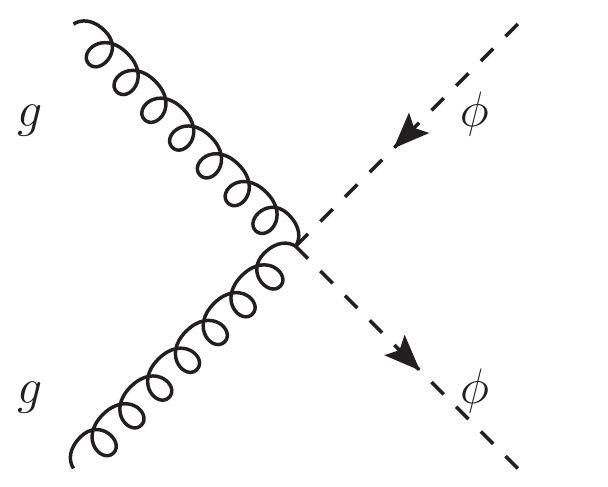}
\includegraphics[width=0.45\textwidth]{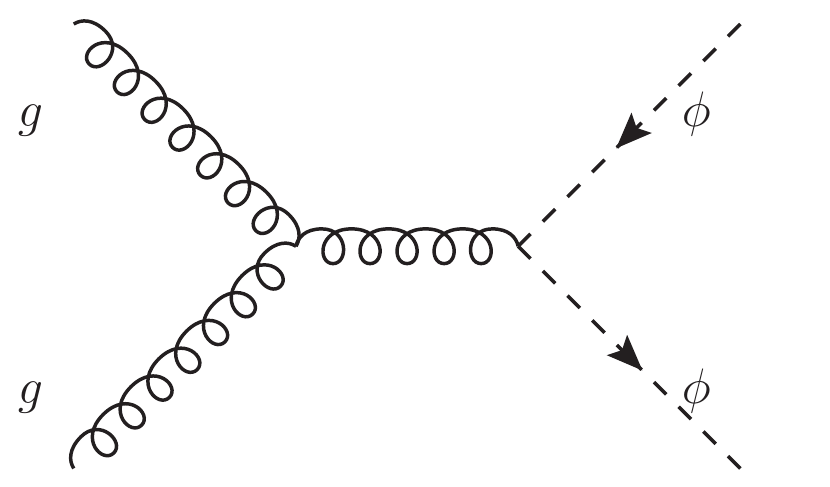}
\\
\includegraphics[width=0.45\textwidth]{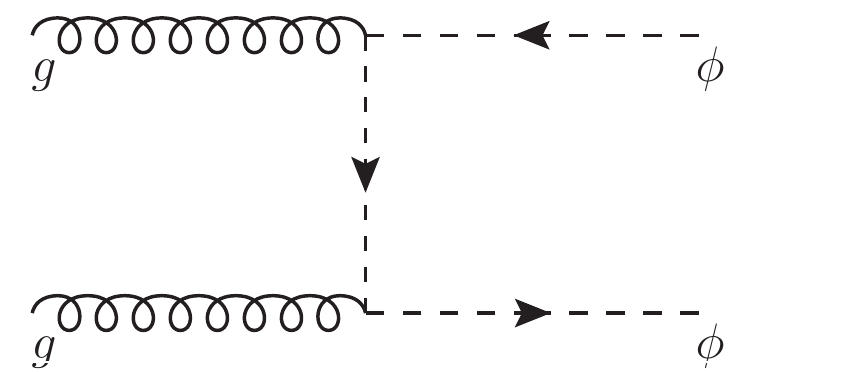}
\caption{Gluon fusion dijet processes \(\sigma \propto \alpha_s^2\)}
\label{Fig:Dijet_diag123}
\end{figure}

\begin{figure}[hb]
\centering
\includegraphics[width=0.45\textwidth]{dijet_4}
\includegraphics[width=0.45\textwidth]{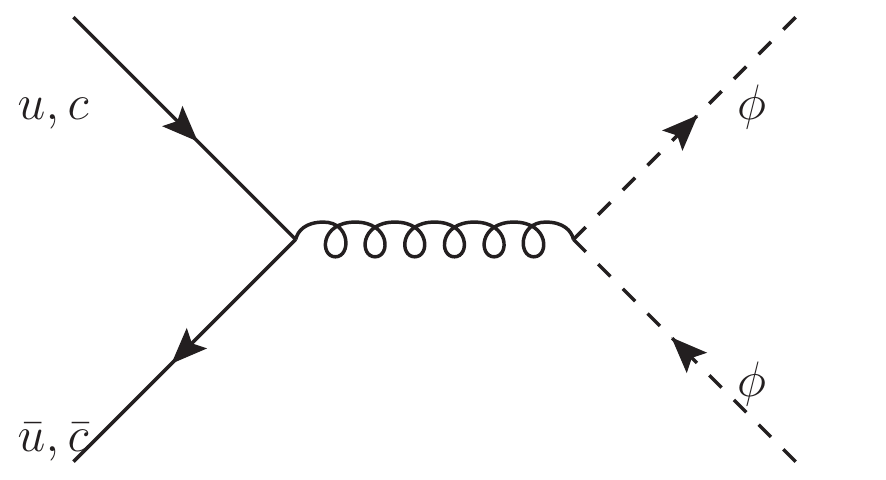}
\caption{The left (right) process has \(\sigma \propto (\lambda \lambda^\dagger)^2 (\alpha_s^2)\) and so the dominance depends on the size of the new couplings -- for couplings which are large enough to be excluded it is usually the left diagram which dominates.}
\label{Fig:Dijet_diag45}
\end{figure}

The dominant processes contributing to the production of on-shell \(\phi\), which decay \(\phi \to q_i \chi_j\) producing a dijet signal. In monojet analyses, this provides a subdominant contribution compared with pure monojet processes \cref{Fig:Monojet_diag1,Fig:Monojet_diag234} in most of the parameter space.

\clearpage

\bibliographystyle{utphys}
\bibliography{Charming_DM}

\end{document}